\documentclass[oneside,11pt]{article}

\usepackage{url,mathrsfs,algorithm}
\usepackage{amsmath,wrapfig,color,bigfoot}
\usepackage{amsfonts}
\usepackage{graphicx}
\usepackage{subfigure}
\newtheorem{theorem}{Theorem}

\begin{document}

\title{\Huge Binary and Multi-Bit Coding for Stable Random Projections}
\author{ \textbf{\Large Ping Li} \vspace{0.05in}\\
         Department of Statistics and Biostatistics\\% \ \ \
         Department of Computer Science\\
       Rutgers University\\
          Piscataway, NJ 08854, USA\\
       \texttt{pingli@stat.rutgers.edu}
}

\date{}
\maketitle

\begin{abstract}

\noindent We develop efficient binary (i.e., 1-bit) and multi-bit coding schemes for estimating the scale parameter of $\alpha$-stable distributions. The work is motivated by the recent work on \textbf{one scan 1-bit compressed sensing} (sparse signal recovery)~\cite{Report:Li_1bitCS15} using $\alpha$-stable random projections, which requires estimating of the scale parameter at  bits-level. Our technique can be naturally  applied to data stream computations for estimating the $\alpha$-th frequency moment. In fact, the method  applies to the general scale family of distributions, not limited to  $\alpha$-stable distributions. \\

\noindent Due to the heavy-tailed nature of $\alpha$-stable distributions, using traditional estimators will potentially need many bits to store each measurement in order to ensure sufficient accuracy. Interestingly, our paper demonstrates that, using a  simple closed-form estimator with merely 1-bit information does not result in a significant loss of accuracy if the parameter is chosen appropriately. For example, when $\alpha=0+$, 1, and 2, the  coefficients of the optimal estimation variances using full (i.e., infinite-bit) information are 1, 2, and 2, respectively. With the 1-bit scheme and appropriately chosen parameters, the corresponding variance coefficients are 1.544, $\pi^2/4$, and 3.066, respectively.  Theoretical tail bounds are  also provided. Using 2 or more bits  per measurements reduces the estimation variance and  importantly, stabilizes the estimate so that  the  variance is not sensitive to parameters. With look-up tables, the computational cost    is minimal.\\

\noindent Extensive simulations are conducted to verify the theoretical results. The estimation procedure is  integrated into the sparse recovery with one scan 1-bit compressed sensing. One interesting observation is that the classical ``Bartlett correction'' (for MLE bias correction) appears particularly effective for our problem when the sample size (number of measurements) is  small.

\end{abstract}

%\newpage

\section{Introduction}

The research problem of interest is  about efficient estimation of the scale parameter of the $\alpha$-stable distribution using binary (i.e., 1-bit) and multi-bit coding  of the samples. That is, given $n$ i.i.d. samples,
\begin{align}
y_j\sim S(\alpha,\Lambda_\alpha), \hspace{0.5in} j=1, 2, ...,  n
\end{align}
from an $\alpha$-stable distribution $S(\alpha,\Lambda_\alpha)$, we hope to estimate the scale parameter $\Lambda_\alpha$ by using only 1-bit or multi-bit information of $|y_j|$.  Here we adopt the parameterization~\cite{Book:Zolotarev_86,Book:Samorodnitsky_94} such that, if $y \sim S(\alpha,\Lambda_\alpha)$, then the characteristic function is $E\left(e^{\sqrt{-1}yt}\right) = e^{-\Lambda_\alpha|t|^\alpha} $. Note that, under this parameterization, when $\alpha=2$, $S(2,\Lambda_2)$ is equivalent to a Gaussian distribution $N(0,\sigma^2=2\Lambda_2)$.  When $\alpha=1$, $S(1,1)$ is the standard Cauchy distribution.

\subsection{Sampling from $\alpha$-stable Distribution}

Although in general there is no closed-form  density  of $S(\alpha,1)$, we can sample from the distribution using a standard procedure provided by ~\cite{Article:Chambers_JASA76}. That is, one can first sample an exponential $w\sim exp(1)$ and a uninform  $u\sim unif(-\pi/2,\pi/2)$ , and then compute
\begin{align}\label{eqn_stable_sample}
s_\alpha = \frac{\sin(\alpha u)}{(\cos u)^{1/\alpha}}
\Big[\frac{\cos(u-\alpha u)}{w}\Big]^{(1-\alpha)/\alpha}\sim S(\alpha,1)
\end{align}
This paper will heavily use the distribution of $|s_\alpha|^\alpha$:
\begin{align}
|s_\alpha|^\alpha = \frac{\left|\sin(\alpha u)\right|^\alpha}{\cos u}
\Big[\frac{\cos(u-\alpha u)}{w}\Big]^{(1-\alpha)}
\end{align}
Intuitively, as $\alpha\rightarrow0$, $1/|s_\alpha|^\alpha$ converges to $exp(1)$ in distribution as formally established by~\cite{Article:Cressie_75}.\\

The use of $\alpha$-stable distributions~\cite{Article:Indyk_JACM06,Proc:Li_SODA08} was studied in the context of estimating frequency moments of data streams~\cite{Proc:Alon_STOC96,Article:Muthukrishnan_05}. The use $\alpha$-stable random projections for sparse signal recovery was established in (e.g.,)~\cite{Proc:CCCS_COLT14}, by using full (i.e., infinite-bit) information of the measurements.  In this paper, the development of binary (1-bit) and multi-bit coding schemes is also motivated by the work recent work on ``one scan 1-bit compressed sensing''~\cite{Report:Li_1bitCS15}.

\subsection{One Scan 1-Bit Compressed Sensing}

In contrast to classical compressed sensing (CS)~\cite{Article:Donoho_CS_JIT06,Article:Candes_Robust_JIT06} and 1-bit compressed sensing~\cite{Proc:Boufounos08,Article:1Bit_IT13,Article:Plan_IT13,Proc:Slawski_NIPS15}, there is a recent line of work on sparse signal recovery based on heavy-tailed designs~\cite{Proc:CCCS_COLT14,Report:Li_1bitCS15}. The main algorithm of ``one scan 1-bit compressed sensing''~\cite{Report:Li_1bitCS15} is summarized in Algorithm~\ref{alg_recovery}. Given $n$ measurements $y_j = \sum_{i=1}^N x_i s_{ij}$, $j=1$ to $n$, where $s_{ij}\sim S(\alpha,1)$ i.i.d. and $x_i$, $i=1$ to $N$, is  a sparse (and possibly dynamic/streaming) vector,  the task is to recover $x$ from only the signs of the measurements, i.e., $sign(y_j)$.  Algorithm~\ref{alg_recovery} provides a simple  recipe for recovering $x$ from $sign(y_j)$ by scanning the coordinates of the vector only once.\\

\begin{algorithm}{\footnotesize
\textbf{Input:} $K$-sparse signal $\mathbf{x}\in\mathbb{R}^{1\times N}$, design matrix $\mathbf{S}\in\mathbb{R}^{N\times M}$ with entries sampled from $S(\alpha,1)$ with small $\alpha$ (e.g., $\alpha=0.05$). We  sample  $u_{ij} \sim uniform(-\pi/2,\ \pi/2)$ and  $w_{ij}\sim exp(1)$ and  compute $s_{ij}$ by (\ref{eqn_stable_sample}).

%\vspace{0.1in}

\textbf{Collect:} Linear measurements:\  $y_j = \sum_{i=1}^N x_i s_{ij}$, $j = 1$ to $M$.

%\vspace{0.1in}

\textbf{Compute:}  For each coordinate $i =1$ to $N$, compute
{\small\begin{align}\notag
&Q_i^{+} = \sum_{j=1}^M\log \left(1+{sgn(y_j)}{sgn(u_{ij})}e^{-\left({K}-1\right)w_{ij}}\right),\hspace{0.2in} Q_i^{-}  = \sum_{j=1}^M\log \left(1-{sgn(y_j)}{sgn(u_{ij})}e^{-\left({K}-1\right)w_{ij}}\right)
\end{align} }

\textbf{Output:} For  $i=1$ to $N$,  report the estimated sign:
$\hat{sgn(x_i)} = \left\{
\begin{array}{ll}
+ 1 & \text{if } Q_i^+>0 \\
- 1 & \text{if } Q_i^->0\\
0 & \text{if } Q_i^+ <0 \text{ and } Q_i^-<0
\end{array}
\right.$

}\caption{\footnotesize Stable measurement collection and the one scan 1-bit algorithm for sign recovery. }
\label{alg_recovery}
\end{algorithm}

 This  efficient recovery procedure, however, requires the knowledge of ``$K$'', which is the $l_\alpha$ norm $\sum_{i=1}^N |x_i|^\alpha$ as $\alpha\rightarrow0+$.  In practice, this $K$ will typically have to estimated  and  the hope is that we do not have to use too many additional measurements just for the task of estimating $K$.  In this paper, we  will  elaborate that only 1 bit or a few bits per measurement can provide accurate estimates of $K$ (as well as the general term $\sum_{i=1}^N |x_i|^\alpha$ for $0<\alpha\leq2$).

Because the samples $y_j$ are heavy-tailed, using traditional estimators, the storage requirement for each sample can be substantial, which consequently would cause  issues in data retrieval, transmission and decoding. It is thus  very  desirable if we just need   1 bit or a few bits for each $|y_j|$.

\section{Estimation of $\Lambda_\alpha$ Using Full (Infinite-Bit) Information}

Given $n$ i.i.d. samples $y_j \sim S(\alpha,\Lambda_\alpha)$, $j=1$ to $n$,  we review various estimators of the scale parameter $\Lambda_\alpha$ using full information (i.e., infinite-bit).  When $\alpha=2$ (i.e., Gaussian),  the  {\em arithmetic mean} estimator is  statistically  optimal, i.e., the (asymptotic) variance reaches the reciprocal of the Fisher Information from classical statistics theory:
\begin{align}
\hat{\Lambda}_{2,f} = \frac{1}{n}\sum_{j=1}^n |y_j|^2,\hspace{0.5in} Var\left(\hat{\Lambda}_{2,f}\right) = \frac{\Lambda^2_2}{n}2
\end{align}
When $\alpha=1$, the MLE $\hat{\Lambda}_{1,f}$ is the solution to the equation
\begin{align}
\sum_{j=1}^n \frac{\hat{\Lambda}_{1,f}^2}{\hat{\Lambda}_{1,f}^2+y_j^2 }= \frac{n}{2},\hspace{0.2in} Var\left(\hat{\Lambda}_{1,f}\right) = \frac{\Lambda_1^2}{n}2 + O\left(\frac{1}{n^2}\right)
\end{align}
The  {\em harmonic mean} estimator~\cite{Proc:Li_SODA08} is suitable for small $\alpha$ and becomes optimal as $\alpha\rightarrow0+$:
\begin{align}\label{eqn_est_hm}
&\hat{\Lambda}_{\alpha,f,hm}
=\frac{-\frac{2}{\pi}\Gamma(-\alpha)\sin\left(\frac{\pi}{2}\alpha\right)}{\sum_{j=1}^n|y_j|^{-\alpha}}\left(n- \left(\frac{-\pi\Gamma(-2\alpha)\sin\left(\pi\alpha\right)}{\left[\Gamma(-\alpha)\sin\left(\frac{\pi}{2}\alpha\right)\right]^2}-1
\right)\right)\\
&{Var}\left(\hat{\Lambda}_{\alpha,f,hm} \right)
=\frac{\Lambda_\alpha^2}{n}\left(\frac{-\pi\Gamma(-2\alpha)\sin\left(\pi\alpha\right)}{\left[\Gamma(-\alpha)\sin\left(\frac{\pi}{2}\alpha\right)\right]^2}-1
\right)+ O\left(\frac{1}{n^2}\right)
\end{align}
where $\Gamma(.)$ is the gamma function. When $\alpha\rightarrow0+$, the variance becomes $\frac{\Lambda_{0+}^2}{n}+ O\left(\frac{1}{n^2}\right)$.

In summary, the optimal variances for $\alpha=0+$, 1, and 2, are respectively
\begin{align}\label{eqn_optVar}
\frac{\Lambda^2_{0+}}{n}1,\hspace{0.5in} \frac{\Lambda^2_{1}}{n}2,\hspace{0.3in}\text{and } \ \  \ \  \frac{\Lambda^2_{2}}{n}2
\end{align}
Our goal is to develop 1-bit and multi-bit schemes to achieve variances which are close to be optimal.

\section{1-Bit Coding and Estimation}

Again, consider $n$ i.i.d. samples $y_j \sim S(\alpha,\Lambda_\alpha)$, $j = 1$ to $n$. In this section, the task is to estimate $\Lambda_\alpha$ using just one bit information of each $|y_j|$, with  a pre-determined threshold. To accomplish this, we consider a threshold  $C$ (which can be a function of $\alpha$)  and compare it with $|y_j|^\alpha$,  $j = 1, 2, ..., n$.  In other word, we store a ``0'' if $|y_j|^\alpha\leq C$ and a ``1'' if $|y_j|^\alpha>C$. Note that we can express $|y_j|^\alpha$ as
\begin{align}\notag
|y_j|^\alpha  \sim  \Lambda_\alpha \left|s_\alpha\right|^\alpha,\hspace{0.2in} s_\alpha \sim S(\alpha,1).
\end{align}

Let $f_\alpha$ and $F_\alpha$ be the pdf and cdf of $|s_\alpha|^\alpha$, respectively. Then we can define $p_1$ and $p_2$ as follows
\begin{align}\notag
&p_1 = \mathbf{Pr}\left(z_\alpha \leq C\right) = F_\alpha\left(C/\Lambda_\alpha\right),\\\notag
&p_2 = \mathbf{Pr}\left(z_\alpha >C\right) = 1- p_1 = 1-F_\alpha\left(C/\Lambda_\alpha\right)
\end{align}
which are needed  for computing the likelihood. Denote
\begin{align}\notag
n_1 = \sum_{j=1}^n 1\{z_j\leq C\},\hspace{0.3in} n_2 = \sum_{j=1}^n 1\{z_j> C\}
\end{align}
The log-likelihood of the $n = n_1+n_2$ observations is
\begin{align}\notag
l =& n_1\log p_1 + n_2 \log p_2 =  n_1\log F_\alpha\left(C/\Lambda_\alpha\right) + n_2\log \left[1-F_\alpha\left(C/\Lambda_\alpha\right) \right]
\end{align}
To seek the MLE (maximum likelihood estimator) of $\Lambda_\alpha$, we need to compute the first derivative $l^\prime = \frac{\partial l}{\partial \Lambda_\alpha}$:
\begin{align}\notag
&l^\prime
 =  n_1\frac{f_\alpha\left(C/\Lambda_\alpha\right)}{F_\alpha\left(C/\Lambda_\alpha\right)}\left(-\frac{C}{\Lambda_\alpha^2}\right) +
n_2\frac{-f_\alpha\left(C/\Lambda_\alpha\right)}{1-F_\alpha\left(C/\Lambda_\alpha\right)}\left(-\frac{C}{\Lambda_\alpha^2}\right)
\end{align}
Setting $l^\prime =0$ yields the MLE solution denoted by $\hat{\Lambda}_\alpha$:
\begin{align}\notag
F_\alpha^{-1}\left(n_1/n\right)  =  C/\Lambda_\alpha\Longrightarrow  \hat{\Lambda}_{\alpha} = C / F_\alpha^{-1}\left(n_1/n\right)
\end{align}
To assess the estimation variance of $\hat{\Lambda}_\alpha$, we resort to  classical  theory of Fisher Information, which says
\begin{align}\notag
Var\left(\hat{\Lambda}_\alpha\right) = \frac{1}{-E\left(l^{\prime\prime}\right)} + O\left(\frac{1}{n^2}\right)
\end{align}
After some algebra, we obtain
\begin{align}\notag
E\left(l^{\prime\prime}\right)=-n\frac{C^2}{\Lambda_\alpha^4}\frac{f_\alpha^2}{F_\alpha(1-F_\alpha)}
\end{align}

For convenience, we introduce $\eta = \frac{\Lambda_\alpha}{C}$, and we summarize the above results in Theorem~\ref{thm_1bit}, which also provides the exact expression of the $O\left(\frac{1}{n}\right)$ bias term using classical statistics results~\cite{Article:Bartlett_53,Article:Shenton_63}.

\begin{theorem}\label{thm_1bit}
Given $n$ i.i.d. samples $y_j\sim S(\alpha,\Lambda_\alpha)$, \ $j=1$ to $n$,  a threshold $C$, and  $n_1 = \sum_{j=1}^n 1\{z_j\leq C\}$, the maximum  likelihood estimator (MLE) of $\Lambda_\alpha$ is
\begin{align}
\hat{\Lambda}_{\alpha} = C / F_\alpha^{-1}\left(n_1/n\right)
\end{align}
Denote $\eta = \frac{\Lambda_\alpha}{C}$.  The asymptotic bias of $\hat{\Lambda}_{\alpha}$ is
\begin{align}\label{eqn_Bias}
E\left(\hat{\Lambda}_\alpha\right) = \Lambda_\alpha +\frac{\Lambda_\alpha}{n}\frac{n_1}{n}\left(1-\frac{n_1}{n}\right)\left(\frac{\eta^2}{f_\alpha^2(1/\eta)} + \frac{\eta f^\prime_\alpha(1/\eta)}{2f^3_\alpha(1/\eta)}\right)+ O\left(\frac{1}{n^2}\right)
\end{align}
and the asymptotic variance of $\hat{\Lambda}_{\alpha}$ is
\begin{align}
&Var\left(\hat{\Lambda}_\alpha\right) = \frac{\Lambda^2_\alpha}{n}V_\alpha\left(\eta\right) +O\left(\frac{1}{n^2}\right)
 \end{align}
where
\begin{align}\label{eqn_Var}
V_\alpha\left(\eta\right) =\eta^2 \frac{F_\alpha(1/\eta)(1-F_\alpha(1/\eta))}{f_\alpha^2(1/\eta)},
\end{align}
where $f_\alpha$ and  $F_\alpha$ are the pdf and  cdf of $|S(\alpha,1)|^\alpha$, respectively, and $f^\prime_\alpha(z) = \frac{\partial f_\alpha(z)}{\partial z}$. \\

\noindent\textbf{Proof}:\hspace{0.2in} See Appendix~\ref{proof_thm_1bit}.$\hfill\Box$
\end{theorem}

%Next, we explicitly and separately compute $\hat{\Lambda}_\alpha$ and the variance for $\alpha=0+$, 1, and 2.

\vspace{-0.in}
\subsection{$\alpha\rightarrow0+$}
\vspace{-0.in}

As $\alpha\rightarrow0+$, we have $1/[s_\alpha]^\alpha \sim exp(1)$. Thus
\begin{align}\notag
F_{0+}(z) = e^{-1/z},\hspace{0.2in}  f_{0+}(z) = \frac{1}{z^2}e^{-1/z},\hspace{0.2in} F^{-1}_{0+}(z) = \frac{1}{\log 1/z}
\end{align}
We can then derive the estimator and its variance as
\begin{align}\notag
&\hat{\Lambda}_{0+} =  \frac{C}{F_{0+}^{-1}\left(n_1/n\right)} = {C}{\log n/n_1},\hspace{0.2in}
Var\left(\hat{\Lambda}_{0+}\right) = \frac{\Lambda^2_{0+}}{n}V_{0+}(\eta) + O\left(\frac{1}{n^2}\right)
\end{align}
where
\begin{align}\notag
V_{0+}\left(\eta\right) =&\eta^2 \frac{F_\alpha(1/\eta)(1-F_\alpha(1/\eta))}{f_\alpha^2(1/\eta)}
= \frac{e^{-\eta}-e^{-2\eta}}{\eta^2 e^{-2\eta}}=\frac{e^{\eta}-1}{\eta^2}
\end{align}
The minimum $V_{0+}\left(\eta\right)$ is 1.544, attained at $\eta = 1.594$. (In this paper, we  keep 3 decimal places.)

\vspace{-0.in}
\subsection{$\alpha=1$}
\vspace{-0.in}

By properties of Cauchy distribution, we know
\begin{align}\notag
F_1(z) = \frac{2}{\pi}\tan^{-1}z,\hspace{0.2in} f_1(z) = \frac{2}{\pi}\frac{1}{1+z^2},\hspace{0.2in} F_1^{-1}(z) = \tan\frac{\pi}{2}z
\end{align}
Thus, we can derive the estimator and variance
\begin{align}\notag
\hat{\Lambda}_1 = \frac{C}{\tan\frac{\pi}{2}\frac{n_1}{n}},\hspace{0.2in} Var\left(\hat{\Lambda}_{1}\right) = \frac{\Lambda^2_{1}}{n}V_1(\eta) + O\left(\frac{1}{n^2}\right)
\end{align}
The minimum of $V_1(\eta)$ is $\frac{\pi^2}{4}$, attained at $\eta =1$. To see this, let $t = 1/\eta$. Then $V_{1}\left(\eta\right) = \frac{1}{t^2} \frac{F_1(t)(1-F_1(t))}{f_1^2(t)}$ and
\begin{align}\notag
\frac{\partial \log V_{1}\left(\eta\right)}{\partial t} =& -\frac{2}{t} + \frac{f_1(t)}{F_1(t)} +\frac{-f_1(t)}{1-F_1(t)} - 2\frac{f^\prime_1(t)}{f_1(t)}\\\notag
=&-\frac{2}{t} +  \frac{4t}{1+t^2}+\frac{\frac{1}{1+t^2}}{\tan^{-1}t} -\frac{\frac{2}{\pi}\frac{1}{1+t^2}}{1-\frac{2}{\pi}\tan^{-1}t} \\\notag
=&\frac{1}{1+t^2}\left[t^2-1 + \frac{1}{\tan^{-1}t} -\frac{1}{\frac{\pi}{2}-\tan^{-1}t} \right]
\end{align}
Setting $\frac{\partial \log V_{1}\left(\eta\right)}{\partial t}=0$, the solution is $t=1$. Hence the optimum is attained at $\eta = 1$.

\subsection{$\alpha=2$}

Since $S(2,1)\sim \sqrt{2}\times N(0,1)$, i.e.,  $|s_\alpha|^2\sim 2 \chi^2_1$, we have
\begin{align}\notag
F_2(z) = F_{\chi^2_1}(z/2),\hspace{0.2in} f_2(z) = f_{\chi^2_1}(z/2)/2,
\end{align}
 where $F_{\chi^2_1}$ and $f_{\chi^2_1}$ are the cdf and pdf of a chi-square distribution with 1 degree of freedom, respectively. The MLE is $\hat{\Lambda}_2 = \frac{C}{F_2^{-1}(n_1/n)}$  and the optimal variance of $\hat{\Lambda}_2$ is $\frac{\Lambda_2^2}{n}3.066$, attained at $\eta=\frac{\Lambda_2}{C} = 0.228$.

\subsection{General $0<\alpha\leq2$}

\begin{figure}[h!]
\begin{center}
\mbox{\includegraphics[width=2.2in]{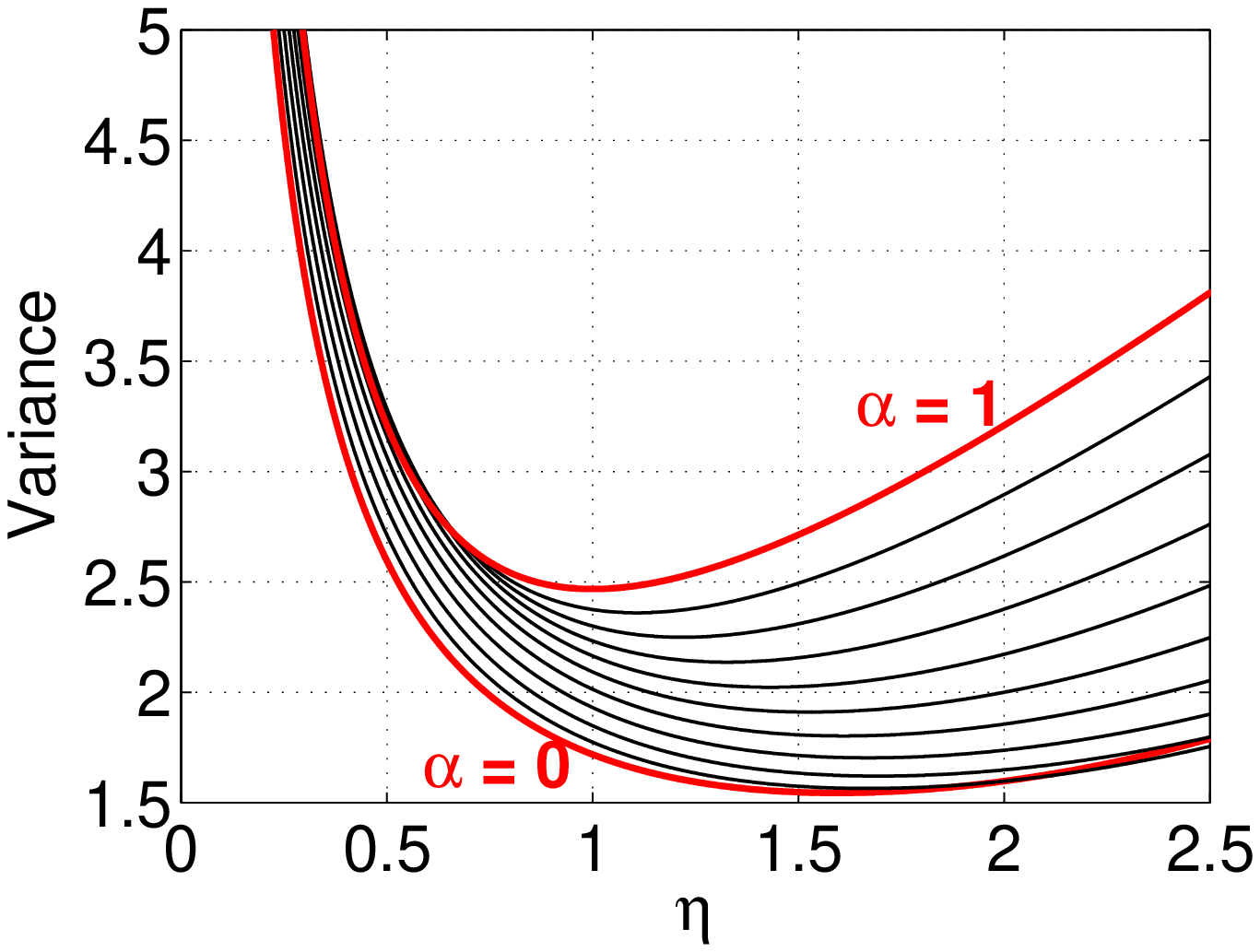}\hspace{0.2in}
\includegraphics[width=2.2in]{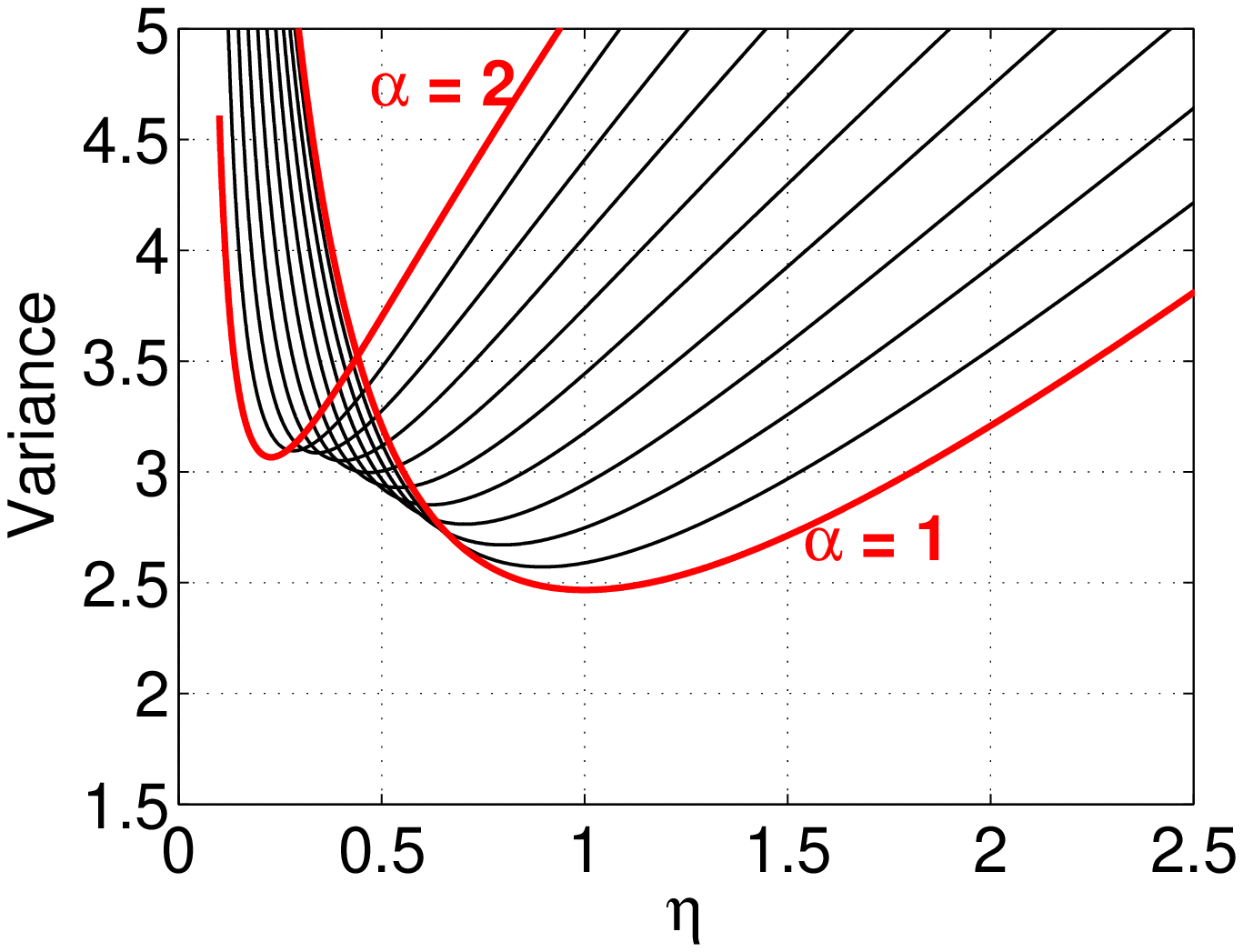}
}
\end{center}
\vspace{-0.3in}
\caption{The variance factor $V_\alpha(\eta)$  in (\ref{eqn_Var}) for $\alpha\in[0,\ 2]$, spaced at 0.1. The lowest point on each curve  corresponds to the optimal variance at that $\alpha$ value. }\label{fig_Var}
\end{figure}

\begin{figure}[h!]
\begin{center}
\mbox{\includegraphics[width=2.2in]{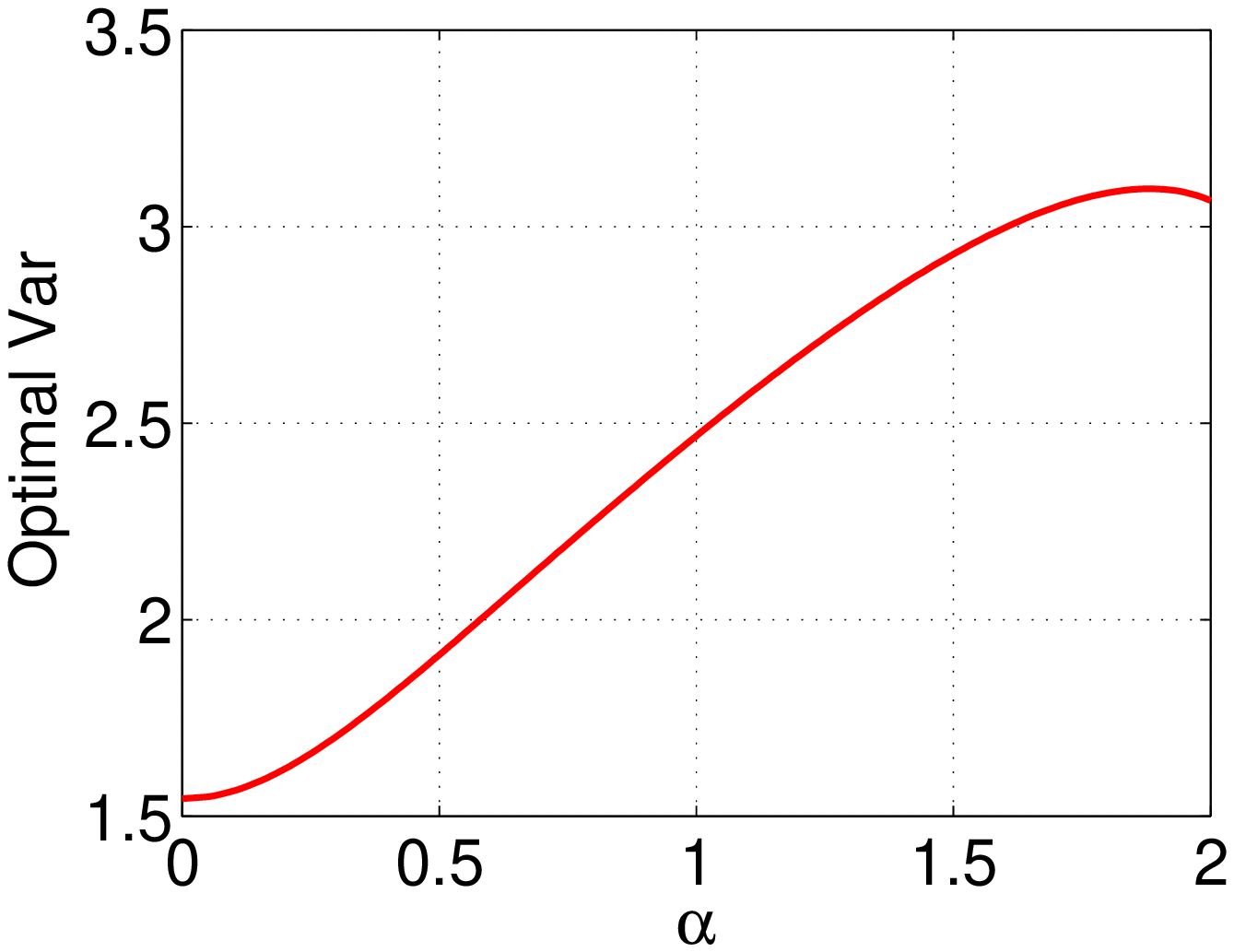}\hspace{0.2in}
\includegraphics[width=2.2in]{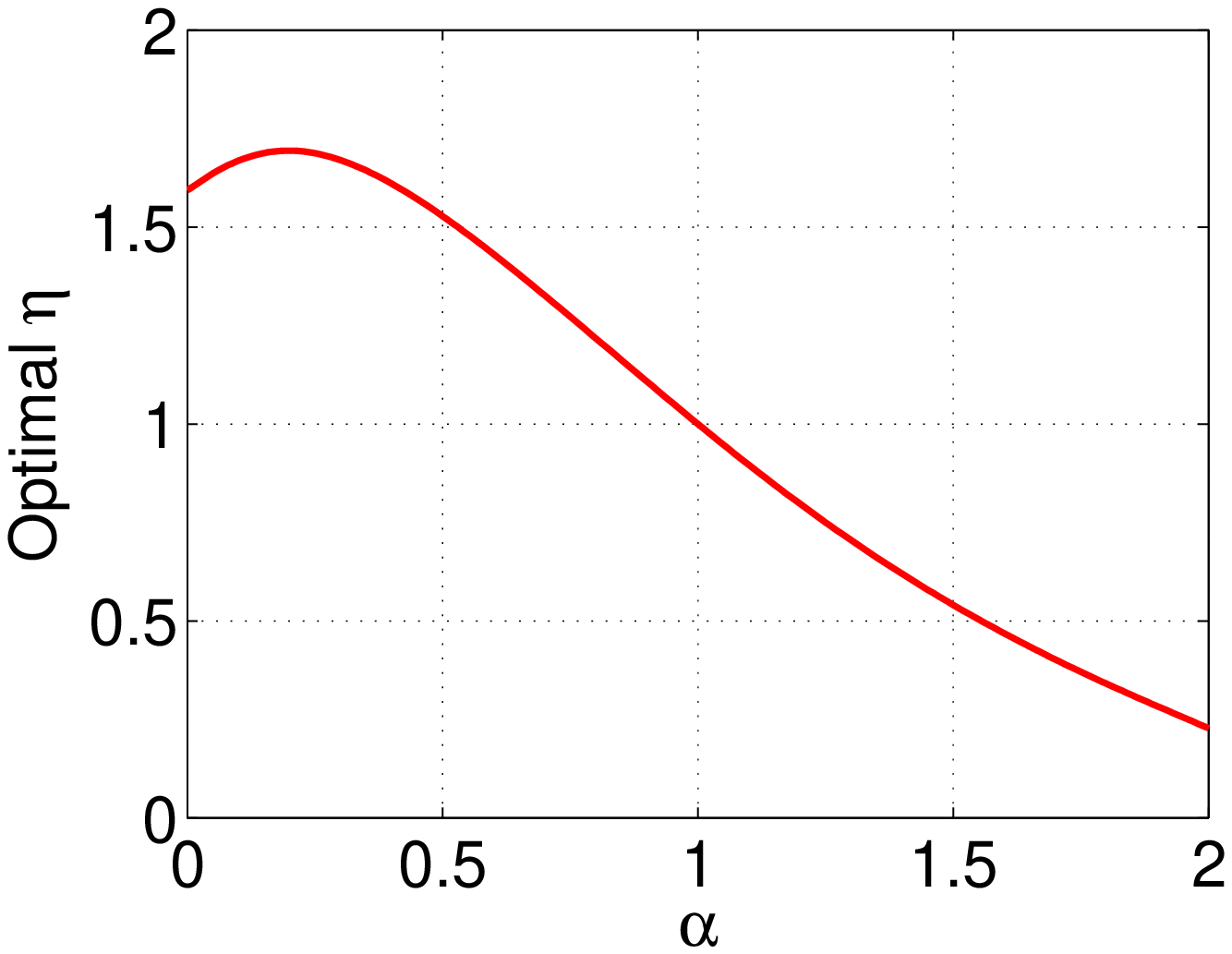}
}
\end{center}
\vspace{-0.3in}
\caption{The optimal variance values $V_\alpha(\eta)$ (left panel) and the corresponding optimal $\eta$ values (right panel). Each point on the curve corresponds to the lowest point of the curve for that $\alpha$ as  in Figure~\ref{fig_Var}.}\label{fig_OptV}
\end{figure}

For general $0<\alpha\leq2$, the cdf $F_\alpha$ and pdf $f_\alpha$ can be computed numerically. Figure~\ref{fig_Var} plots  $V_\alpha(\eta)$ for  $\alpha$ from 0 to 2. The lowest point on each curve corresponds to the optimal (smallest) $V_\alpha(\eta)$. Figure~\ref{fig_OptV} plots the optimal $V_\alpha$ values (left panel) and optimal $\eta$ values (right panel).

Figure~\ref{fig_Var} suggests  that the 1-bit scheme performs reasonably well.  The optimal variance coefficient $V_\alpha$ is not  much larger than the  variance using full information. For example, when $\alpha=1$, the optimal variance coefficient using full information is 2 (i.e., see (\ref{eqn_optVar})),  while the optimal variance coefficient of the 1-bit scheme is just $\frac{\pi^2}{4} = 2.467$ which is only about $20\%$ larger. Furthermore, we can see that, at least when $\alpha\leq1$, $V_\alpha(\eta)$ is not very sensitive to $\eta$ in a wide range of $\eta$ values, which is practically important, because an optimal choice of $\eta$ requires knowing $\Lambda_\alpha$ and is general not achievable. The best we can hope for is that the estimate is not  sensitive to the choice of $\eta$.

\subsection{Error Tail Bounds}

\begin{theorem}\label{thm_bounds}
\begin{align}\notag
&\mathbf{Pr}\left(\hat{\Lambda}_\alpha\geq(1+\epsilon)\Lambda_\alpha\right)\leq \exp\left(-n\frac{\epsilon^2}{G_{R,\alpha,C,\epsilon}}\right),\hspace{0.05in} \epsilon\geq 0\\\notag
&\mathbf{Pr}\left(\hat{\Lambda}_\alpha\leq (1-\epsilon)\Lambda_\alpha\right)\leq \exp\left(-n\frac{\epsilon^2}{G_{L,\alpha,C,\epsilon}}\right),\hspace{0.05in} 0\leq\epsilon\leq1
\end{align}
where $G_{R,\alpha,C,\epsilon}$ and $G_{L,\alpha,C,\epsilon}$ are computed as follows:
\begin{align}\label{eqn_GR}
\frac{\epsilon^2}{G_{R,\alpha,C,\epsilon}} =& -F_\alpha(1/(1+\epsilon)\eta)\log\left[\frac{F_\alpha(1/\eta)}{F_\alpha(1/(1+\epsilon)\eta)}\right]\\\notag
&-(1-F_\alpha(1/(1+\epsilon)\eta))\log\left[\frac{1-F_\alpha(1/\eta)}{1-F_\alpha(1/(1+\epsilon)\eta)}\right]\\\label{eqn_GL}
\frac{\epsilon^2}{G_{L,\alpha,C,\epsilon}} =& -F_\alpha(1/(1-\epsilon)\eta)\log\left[\frac{F_\alpha(1/\eta)}{F_\alpha(1/(1-\epsilon)\eta)}\right]\\\notag
&-(1-F_\alpha(1/(1-\epsilon)\eta))\log\left[\frac{1-F_\alpha(1/\eta)}{1-F_\alpha(1/(1-\epsilon)\eta)}\right]
\end{align}
\noindent \textbf{Proof:}\hspace{0.2in} See Appendix~\ref{proof_thm_bounds}.  The proof is based on Chernoff's original tail bounds~\cite{Article:Chernoff_52} for the binomial distribution.$\hfill\Box$
\end{theorem}
To ensure the error $\mathbf{Pr}\left(\hat{\Lambda}_\alpha\geq(1+\epsilon)\Lambda_\alpha\right) + \mathbf{Pr}\left(\hat{\Lambda}_\alpha\leq(1-\epsilon)\Lambda_\alpha\right)\leq \delta,\hspace{0.1in} 0\leq \delta\leq 1$,  it suffices that
\begin{align}\label{eqn_two_error_sum}
\exp\left(-n\frac{\epsilon^2}{G_{R,\alpha,C,\epsilon}}\right) + \exp\left(-n\frac{\epsilon^2}{G_{L,\alpha,C,\epsilon}}\right)\leq \delta
\end{align}
for which it suffices
\begin{align}\label{eqn_sample_complexity}
&n \geq \frac{G_{\alpha,C,\epsilon}}{\epsilon^2}\log 2/\delta,\hspace{0.25in} \text{where}\\\notag
&G_{\alpha,C,\epsilon} = \max\{G_{R,\alpha,C,\epsilon},\  G_{L,\alpha,C,\epsilon}\}
\end{align}

\begin{figure}[h!]
\begin{center}
\includegraphics[width=2.45in]{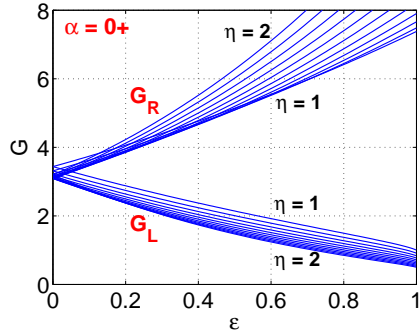}
\end{center}
\vspace{-0.3in}
\caption{The tail bound constants $G_{R,0+,C,\epsilon}$ (\ref{eqn_GR}) (upper group) and $G_{L,0+,C,\epsilon}$ (\ref{eqn_GL}) (lower group), for $\eta = 1$ to 2 spaced at 0.1. Recall $\eta = \frac{\Lambda_\alpha}{C}$.  }\label{fig_G0}\vspace{-0.in}
\end{figure}

Obviously, it will be even more precise to numerically compute $n$ from (\ref{eqn_two_error_sum}) instead of using the convenient sample complexity bound (\ref{eqn_sample_complexity}). Figure~\ref{fig_G0} provides the tail bound constants for $\alpha = 0+$, i.e., $G_{R,0+,C,\epsilon}$  and $G_{L,0+,C,\epsilon}$ at selected $\eta$ values ranging from 1 to 2. %Recall that when $\alpha = 0+$, the optimal choice of $\eta$ is 1.594 and the optimal variance coefficient $V_{0+}(\eta)$ is 1.544.

\subsection{Bias-Correction}

Bias-correction for MLE is important for small sample size $n$.  In Theorem~\ref{thm_1bit}, Eq. (\ref{eqn_Bias}) says
\begin{align}\notag
E\left(\hat{\Lambda}_\alpha\right) = \Lambda_\alpha +\frac{\Lambda_\alpha}{n}\frac{n_1}{n}\left(1-\frac{n_1}{n}\right)\left(\frac{\eta^2}{f_\alpha^2(1/\eta)} + \frac{\eta f^\prime_\alpha(1/\eta)}{2f^3_\alpha(1/\eta)}\right)+ O\left(\frac{1}{n^2}\right)
\end{align}
which naturally provides a bias-correction for $\hat{\Lambda}_\alpha$, known as the ``Bartlett correction'' in statistics.  To do so, we will need to use the estimate $\hat{\Lambda}_\alpha$ to compute the $\eta$. Since $\hat{\Lambda}_\alpha = C/F^{-1}_\alpha(n_1/n)$, we have $\hat{\Lambda}_\alpha/C = 1/F^{-1}_\alpha(n_1/n)$. The bias-corrected estimator, denoted by $\hat{\Lambda}_{\alpha,c}$ is
\begin{align}
\hat{\Lambda}_{\alpha,c} =& \frac{\hat{\Lambda}_\alpha}{1+\frac{1}{n}\frac{n_1}{n}\left(1-\frac{n_1}{n}\right)\left(
\frac{\hat{\eta}^2}{f_\alpha^2(1/\hat{\eta})} + \frac{\hat{\eta} f^\prime_\alpha(1/\hat{\eta})}{2f^3_\alpha(1/\hat{\eta})}\right)},\hspace{0.2in} \text{where } \
\hat{\eta} = 1/F^{-1}_\alpha(n_1/n)
\end{align}
which, when $\alpha=0+$, $\alpha=1$, and $\alpha=2$, becomes respectively
\begin{align}
&\hat{\Lambda}_{0+,c} =\frac{C\log n/n_1}{1+ \frac{1/n_1-1/n}{2\log n/n_1}  }\\
&\hat{\Lambda}_{1,c} = \frac{\frac{C}{\tan\frac{\pi}{2}\frac{n_1}{n}}}{1+\frac{1}{n}\frac{\pi^2}{4}\frac{n_1}{n}\left(1- \frac{n_1}{n}\right)\left(1+\frac{1}{\tan^2 \frac{\pi}{2}\frac{n_1}{n}}\right)
}\\
&\hat{\Lambda}_{2,c} = \frac{\frac{C}{2F^{-1}_{\chi^2_1}(n_1/n)}}{1 +  \frac{\pi}{n}\frac{n_1}{n}\left(1-\frac{n_1}{n}\right)  \left(\frac{3}{F^{-1}_{\chi^2_1}(n_1/n) }-1\right) e^{F^{-1}_{\chi^2_1}(n_1/n)}}
\end{align}
See the detailed derivations in Appendix~\ref{proof_thm_1bit}, together with the proof of Theorem~\ref{thm_1bit}.

\section{Experiments on 1-Bit Coding and Estimation}

We conduct extensive simulations to (i) verify the 1-bit variance formulas of the MLE, and (ii)  apply the 1-bit estimator in Algorithm~\ref{alg_recovery} for one scan 1-bit compressed sensing~\cite{Report:Li_1bitCS15}.

\subsection{Bias and Variance of the Proposed Estimators}

\begin{figure}[h!]
\begin{center}
\mbox{
\includegraphics[width=2.2in]{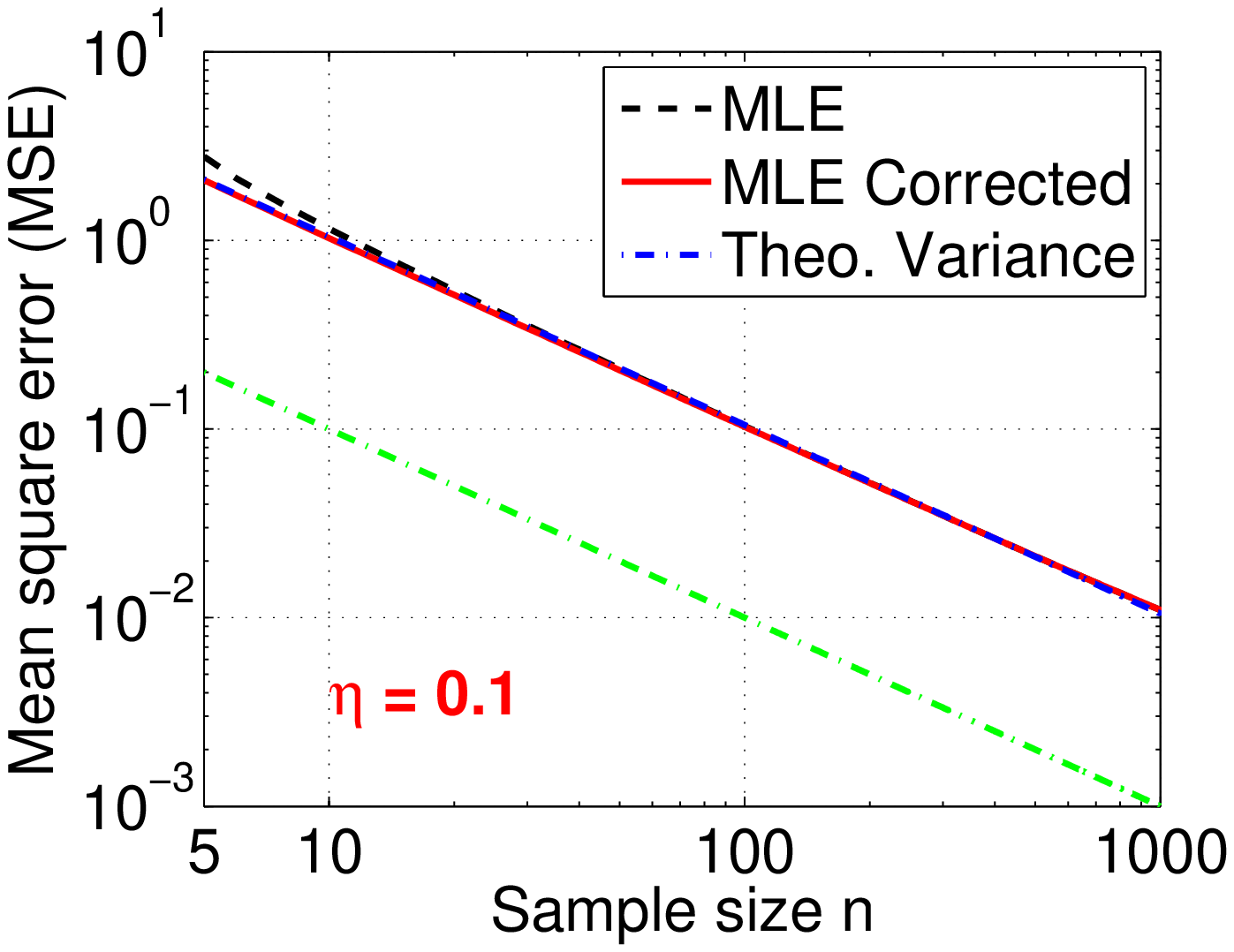}\hspace{0in}
\includegraphics[width=2.2in]{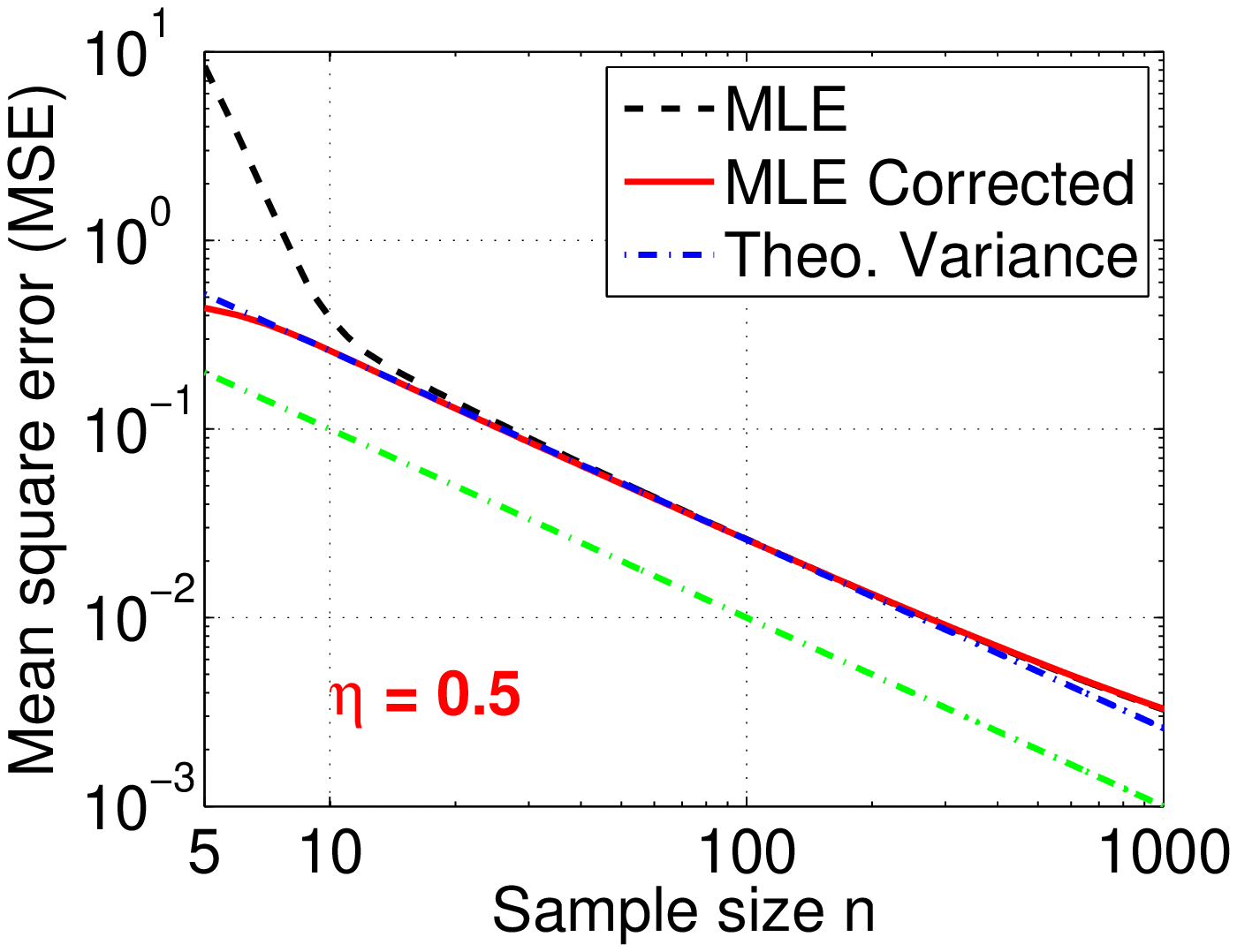}\hspace{0in}
\includegraphics[width=2.2in]{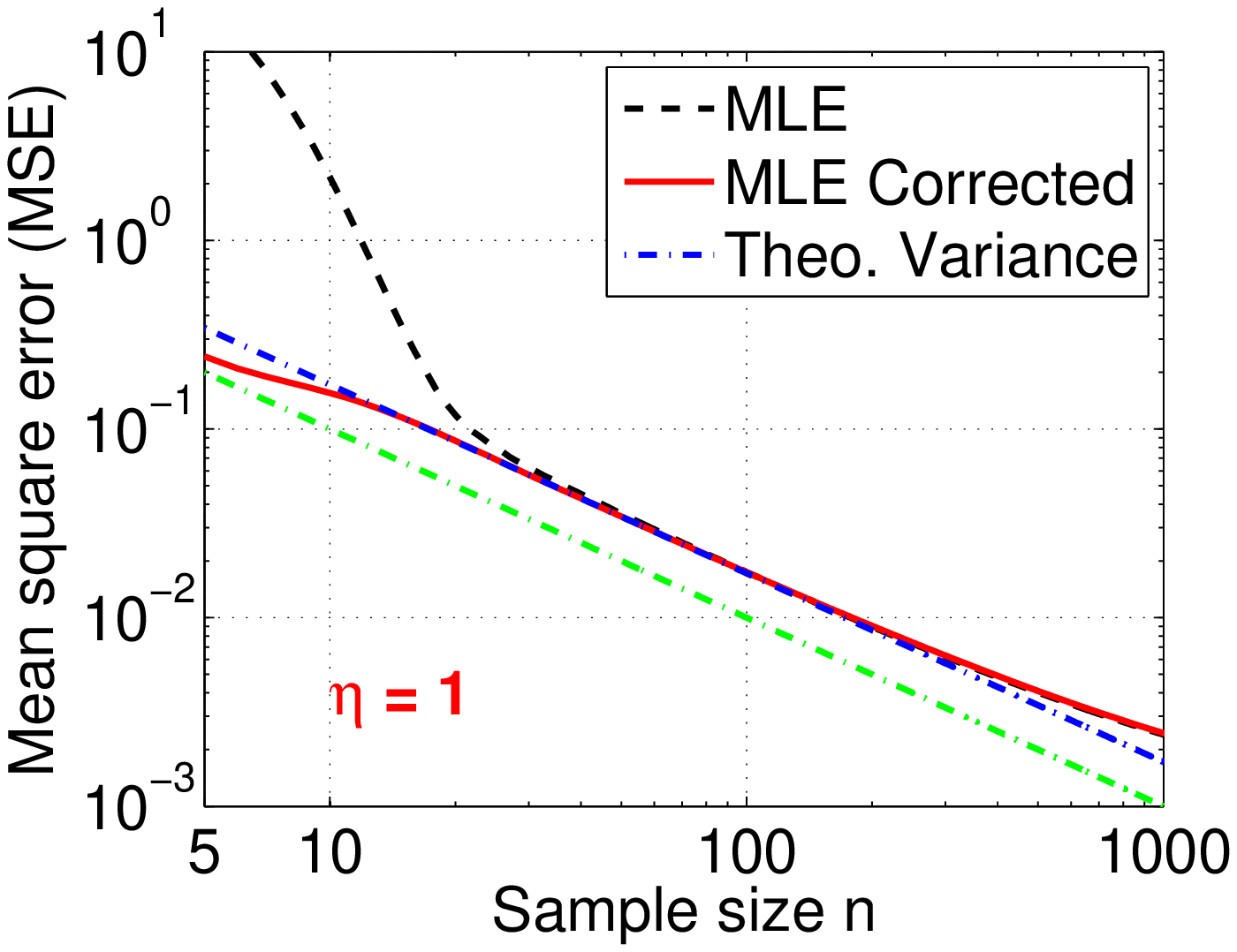}
}

\mbox{
\includegraphics[width=2.2in]{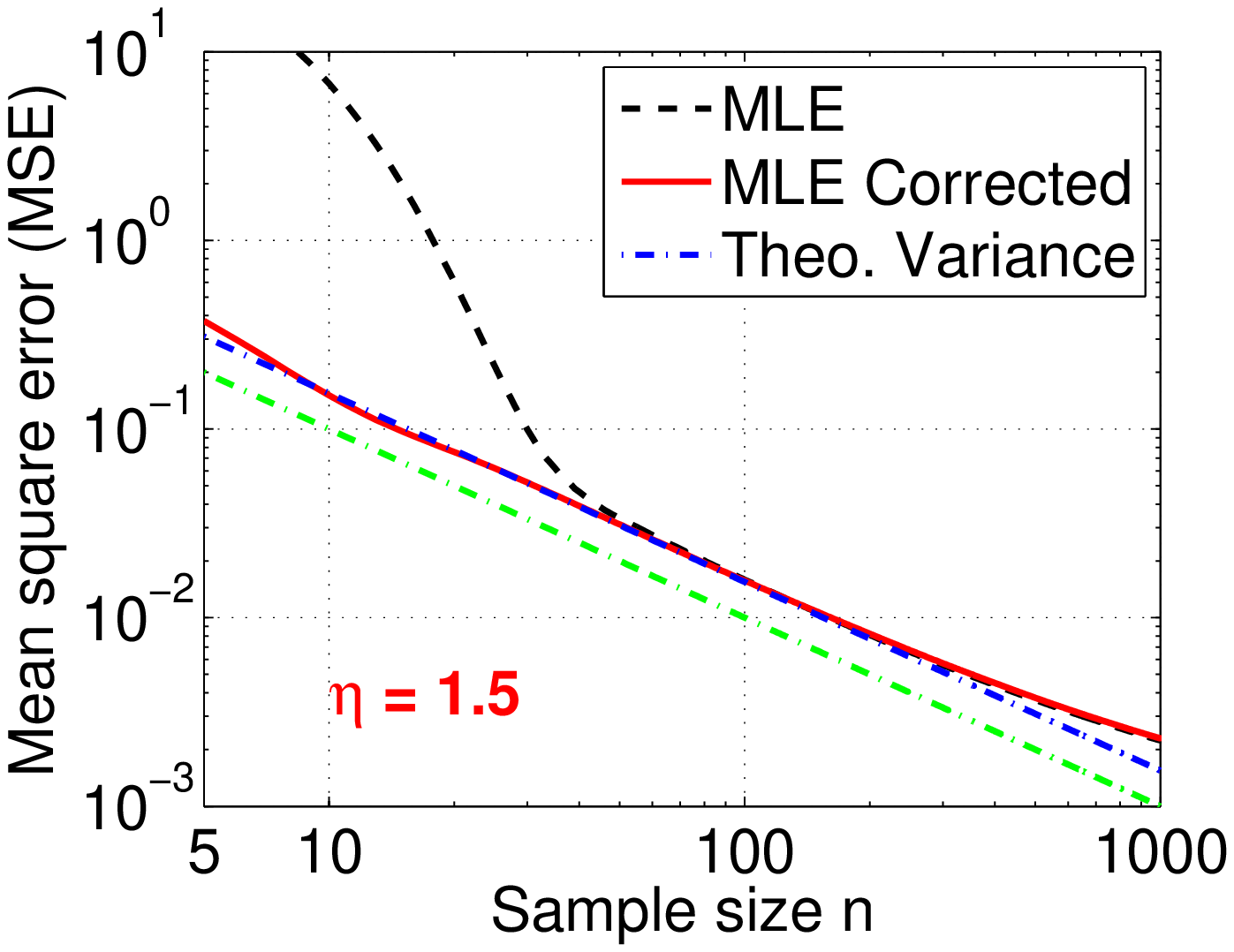}\hspace{0in}
\includegraphics[width=2.2in]{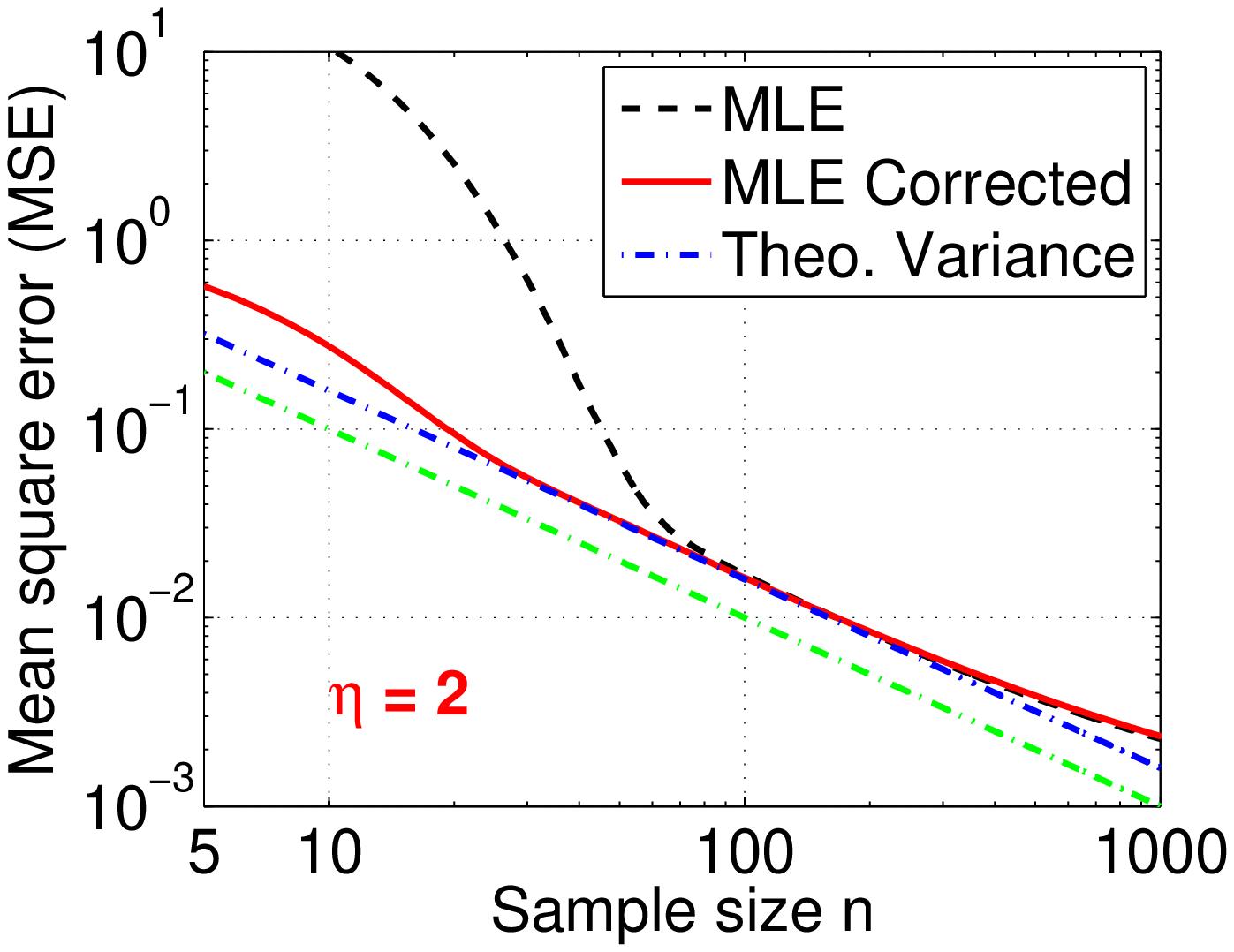}\hspace{0in}
\includegraphics[width=2.2in]{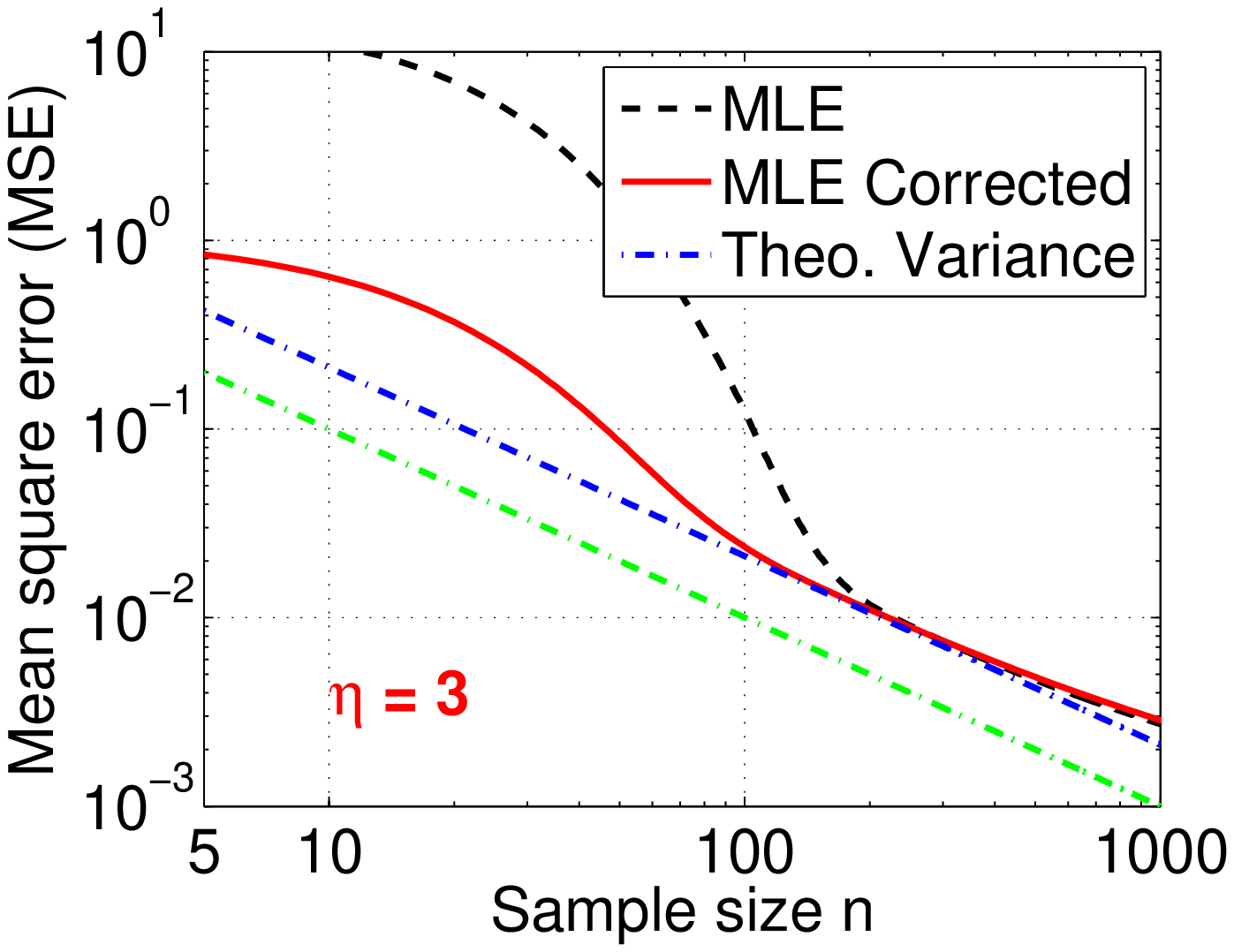}
}

\end{center}
\vspace{-0.3in}
\caption{Empirical Mean square errors of $\hat{\Lambda}_{0+}$ (dashed curves) and $\hat{\Lambda}_{0+,c}$ (solid curves) from $10^6$ simulations of $S(\alpha,1)$ for $\alpha=0.05$, at each sample size $n$. Each panel present results for a different $\eta = \frac{\Lambda_\alpha}{C}$. For both estimators, the empirical MSEs converge to the theoretical asymptotic variances (\ref{eqn_Var}) (dashed dot curves and blue if color is available) when $n$ is large enough.  In each panel, the lowest curve (dashed dot and green if color is available) represents the theoretical variance using full (infinite-bit) information, i.e., $1/n$ in this case.  For small $n$, the bias-correction step important.  Note that the small (and exaggerated) discrepancies at  large $n$ are due to use of $\alpha=0.05$ to generate samples and the estimators based on $\alpha=0+$. Also recall that $\eta = 1.594$ is the optimal $\eta$. }\label{fig_Mse1bitP0}\vspace{-0.1in}
\end{figure}

Figure~\ref{fig_Mse1bitP0} provides the simulations for verifying the 1-bit estimator $\hat{\Lambda}_{0+}$ and its bias-corrected version $\hat{\Lambda}_{0+,c}$ using small $\alpha$ (i.e., 0.05). Basically, for each sample size $n$, we generate $10^6$ samples from $S(\alpha,1)$, which are  quantized according a pre-selected threshold $C$. Then we apply both $\hat{\Lambda}_{0+}$ and  $\hat{\Lambda}_{0+,c}$  and report the empirical  mean square error (MSE = variance + bias$^2$) from $10^6$ repetitions. For thorough evaluations, we conduct simulations for a wide range of $n \in[5,1000]$. \\

The results are presented in log-log scale, which  exaggerates the portion for small $n$ and the y-axis for large $n$. The plots confirm that when $n$ is not too small (e.g., $n>100$), the bias of MLE estimate varnishes and the asymptotic variance formula (\ref{eqn_Var}) matches the mean square error. For small $n$ (e.g., $n<100$), the bias correction becomes  important.

\begin{figure}[h!]
\begin{center}
\mbox{
\includegraphics[width=2.2in]{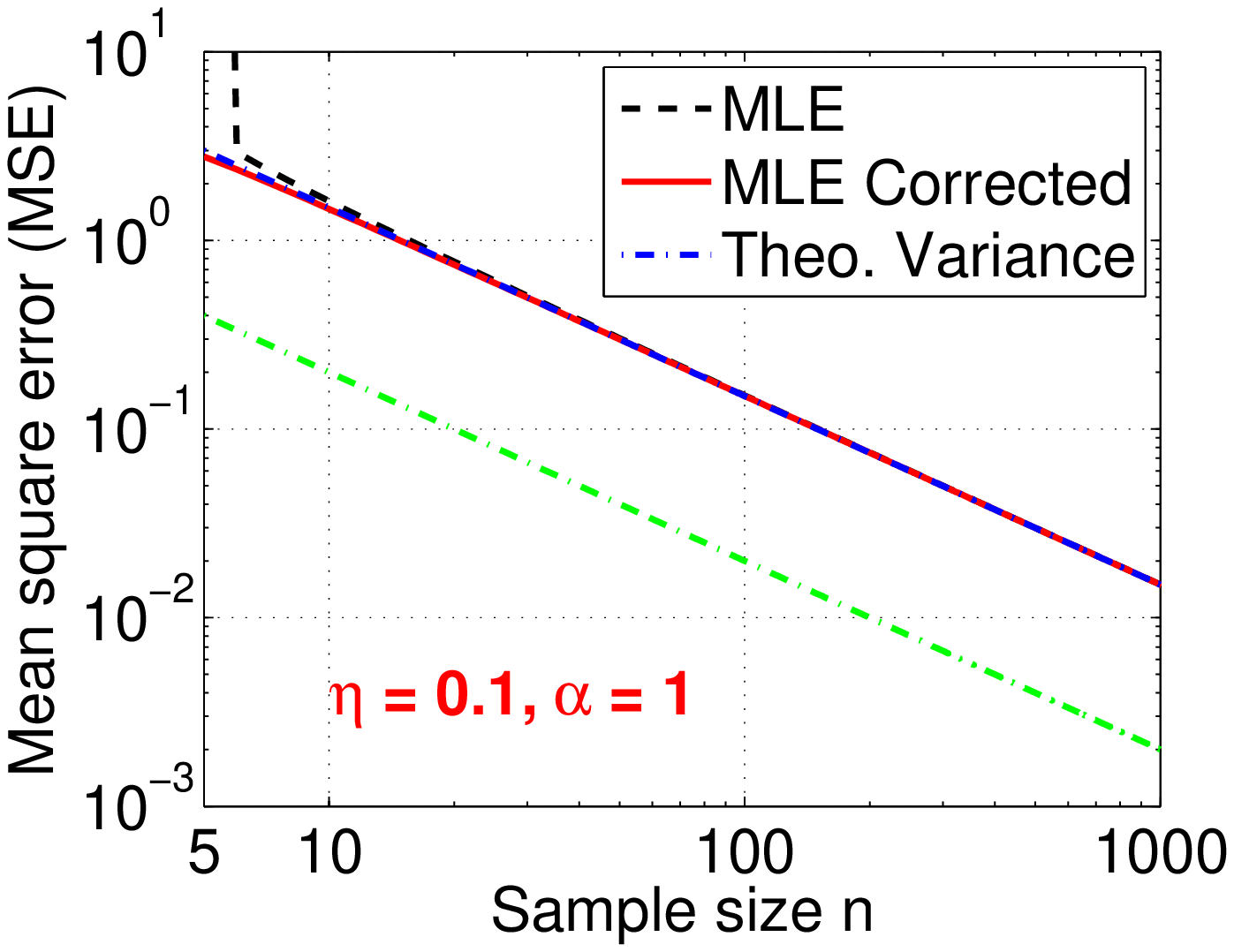}\hspace{0in}
\includegraphics[width=2.2in]{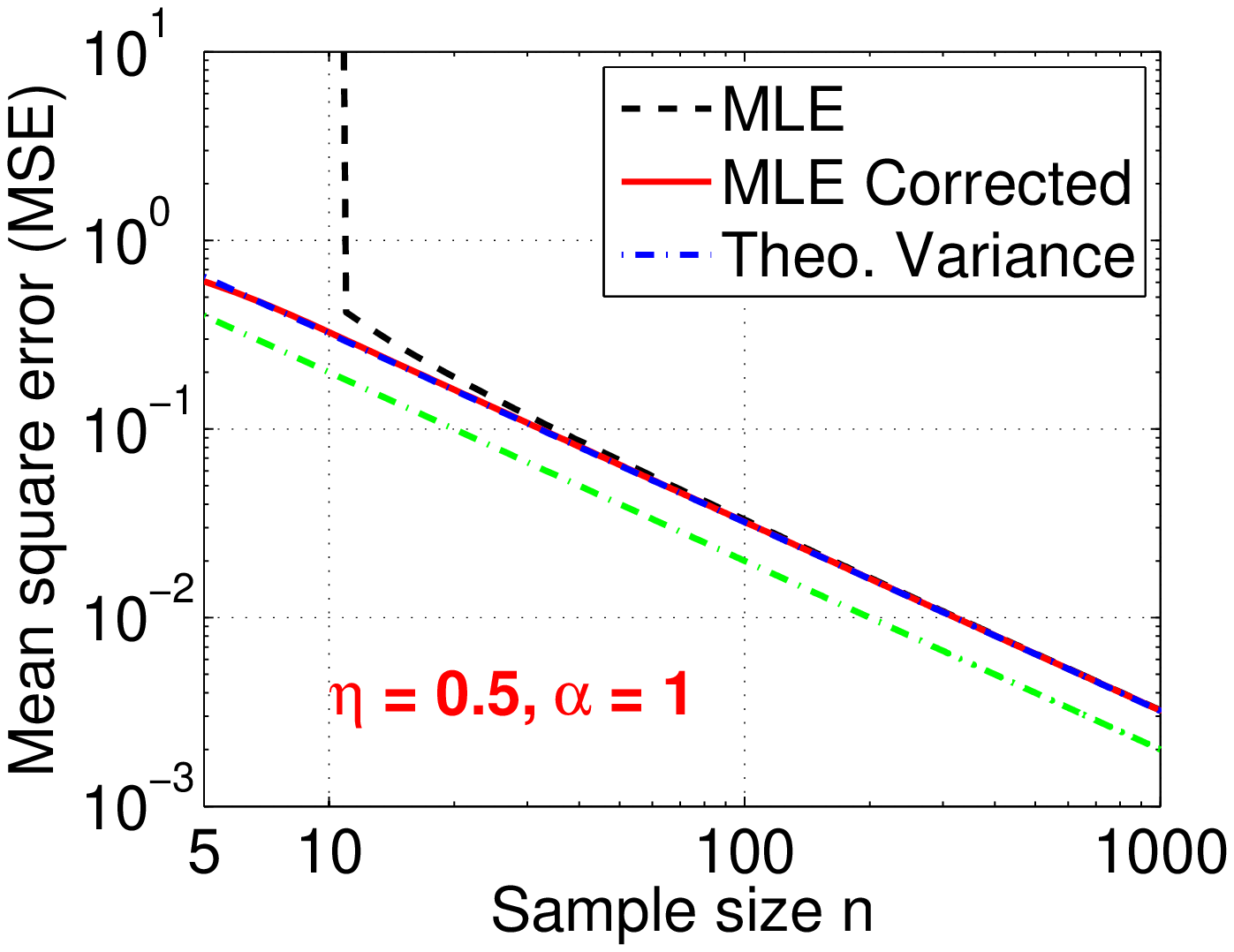}\hspace{0in}
\includegraphics[width=2.2in]{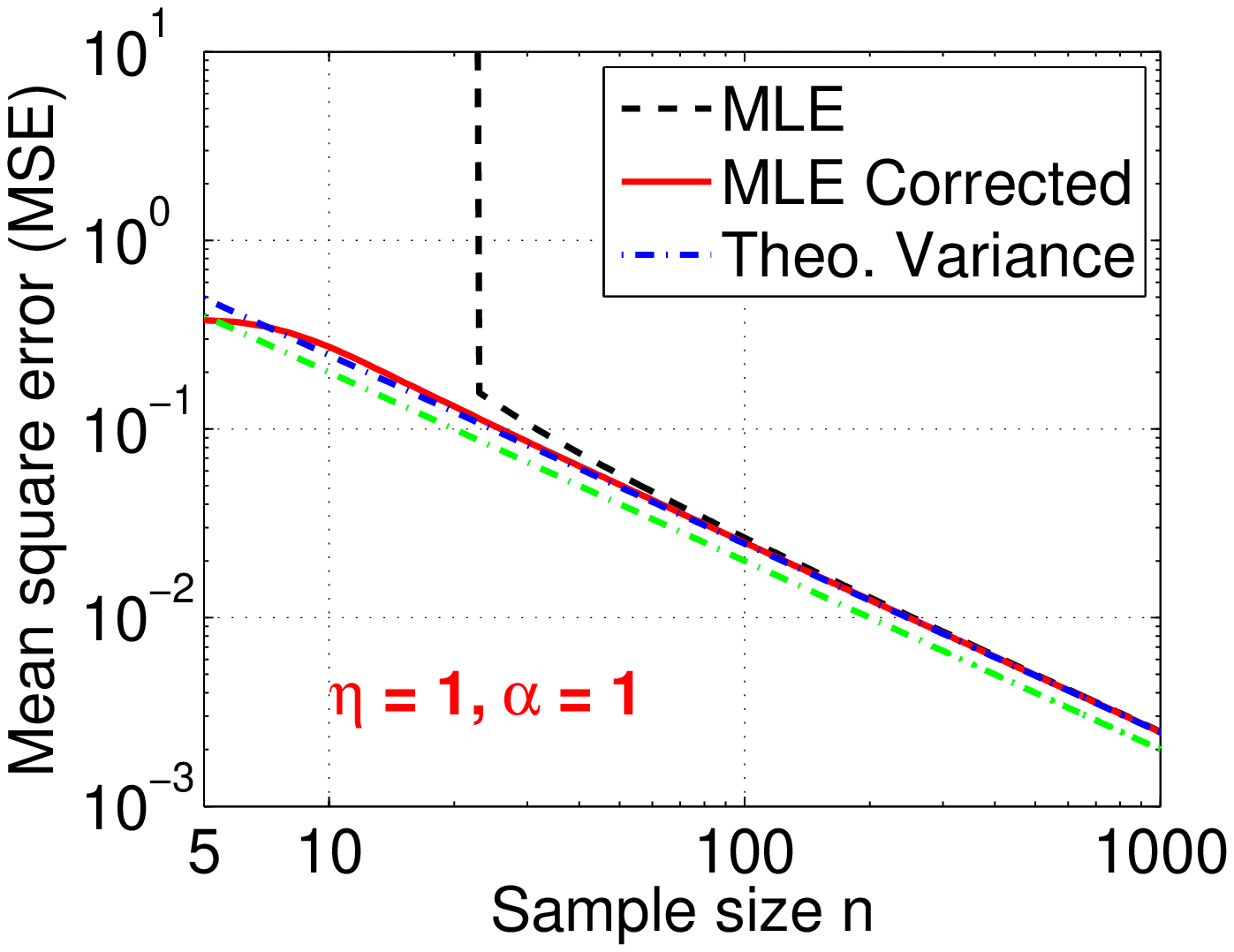}
}

\mbox{
\includegraphics[width=2.2in]{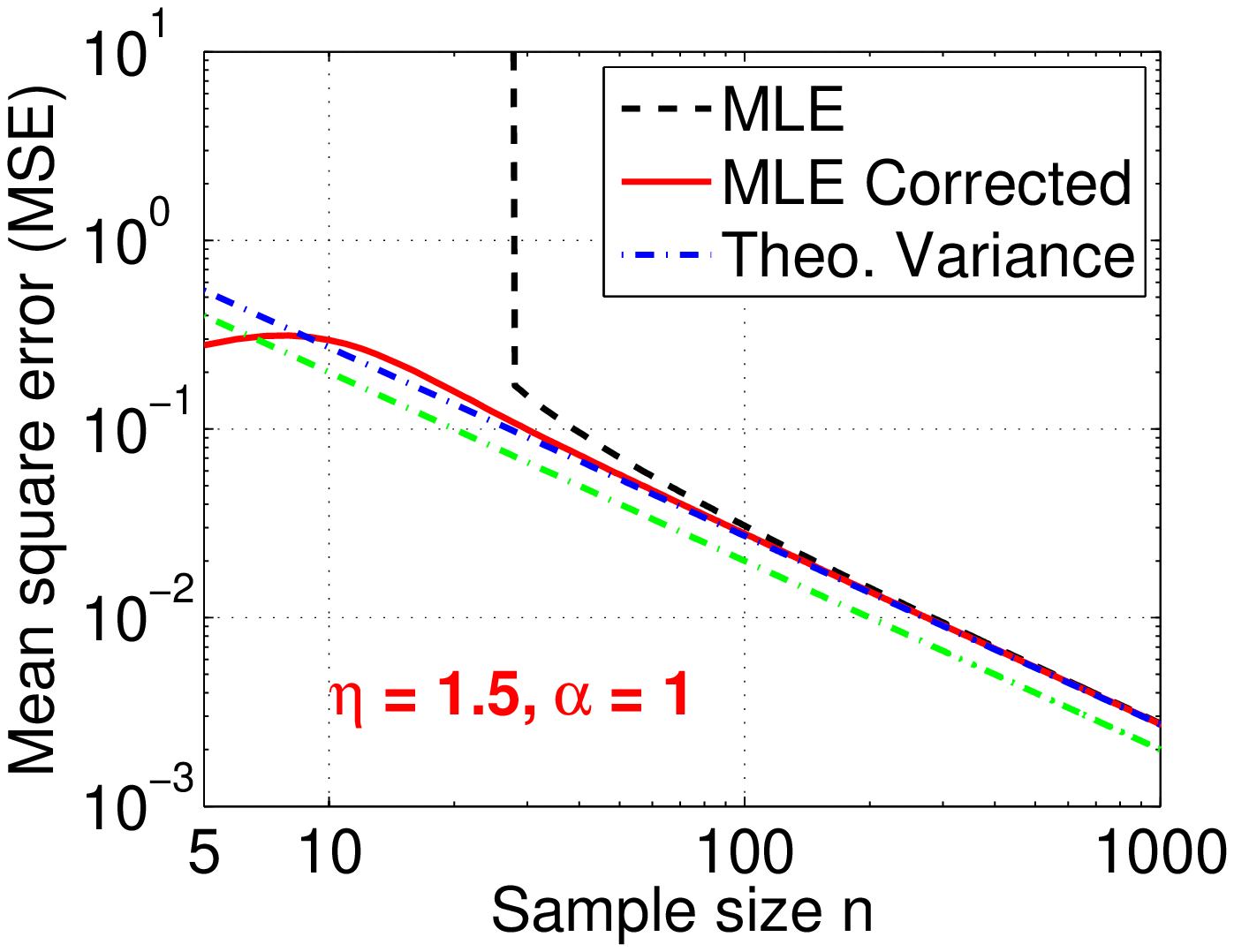}\hspace{0in}
\includegraphics[width=2.2in]{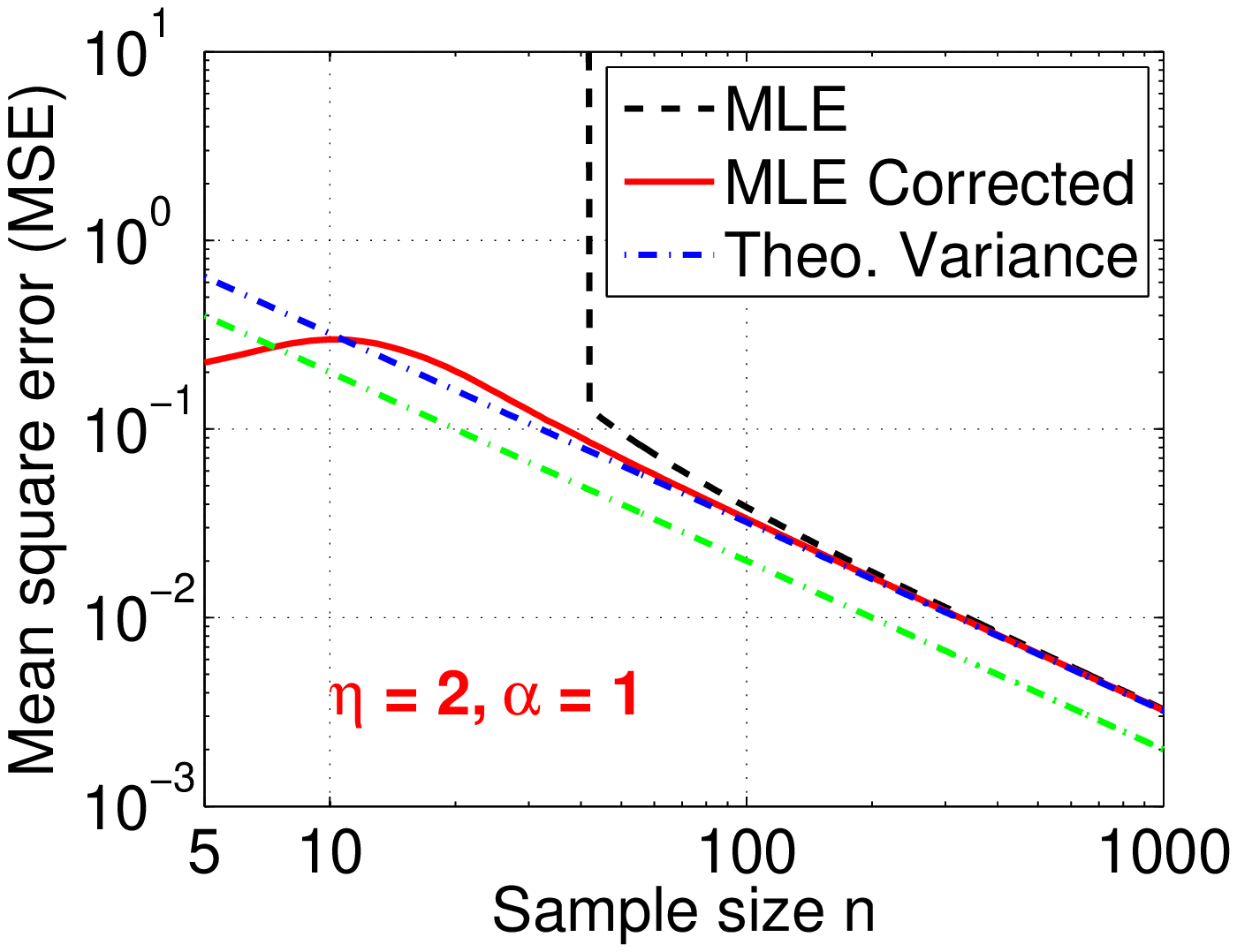}\hspace{0in}
\includegraphics[width=2.2in]{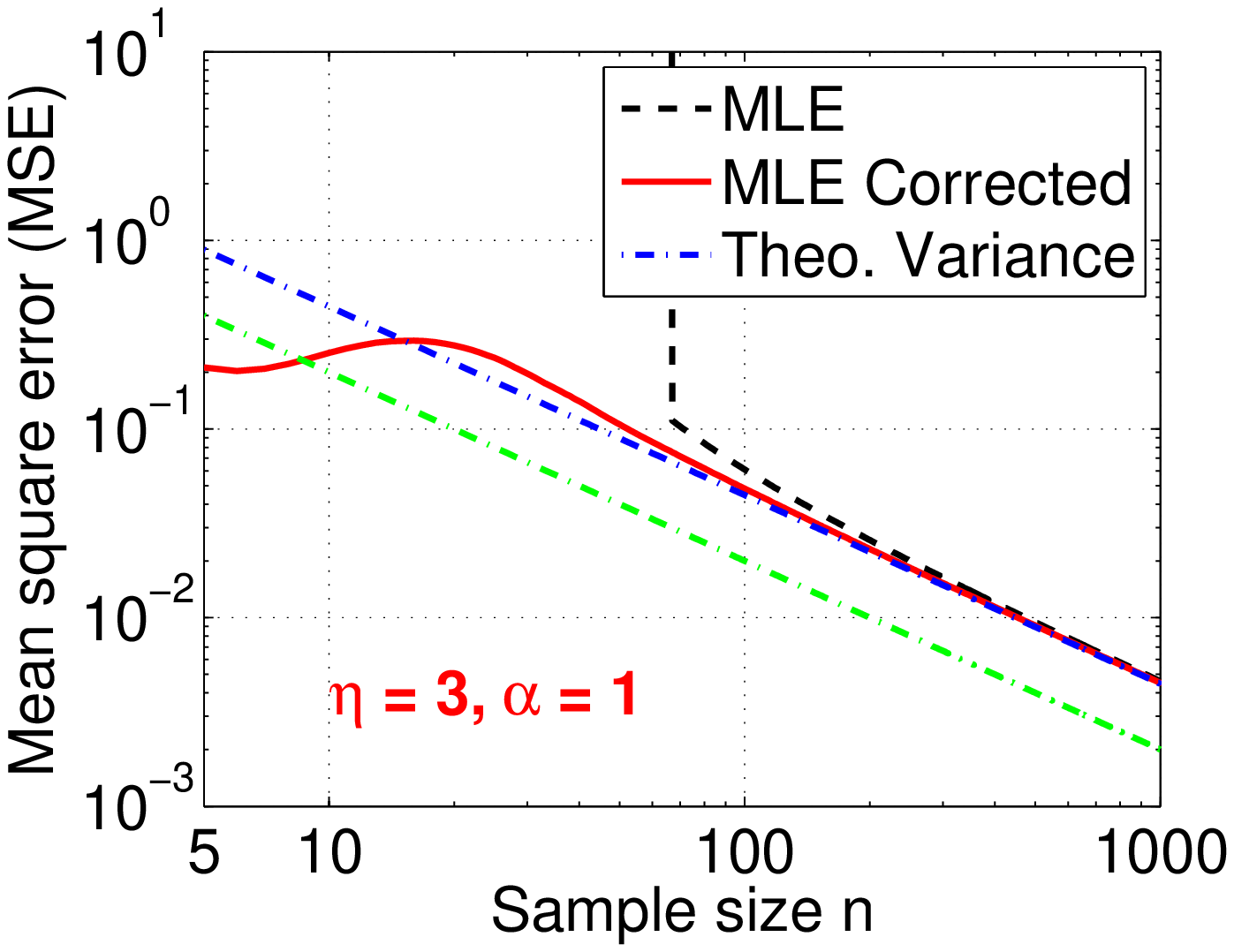}
}

\end{center}
\vspace{-0.3in}
\caption{Mean square errors of $\hat{\Lambda}_{1}$ (dashed curves) and $\hat{\Lambda}_{1,c}$ (solid curves) for $\alpha=1$. Note that the lowest curve (dashed dot and green if color is available) in each panel represents the optimal variance using full (i.e., infinite-bit) information, which is $2/n$ for $\alpha=1$. }\label{fig_Mse1bitP1}\vspace{-0.3in}
\end{figure}

Note that when $n$ is large (i.e., when errors are very small), the plots show some discrepancies. This is due to the fact that we have to use small $\alpha$ for the simulations but the estimators $\hat{\Lambda}_{0+}$ and  $\hat{\Lambda}_{0+,c}$ are based on $\alpha=0+$. The differences are very small and only become visible when the estimation errors are so small (due to the exaggeration of the log-scale). To remove this effect, we also conduct similar simulations for $\alpha=1$ and present the results in Figure~\ref{fig_Mse1bitP1}, which does not show the discrepancies at large $n$. We can see that the bias-correction step is also important for $\alpha=1$. \\

We should mention that, for numerical issue, we added a  small real number ($10^{-6}$) to $n_1$.  We did not further investigate various smoothing techniques as it appears that this Bartlett-correction procedure already serves the purpose well.\vspace{0.2in}

\newpage

%\newpage

\subsection{One Scan 1-Bit Compressed Sensing}

Next, we integrate $\hat{\Lambda}_{0+,c}$ into the sparse recovery procedure in Algorithm~\ref{alg_recovery}, by   replacing $K$ with $\hat{\Lambda}_{0+,c}$ for computing $Q^+_i$ and $Q_i^-$~\cite{Report:Li_1bitCS15}. We report the sign recovery errors  $\sum_{i}|\hat{sgn(x_i)} - sgn(x_i)|/K$ from $10^4$ simulations. In this study, we let $N=1000$, $K=20$, and sample the nonzero coordinates from $N(0,5^2$). For estimating $K$, we use  $n\in\{20,\ 50, 100\}$ samples with $\eta \in\{0.2,\ 0.5,\ 1.5,\ 2,\ 3\}$. Recall $\eta = 1.5$ is close to be optimal (1.594) for $\hat{\Lambda}_{0+}$. \\

\begin{figure}[h!]
\begin{center}

\mbox{
\includegraphics[width=2.2in]{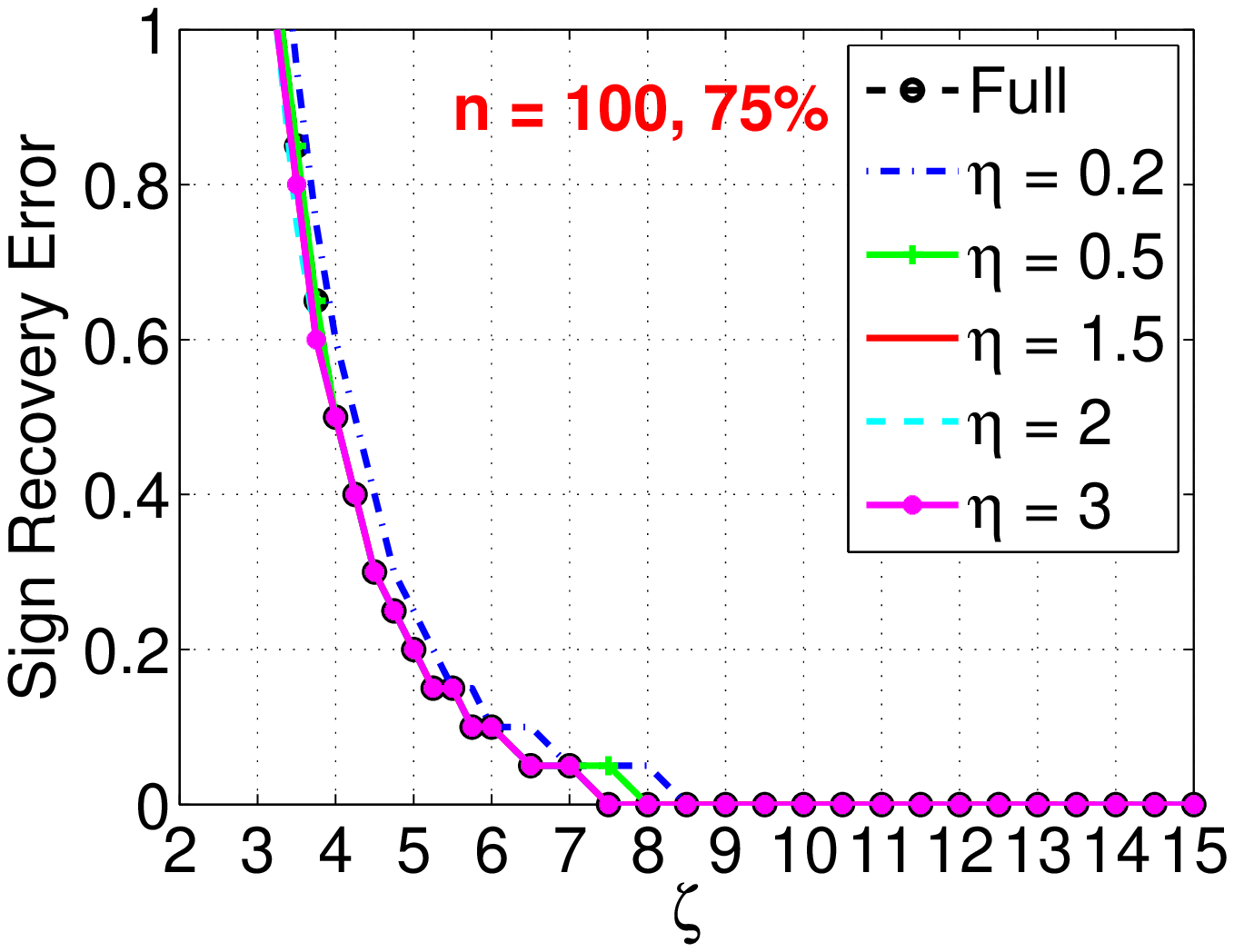}\hspace{0in}
\includegraphics[width=2.2in]{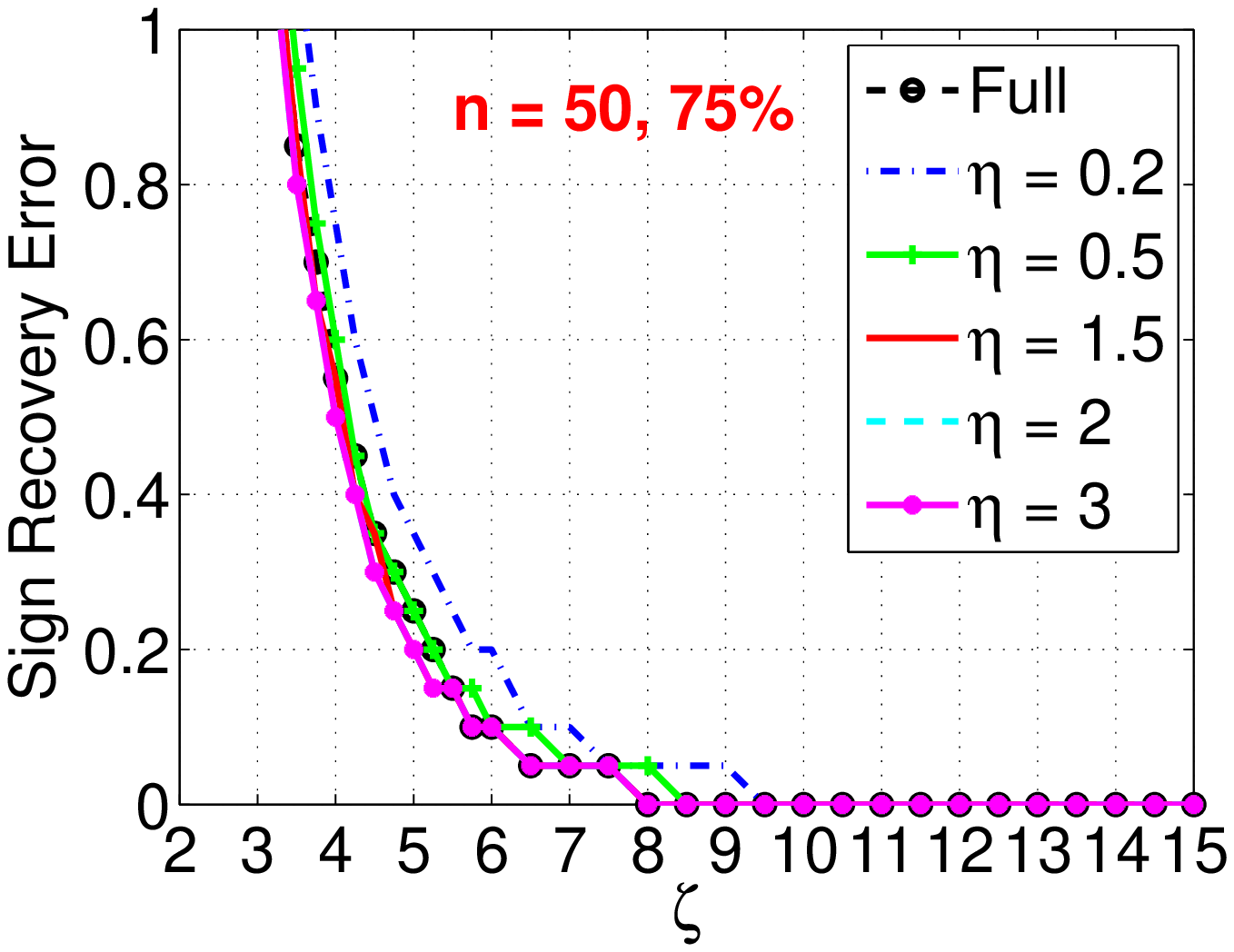}\hspace{0in}
\includegraphics[width=2.2in]{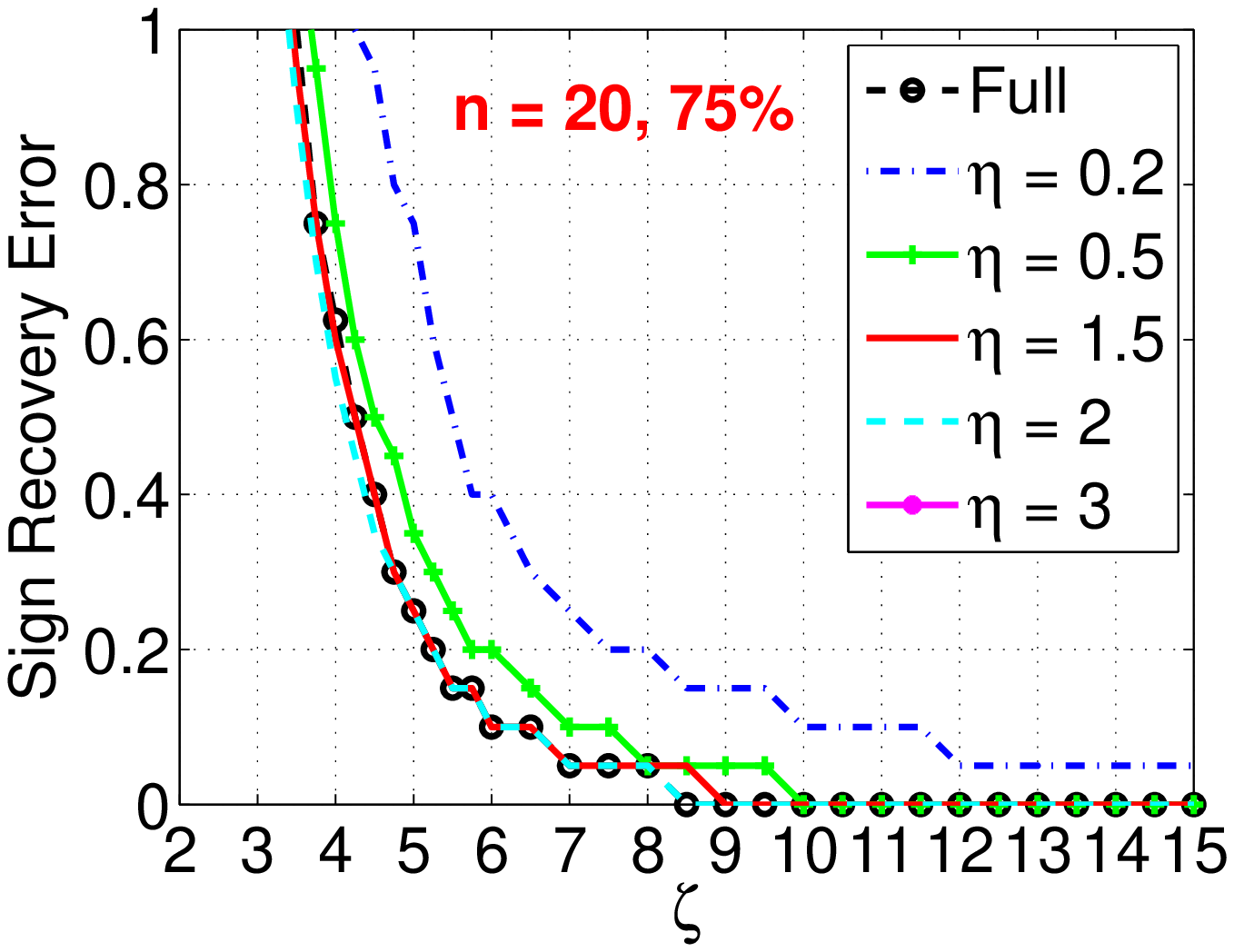}
}

\mbox{
\includegraphics[width=2.2in]{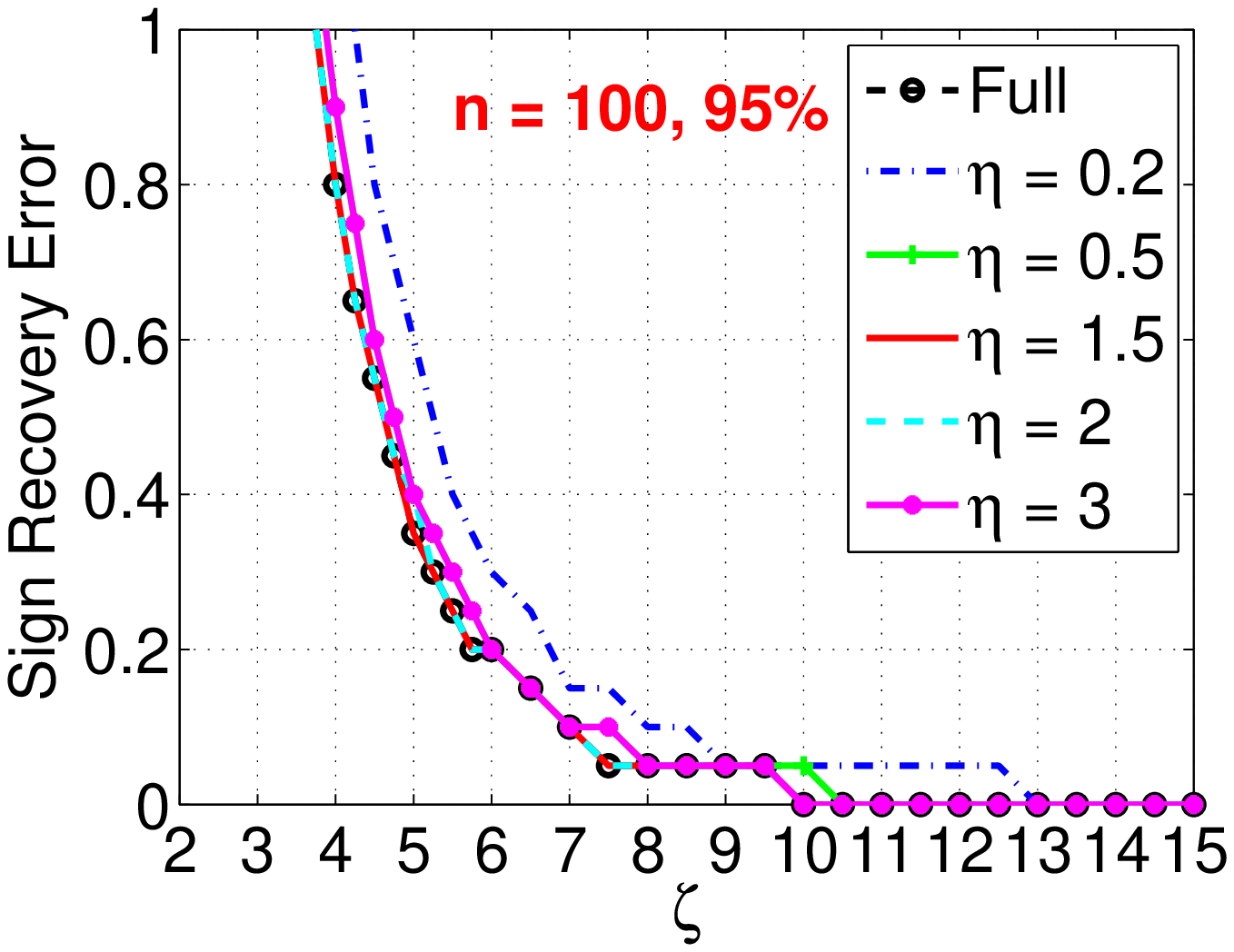}\hspace{0in}
\includegraphics[width=2.2in]{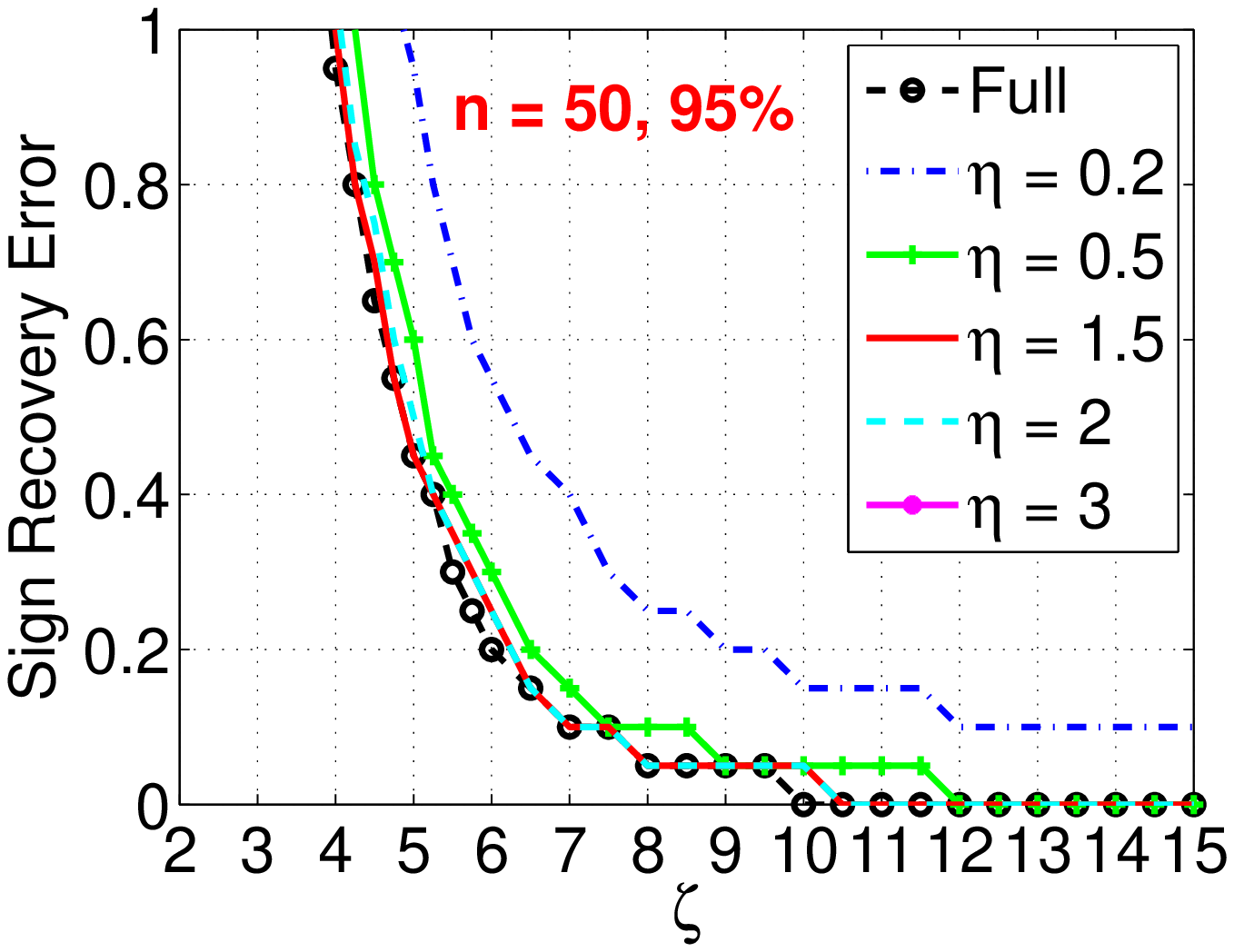}\hspace{0in}
\includegraphics[width=2.2in]{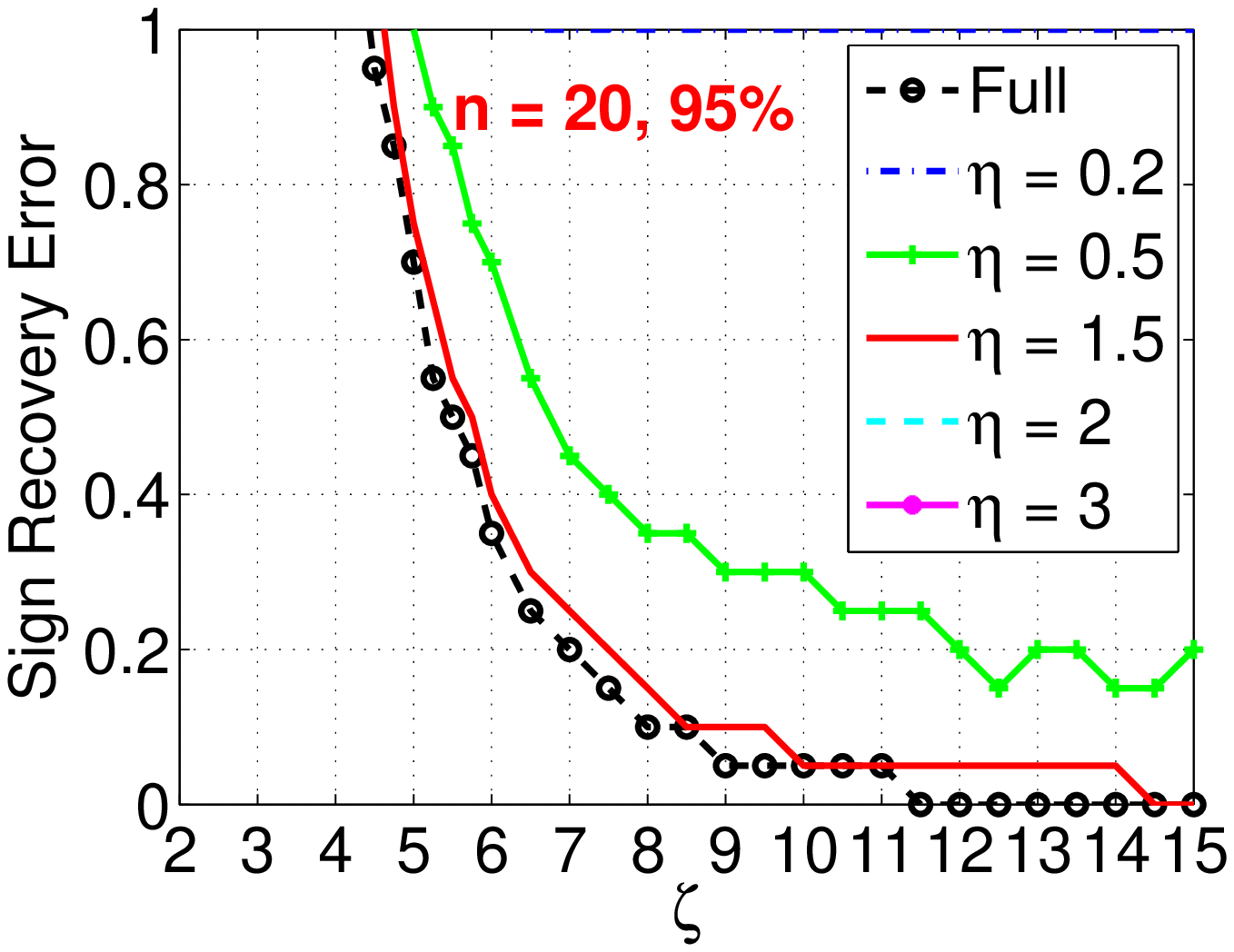}
}

\end{center}
\vspace{-0.3in}
\caption{Sign recovery error: $\sum_{i}|\hat{sgn(x_i)} - sgn(x_i)|/K$, using Algorithm~\ref{alg_recovery} and estimated $K$ in computing $Q_i^+$ and $Q_i^-$ in Algorithm~\ref{alg_recovery}. In this study, $N=1000$, $K=20$, and the nonzero entries are generated from $N(0,5^2$). The number of measurements for recovery is $M = \zeta K \log (N/0.01)$ and we use $n$ samples to estimate $K$ for $n \in \{20,\, 50,\ 100\}$. We report $75\%$ (upper panels) and $95\%$  (bottom panels) quantiles of the sign recovery errors, from $10^4$ repetitions.  We estimate $K$ using the full information (i.e., the estimator (\ref{eqn_est_hm})) as well as 1-bit  estimator $\hat{\Lambda}_{0+,c}$ with selected values of $\eta\in\{0.2,\, 0.5,\ 1.5,\ 2,\ 3\}$. When $n=100$, except for $\eta=0.2$ (which is too small), the performance of $\hat{\Lambda}_{0+,c}$ is fairly stable with no essential difference from the estimator using full information. The performance of $\hat{\Lambda}_{0+,c}$  deteriorates with decreasing $n$. But even when $n=20$, the performance of  $\hat{\Lambda}_{0+,c}$ at $\eta=1.5$ (which is close to be optimal) is still very good. Note that, when a curve does not show in the panel (e.g., $n=50$, $\eta=3$, and $95\%$), it basically means the error is too large to fit in.  }\label{fig_SRC1bit}
\end{figure}

Figure~\ref{fig_SRC1bit} reports the sign recovery errors at $75\%$ quantile (upper panels) and $95\%$ quantile (bottom panels). The number of measurements for sparse recovery is chosen according to $M = \zeta K \log (N/0.01)$, although we only use $n\in\{20,\ 50, 100\}$  samples to estimate $K$. For comparison, Figure~\ref{fig_SRC1bit} also reports the results for estimating $K$ using $n$ full (i.e., infinite-bit) samples.  \\

When $n=100$, except for $\eta=0.2$ (which is too small), the performance of $\hat{\Lambda}_{0+,c}$ is fairly stable with no essential difference from the estimator using full information. The performance of $\hat{\Lambda}_{0+,c}$  deteriorates with decreasing $n$. But even for $n=20$,   $\hat{\Lambda}_{0+,c}$ at $\eta=1.5$ still performs well.

\section{2-Bit Coding and Estimation }

As shown by theoretical analysis and simulations, the performance of 1-bit coding and estimation is fairly good and stable for a wide range of threshold values. Nevertheless, it is desirable to further stabilize the estimates (and lower the variance) by using  more bits.

With the  2-bit scheme, we need to  introduce 3 threshold values: $C_1\leq C_2\leq C_3$.  We define
\begin{align}\notag
&p_1 = \mathbf{Pr}\left(z_j \leq C_1\right) = F_\alpha\left(C_1/\Lambda_\alpha\right)\\\notag
&p_2 = \mathbf{Pr}\left(C_1<z_j \leq C_2\right) = F_\alpha\left(C_2/\Lambda_\alpha\right) - F_\alpha\left(C_1/\Lambda_\alpha\right)\\\notag
&p_3 = \mathbf{Pr}\left(C_2<z_j \leq C_3\right) = F_\alpha\left(C_3/\Lambda_\alpha\right) - F_\alpha\left(C_2/\Lambda_\alpha\right)\\\notag
&p_4 = \mathbf{Pr}\left(z_j > C_3\right) = 1 - F_\alpha\left(C_3/\Lambda_\alpha\right)
\end{align}
and
\begin{align}\notag
&n_1 =  \sum_{j=1}^n 1\{z_j\leq C_1\},\hspace{0.15in}
n_2 =  \sum_{j=1}^n 1\{C_1<z_j\leq C_2\}\\\notag
&n_3 =  \sum_{j=1}^n 1\{C_2<z_j\leq C_3\},\hspace{0.15in}
n_4 =  \sum_{j=1}^n 1\{z_j> C_3\}
\end{align}
The log-likelihood of these $n = n_1+n_2+n_3+n_4$ observations can be expressed as
\begin{align}\notag
l =& n_1\log p_1 + n_2 \log p_2 +  n_3\log p_3 + n_4 \log p_4\\\notag
=& n_1\log F_\alpha\left(C_1/\Lambda_\alpha\right) + n_2\log \left[F_\alpha\left(C_2/\Lambda_\alpha\right)-F_\alpha\left(C_1/\Lambda_\alpha\right) \right]+ \\\notag
& n_3\log \left[F_\alpha\left(C_3/\Lambda_\alpha\right)-F_\alpha\left(C_2/\Lambda_\alpha\right) \right]+ n_4\log \left[1-F_\alpha\left(C_3/\Lambda_\alpha\right) \right],
\end{align}
from which we can derive the MLE and variance as presented in Theorem~\ref{thm_2bit}.

\begin{theorem}\label{thm_2bit}
Given $n$ i.i.d. samples $y_j \sim S(\alpha,1)$, $j =1$ to $n$, three thresholds $0<C_1\leq C_2\leq C_3$,   $n_1 =  \sum_{j=1}^n 1\{z_j\leq C_1\}$,  $n_2 =  \sum_{j=1}^n 1\{C_1<z_j\leq C_2\}$,  $n_3 =  \sum_{j=1}^n 1\{C_2<z_j\leq C_3\}$,  $n_4=  \sum_{j=1}^n 1\{z_j> C_3\}$, and
\begin{align}\notag
\eta_1 = \frac{\Lambda_\alpha}{C_1}, \hspace{0.2in} \eta_2 = \frac{\Lambda_\alpha}{C_2},\hspace{0.2in} \eta_3 = \frac{\Lambda_\alpha}{C_3}
\end{align}
the MLE, denoted by $\hat{\Lambda}_\alpha$, is the solution to the following equation:
\begin{align}\notag
0 =&  n_1\frac{C_1f_\alpha\left(1/\eta_1\right)}{F_\alpha\left(1/\eta_1\right)} +n_2\frac{C_2f_\alpha\left(1/\eta_2\right)-C_1f_\alpha\left(1/\eta_1\right)}{F_\alpha\left(1/\eta_2\right)-F_\alpha\left(1/\eta_1\right)}\\\notag
&+
n_3\frac{C_3f_\alpha\left(1/\eta_3\right)-C_2f_\alpha\left(1/\eta_2\right)}{F_\alpha\left(1/\eta_3\right)-F_\alpha\left(1/\eta_2\right)}
+n_4\frac{-C_3f_\alpha\left(1/\eta_3\right)}{1-F_\alpha\left(1/\eta_3\right)}
\end{align}
The asymptotic variance  of the MLE is
\begin{align}\notag
Var\left(\hat{\Lambda}_\alpha\right) = \frac{\Lambda^2_\alpha}{n}V_\alpha(\eta_1,\eta_2,\eta_3) + O\left(\frac{1}{n^2}\right)
\end{align}
where the variance factor can be expressed as
\begin{align}\notag
\frac{1}{V_\alpha(\eta_1,\eta_2,\eta_3)} = &
\frac{1}{\eta_1^2}\frac{f^2_\alpha\left(1/\eta_1\right)}{F_\alpha\left(1/\eta_1\right)}+\frac{1}{\eta_3^2}\frac{f^2_\alpha\left(1/\eta_3\right)}{1-F_\alpha\left(1/\eta_3\right)}
+\frac{\left[f_\alpha\left(1/\eta_2\right)/\eta_2-f_\alpha\left(1/\eta_1\right)/\eta_1\right]^2}{F_\alpha\left(1/\eta_2\right)-F_\alpha\left(1/\eta_1\right)}\\\notag
+&\frac{\left[f_\alpha\left(1/\eta_3\right)/\eta_3-f_\alpha\left(1/\eta_2\right)/\eta_2\right]^2}{F_\alpha\left(1/\eta_3\right)-F_\alpha\left(1/\eta_2\right)}
\end{align}
The asymptotic bias is
\begin{align}\notag
E\left(\hat{\Lambda}_\alpha\right) =\Lambda_\alpha\left(1+\frac{1}{nB}-\frac{D}{2nB^2}\right)+ O\left(\frac{1}{n^2}\right)
\end{align}
where
\begin{align}\notag
B = &\frac{\left(-\frac{C_1}{\Lambda_\alpha}\right)^2f_1^2}{F_1}
+\frac{ \left[\left(-\frac{C_2}{\Lambda_\alpha}\right)f_2 -\left(-\frac{C_1}{\Lambda_\alpha}\right)f_1 \right]^2}{F_2-F_1}
+\frac{ \left[\left(-\frac{C_3}{\Lambda_\alpha}\right)f_3 -\left(-\frac{C_2}{\Lambda_\alpha}\right)f_2 \right]^2}{F_3-F_2}
+\frac{\left(-\frac{C_3}{\Lambda_\alpha}\right)^2f_3^2}{1-F_3}
\end{align}
and
\begin{align}\notag
D=& \frac{\left(-\frac{C_1}{\Lambda_\alpha}\right)^3f_1f_1^\prime}{F_1}
+ \frac{\left[\left(-\frac{C_2}{\Lambda_\alpha}\right)f_2 -\left(-\frac{C_1}{\Lambda_\alpha}\right)f_1 \right]
\left[ \left(-\frac{C_2}{\Lambda_\alpha}\right)^2f_2^\prime - \left(-\frac{C_1}{\Lambda_\alpha}\right)^2f_1^\prime\right]}{F_2-F_1}\\\notag
+& \frac{\left[\left(-\frac{C_3}{\Lambda_\alpha}\right)f_3 -\left(-\frac{C_2}{\Lambda_\alpha}\right)f_2 \right]
\left[\left(-\frac{C_3}{\Lambda_\alpha}\right)^2f_3^\prime - \left(-\frac{C_2}{\Lambda_\alpha}\right)^2f_2^\prime\right]}{F_3-F_2}
+\frac{\left(-\frac{C_3}{\Lambda_\alpha}\right)^3f_3  f_3^\prime}{1-F_3}
\end{align}
\textbf{Proof:}\hspace{0.2in} See Appendix~\ref{proof_thm_2bit}.$\hfill\Box$
\end{theorem}

The asymptotic bias formula in Theorem~\ref{thm_2bit} leads to a bias-corrected estimator
\begin{align}
\hat{\Lambda}_{\alpha,c}=\frac{\hat{\Lambda}_\alpha}{1+\frac{1}{nB}-\frac{D}{2nB^2}}
\end{align}
Note that, with a slight abuse of notation, we still use $\hat{\Lambda}_\alpha$ to denote the MLE of the 2-bit scheme and we rely on the number of parameters (e.g., $\eta_1$, $\eta_2$, $\eta_3$) to differentiate $V_\alpha$ for different schemes.

\subsection{$\alpha\rightarrow0+$}

\begin{figure}[h!]
\begin{center}
\mbox{
\includegraphics[width=2.2in]{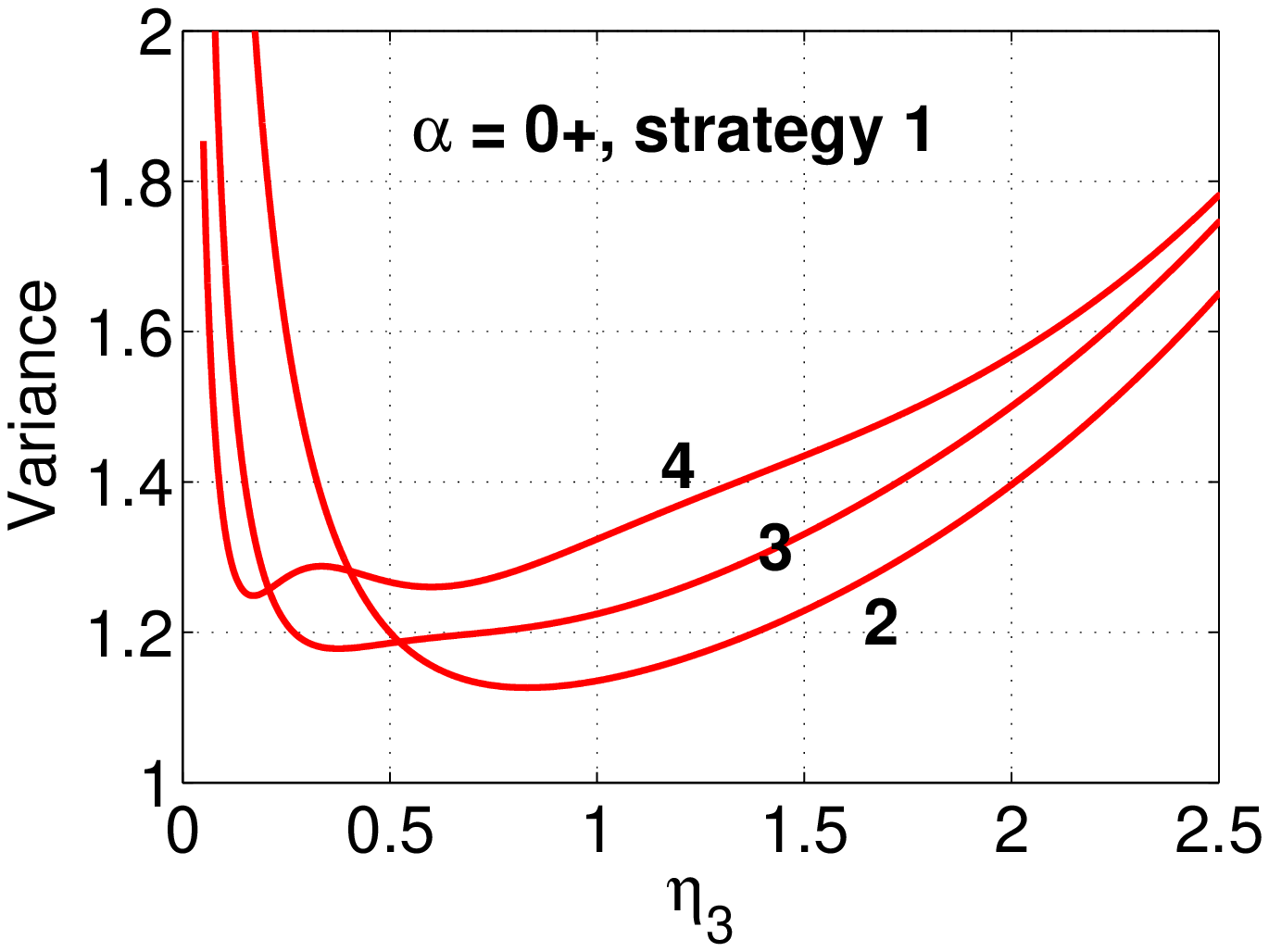}\hspace{0.2in}
\includegraphics[width=2.2in]{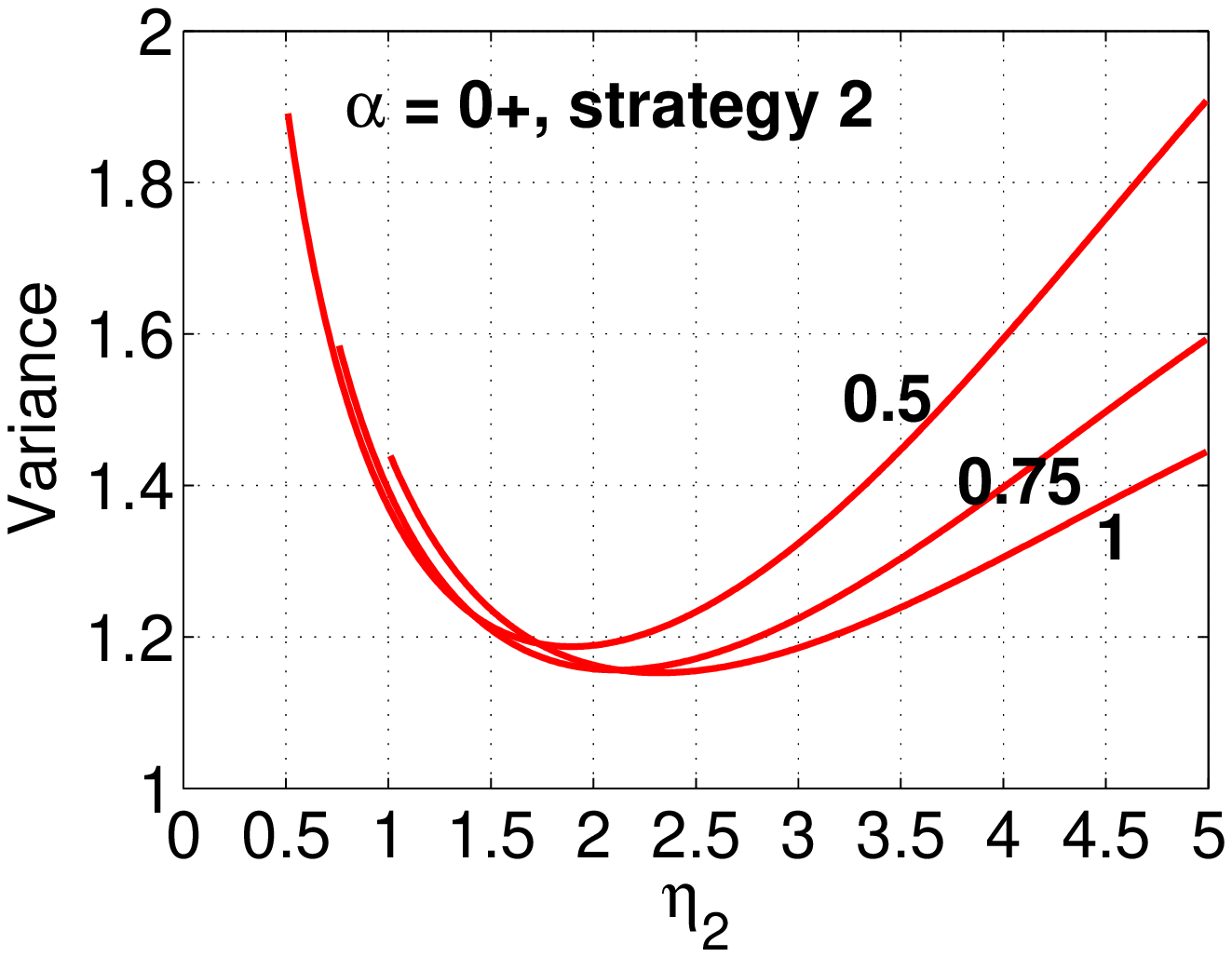}
}
\end{center}
\vspace{-0.3in}
\caption{\textbf{Left} (strategy 1): \ $V_{0+}\left(\eta_1,\eta_2,\eta_3\right)$ for $\eta_2 = t \eta_3$, $\eta_1 = t\eta_2$, at $t=2, 3, 4$, with varying $\eta_3$.
\textbf{Right} (strategy 2): \ $V_{0+}$ for fixed $\eta_1 = 5$, $\eta_3\in\{0.5,\ 0.75,\ 1\}$, and $\eta_2$ varying between $\eta_3$ and $\eta_1$. }\label{fig_V3Al0}
\end{figure}

In this case, we can slightly simplify the expression:
\begin{align}\notag
V_\alpha(\eta_1,\eta_2,\eta_3)
=&\frac{1}{\frac{(\eta_1-\eta_2)^2}{e^{\eta_1}-e^{\eta_2}} +\frac{(\eta_2-\eta_3)^2}{e^{\eta_2}-e^{\eta_3}} +\frac{\eta_3^2}{e^{\eta_3}-1}}
\end{align}
Numerically, the minimum of $V_{0+}(\eta_1,\eta_2,\eta_3)$ is 1.122, attained at $\eta_1 = 3.365, \eta_2 = 1.771, \eta_3 = 0.754$. The value 1.122 is substantially smaller than 1.544 which is the minimum variance  coefficient of the 1-bit scheme. Figure~\ref{fig_V3Al0} illustrates that, with the 2-bit scheme, the variance is less sensitive to the choice of the thresholds, compared to the 1-bit scheme.\\

In practice, there are at least two simple strategies for selecting the parameters $\eta_1\geq \eta_2\geq \eta_3$:
\begin{itemize}
\item {\em Strategy 1}:  First select a ``small'' $\eta_3$, then let $\eta_2 = t \eta_3$ and $\eta_1 = t\eta_2$, for some $t>1$.
\item {\em Strategy 2}:  First select a ``small'' $\eta_3$  and a ``large'' $\eta_1$, then select a ``reasonable'' $\eta_2$ in between.
\end{itemize}

See the plots for  examples of the two strategies in Figure~\ref{fig_V3Al0}. We  re-iterate that for the task of estimating $\Lambda_\alpha$ using only a few bits, we must
choose parameters (thresholds) beforehand. While in general the optimal results are not attainable,  as long as the chosen parameters fall in a ``reasonable'' range (which is fairly wide),  the estimation variance will not be  far away from the  optimal value.

\newpage

\subsection{$\alpha=1$}

\begin{figure}[h!]
\begin{center}
\mbox{
\includegraphics[width=2.2in]{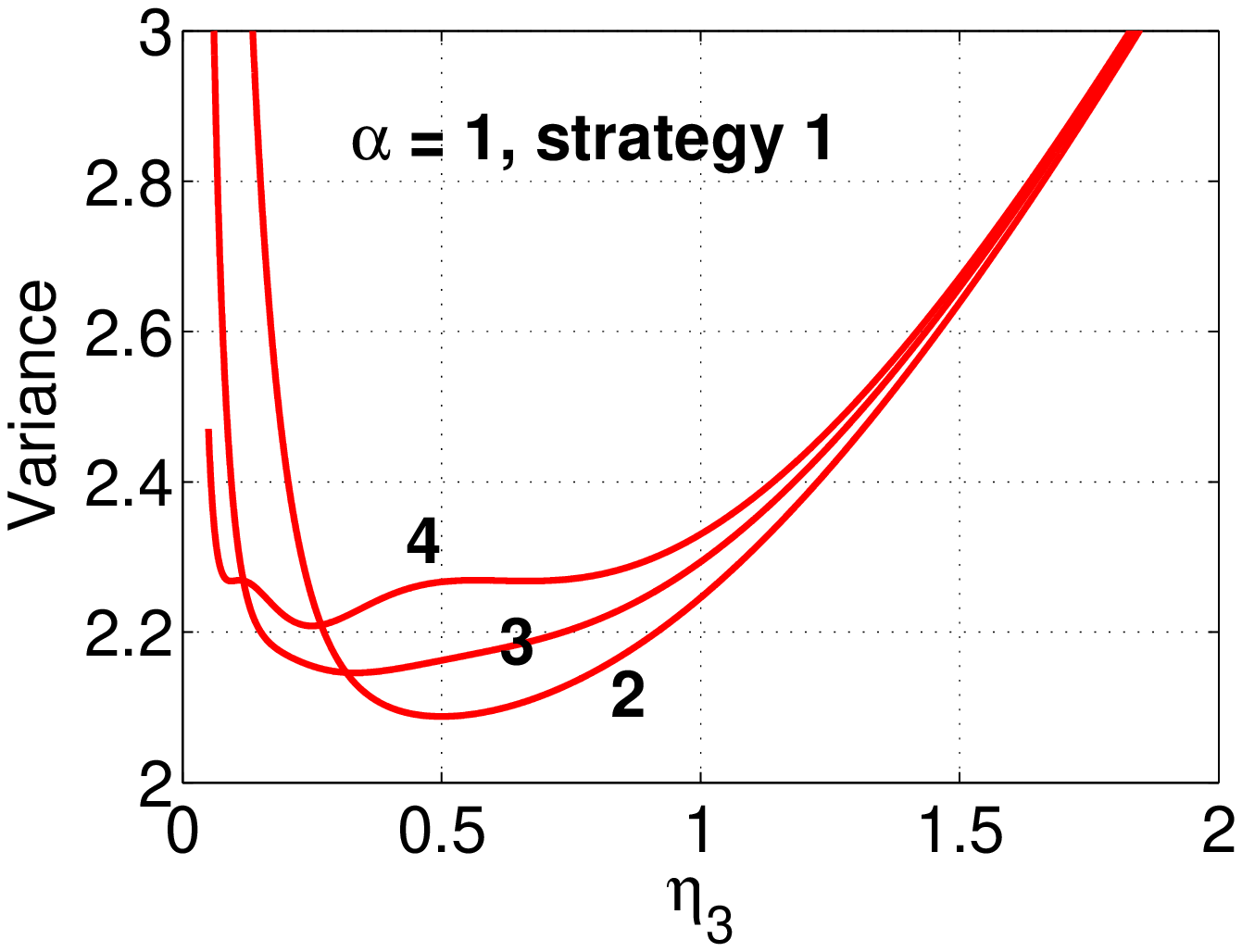}\hspace{0.2in}
\includegraphics[width=2.2in]{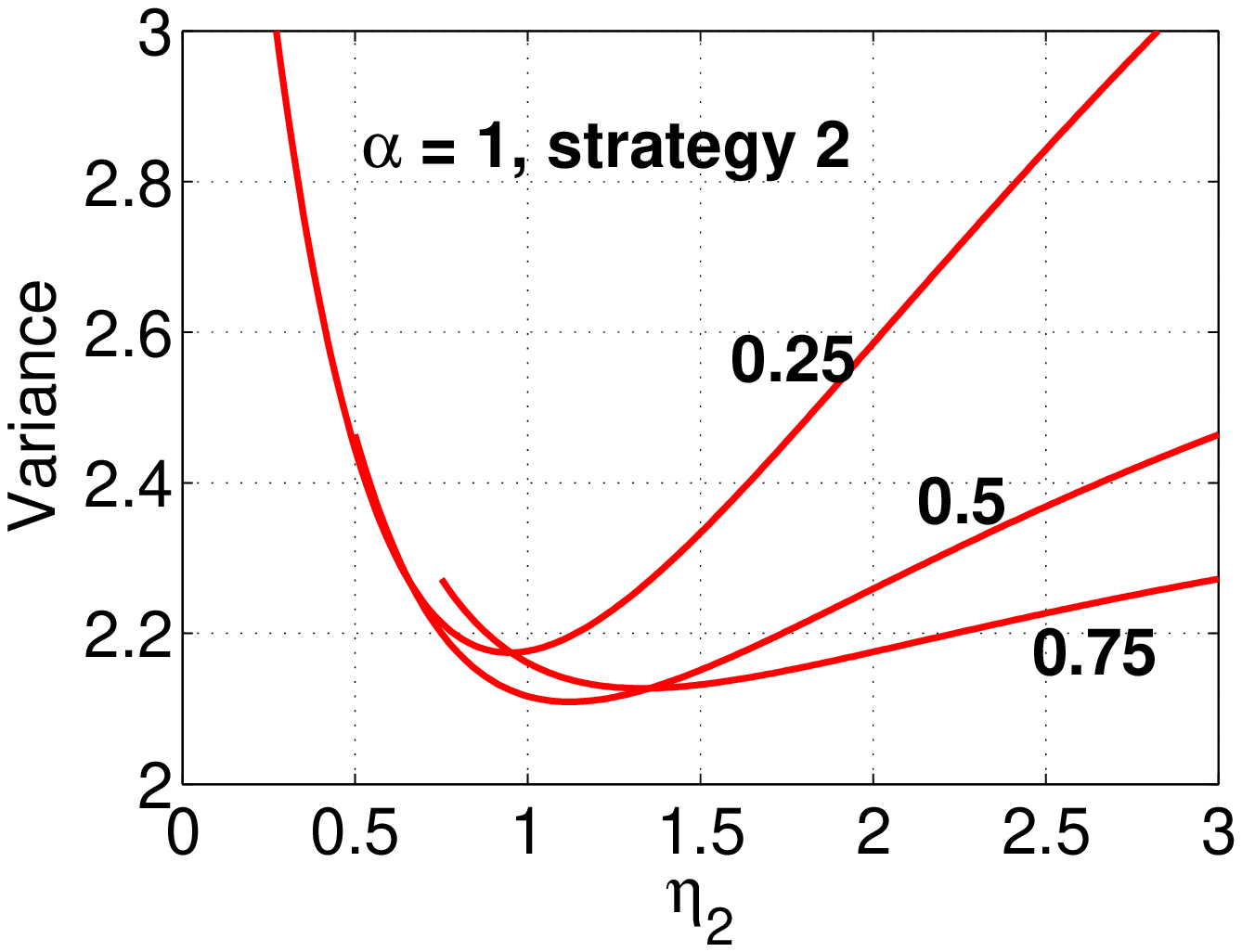}
}
\end{center}
\vspace{-0.3in}
\caption{\textbf{Left} (strategy 1): \ $V_{1}\left(\eta_1,\eta_2,\eta_3\right)$ for $\eta_2 = t \eta_3$, $\eta_1 = t\eta_2$, at $t=2, 3, 4$, with varying   $\eta_3$.
\textbf{Right} (strategy 2): \ $V_{1}$ for fixed $\eta_1 = 3$, $\eta_3\in\{0.25,\ 0.5,\  0.75\}$, and $\eta_2$ varying between $\eta_3$ and $\eta_1$. }\label{fig_V3Al1}
\end{figure}

Numerically, the minimum of $V_1(\eta_1,\eta_2,\eta_3)$ is 2.087, attained at $\eta_1 = 1.927,\hspace{0.05in}    \eta_2 = 1.000,\,\hspace{0.05in}  \eta_3 = 0.519$. \ Note that the value $2.087$ is very close to the optimal variance coefficient 2 using full information.  Figure~\ref{fig_V3Al1} plots the examples of $V_1(\eta_1,\eta_2,\eta_3)$  for both ``strategy 1''  and ``strategy 2''.

\subsection{$\alpha=2$}

\begin{figure}[h!]
\begin{center}
\mbox{
\includegraphics[width=2.2in]{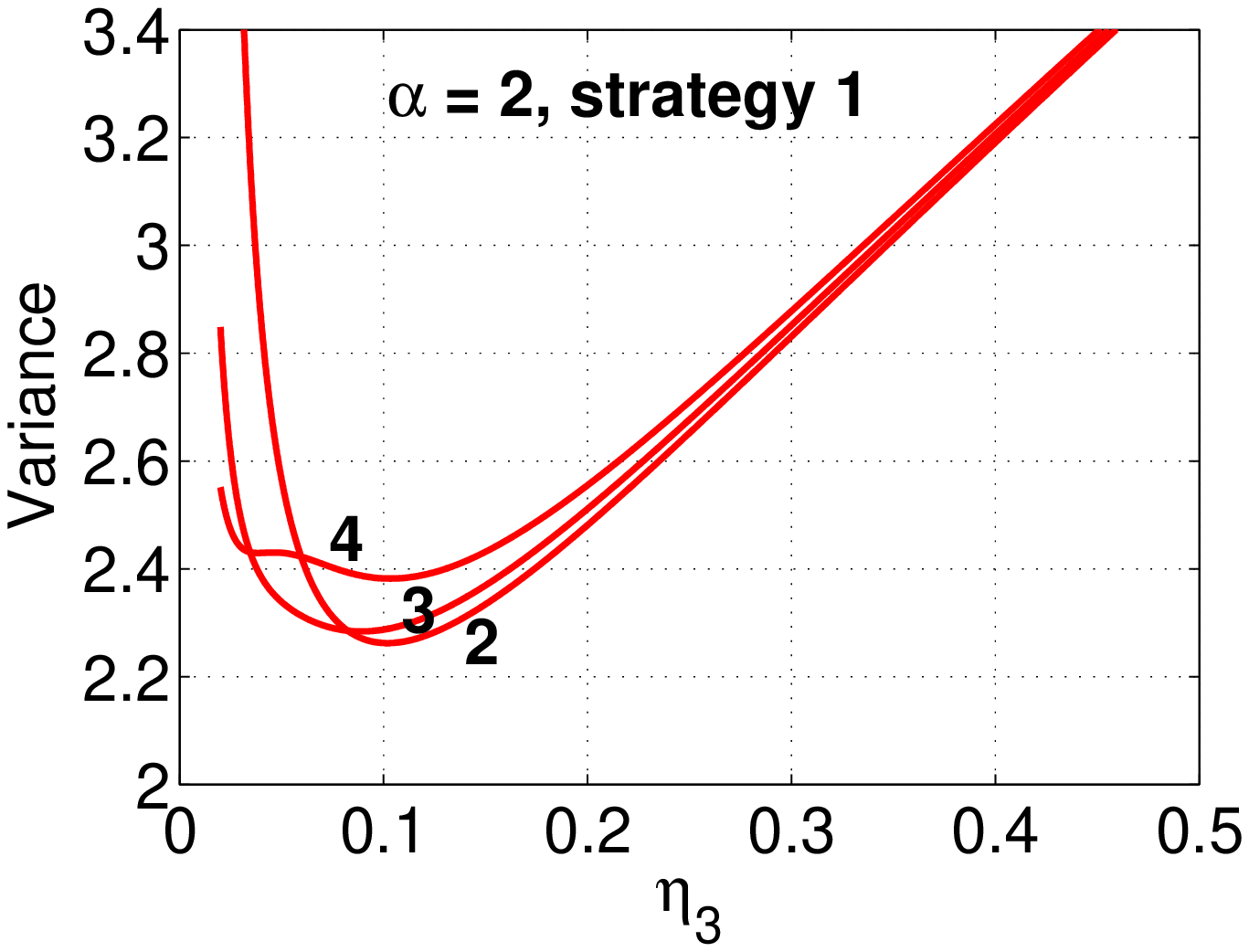}\hspace{0.2in}
\includegraphics[width=2.2in]{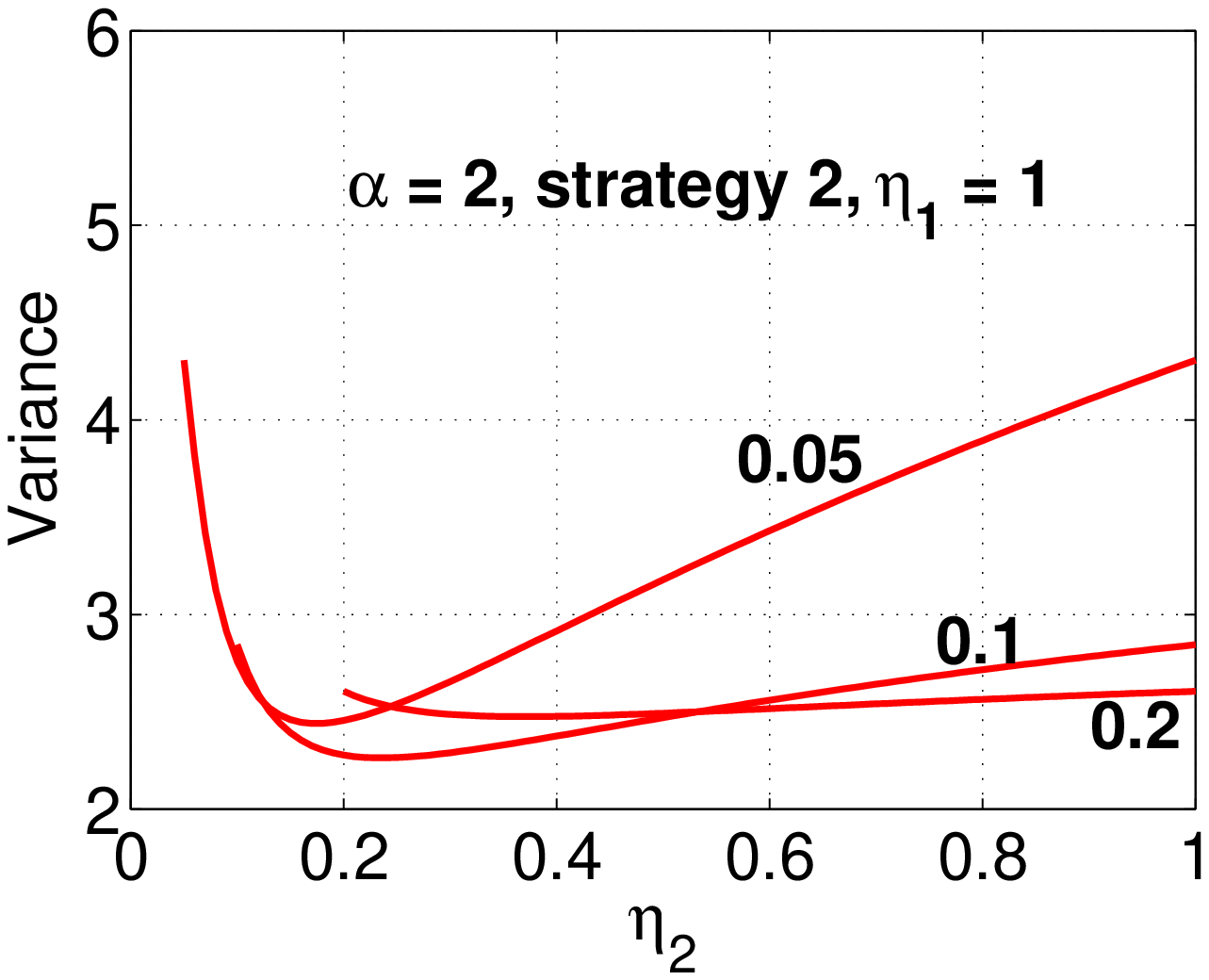}
}

%\mbox{
%\includegraphics[width=2.2in]{fig/Var3Al2Method2Eta2.eps}\hspace{0.2in}
%\includegraphics[width=2.2in]{fig/Var3Al2Method2Eta1.eps}
%}

\end{center}
\vspace{-0.3in}
\caption{
\textbf{Left} (strategy 1): \ $V_{1}\left(\eta_1,\eta_2,\eta_3\right)$ for $\eta_2 = t \eta_3$, $\eta_1 = t\eta_2$, at $t=2, 3, 4$, with varying   $\eta_3$.
\textbf{Right} (strategy 2): \ $V_{1}$ for fixed $\eta_1 = 1$, $\eta_3\in\{0.05,\ 0.1,\  0.2\}$, and $\eta_2$ varying between $\eta_3$ and $\eta_1$.
 }\label{fig_V3Al2}
\end{figure}

Numerically, the minimum of $V_2(\eta_1,\eta_2,\eta_3)$ is 2.236, attained at
$\eta_1 = 0.546,\hspace{0.1in}    \eta_2 = 0.195,\hspace{0.1in}  \eta_3 = 0.093$.  Figure~\ref{fig_V3Al2} presents  examples of $V_2(\eta_1,\eta_2,\eta_3)$ for both strategies for choosing $\eta_1$, $\eta_2$, and $\eta_3$.

\subsection{Simulations}

\begin{figure}[h!]
\begin{center}
\mbox{
\includegraphics[width=2.2in]{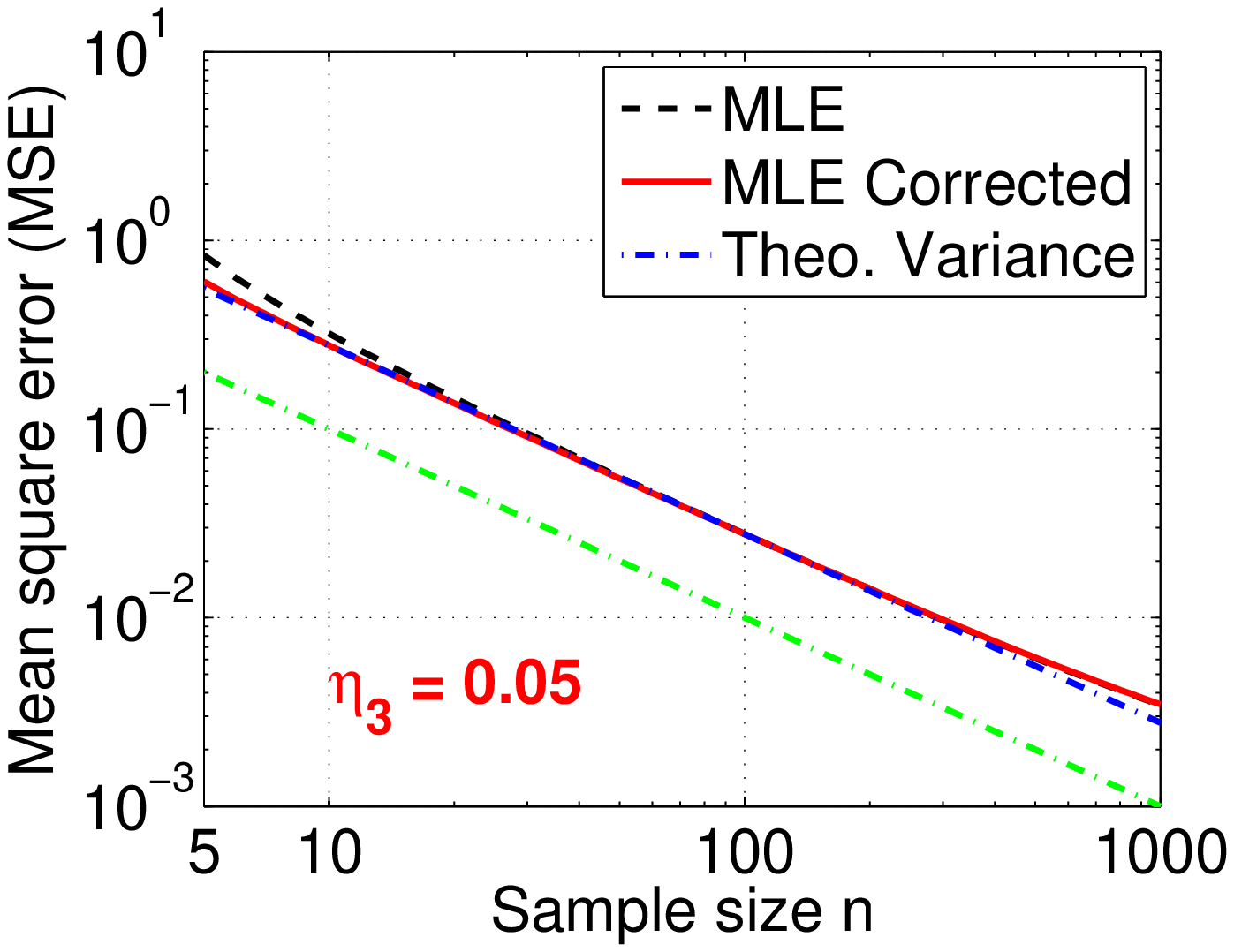}\hspace{0in}
\includegraphics[width=2.2in]{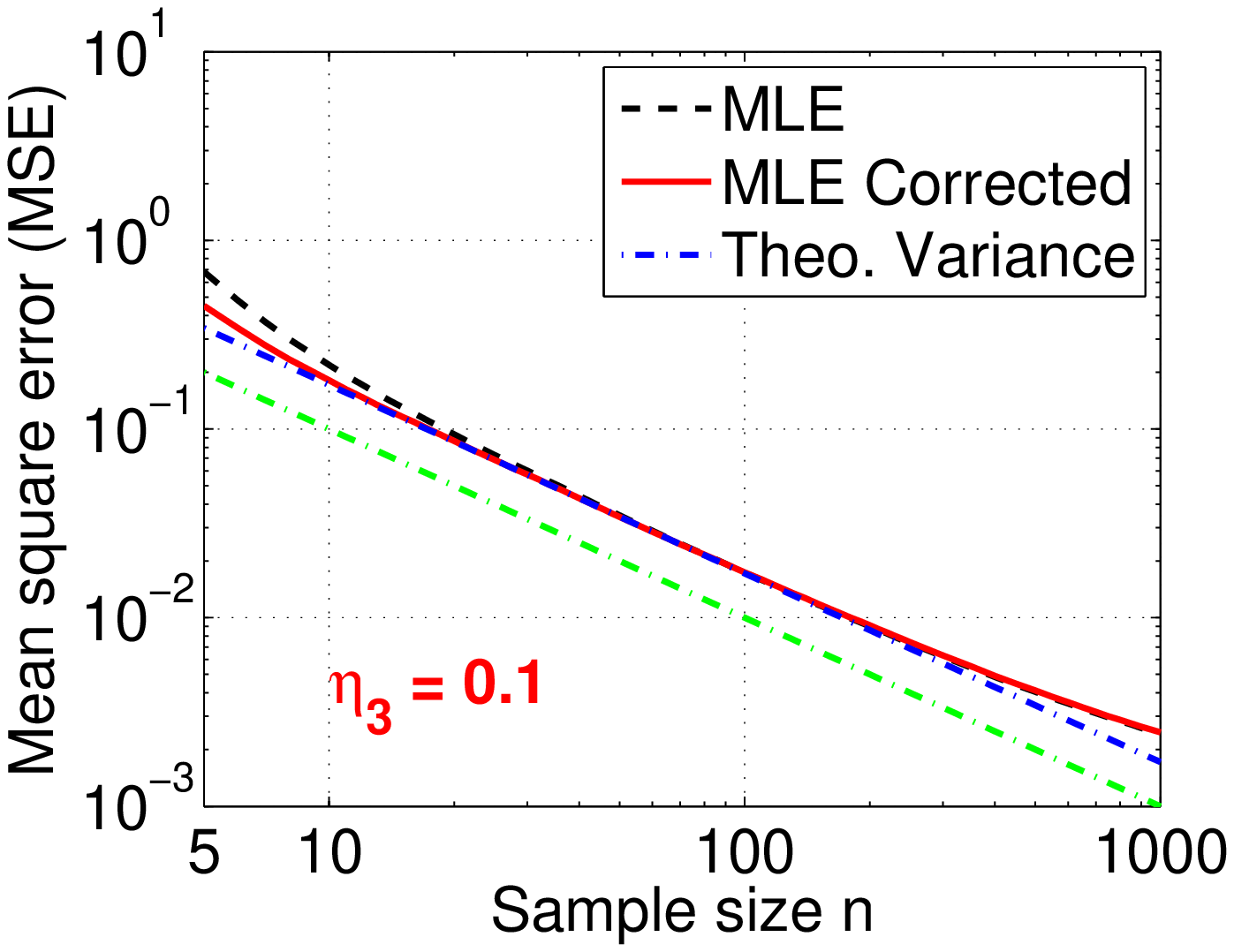}\hspace{0in}
\includegraphics[width=2.2in]{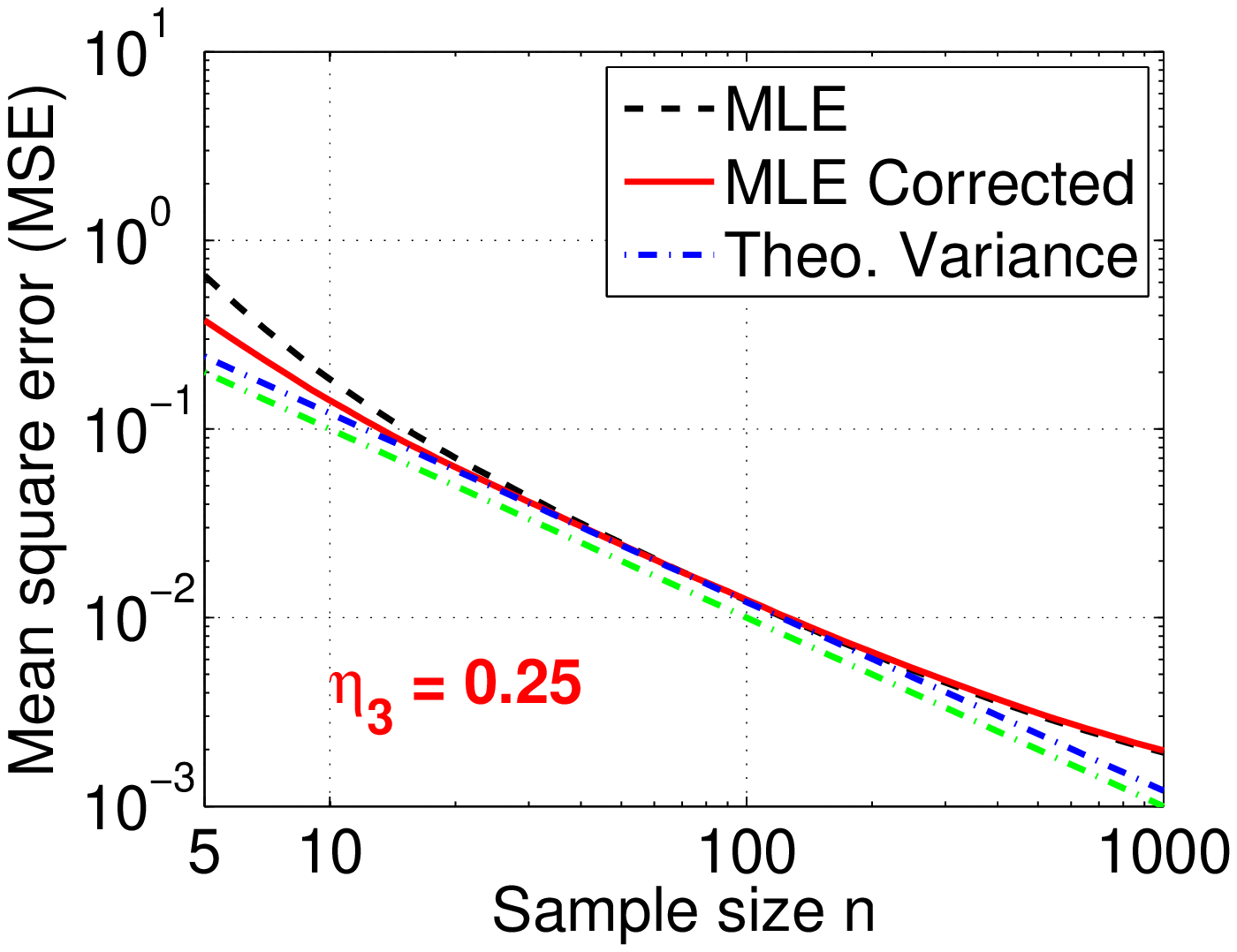}
}

\mbox{
\includegraphics[width=2.2in]{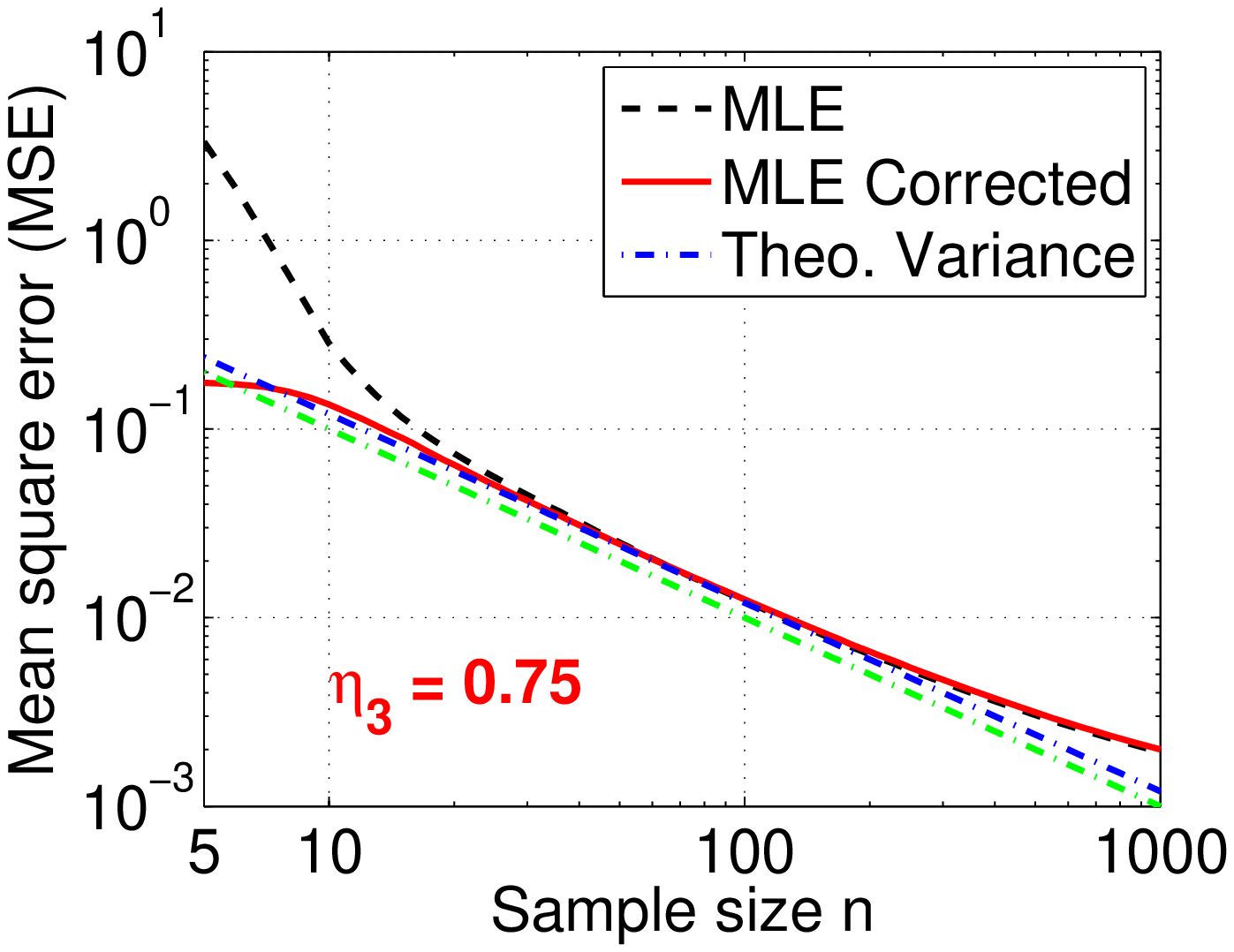}\hspace{0in}
\includegraphics[width=2.2in]{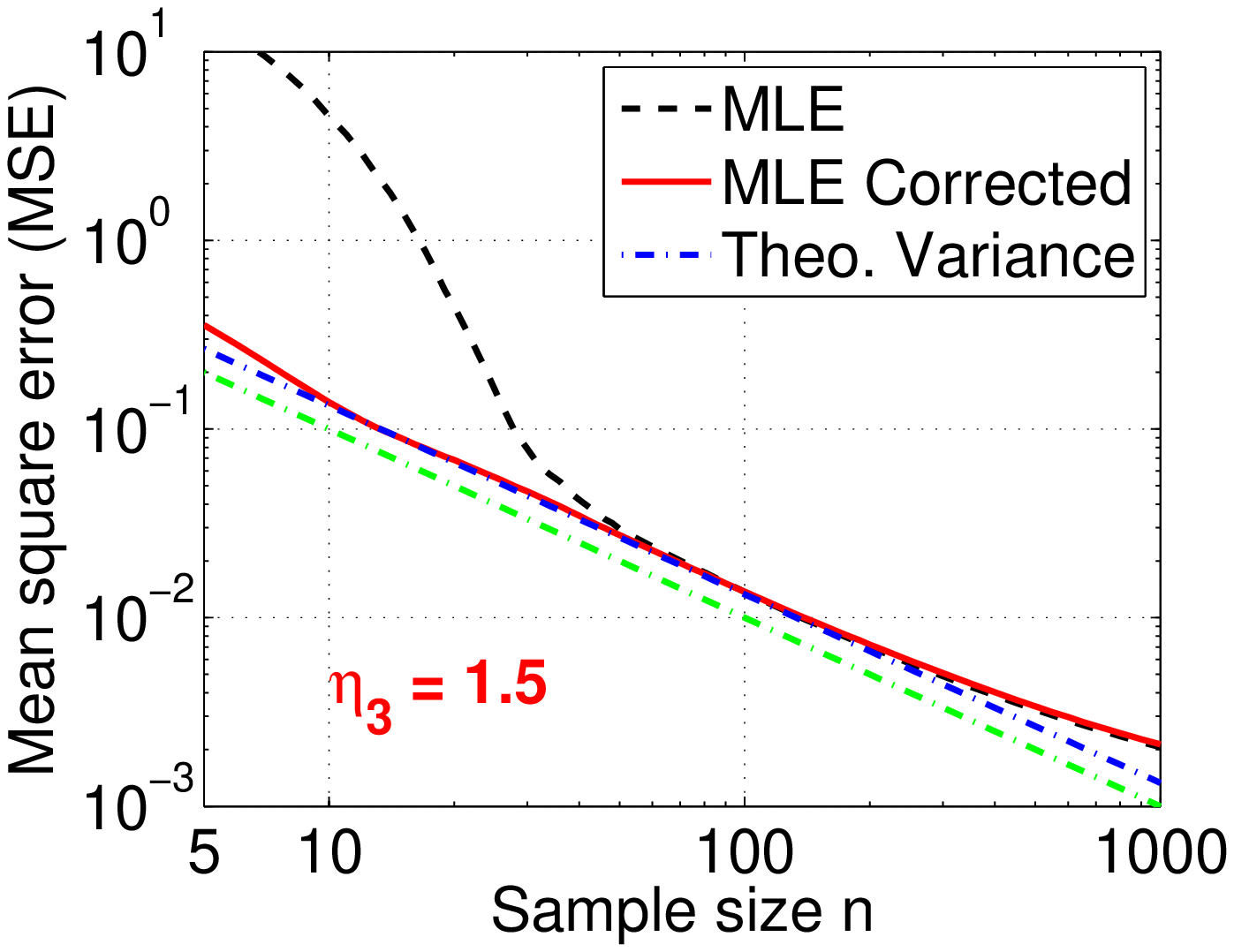}\hspace{0in}
\includegraphics[width=2.2in]{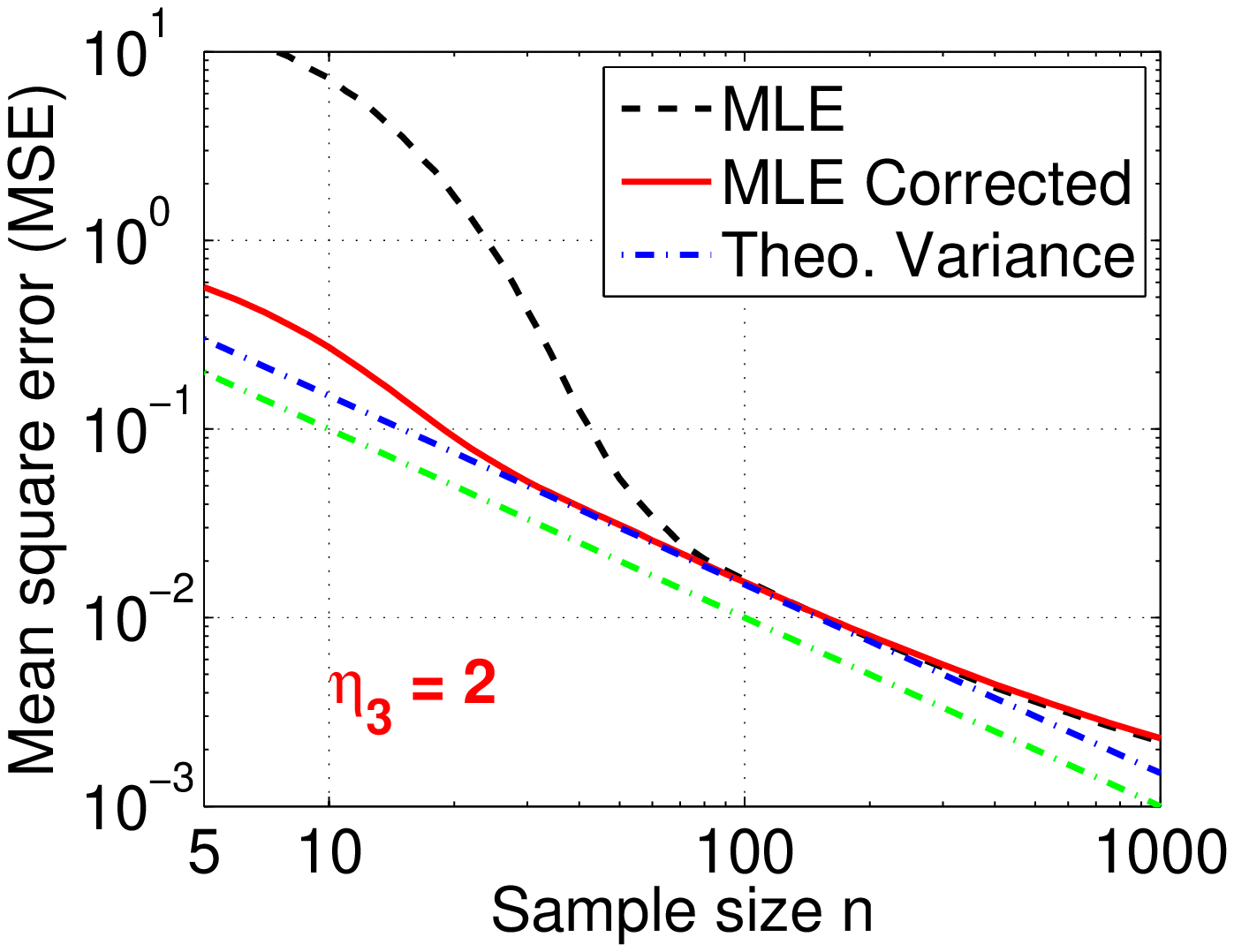}
}

\end{center}
\vspace{-0.3in}
\caption{Empirical Mean square errors of the 2-bit estimators: $\hat{\Lambda}_{0+}$ (dashed curves) and $\hat{\Lambda}_{0+,c}$ (solid curves), for $10^6$ simulations at each sample size $n$. We use $\alpha=0.05$ to generate stable samples $S(\alpha,1)$ and we consider 6 different $\eta_3 = \frac{\Lambda_\alpha}{C_3}$ values presented in 6 panels. We always let $\eta_2 = 3 \eta_3$ and $\eta_1 = 3\eta_2$.  For both estimators, the empirical MSEs converge to the theoretical asymptotic variances (\ref{eqn_Var}) (dashed dot curves and blue if color is available) when $n$ is not small.  In each panel, the lowest curve (dashed dot and green if color is available) represents the theoretical variances using full (infinite-bit) information, i.e., $1/n$ in this case.  When $n$ is small, the bias-correction step important.  Note that the small (and exaggerated) discrepancies at  large $n$ are due to the fact that we use  $\alpha=0.05$ to  simulate the data and use estimators based on $\alpha=0+$.  }\label{fig_Mse2bitP0}
\end{figure}

Figure~\ref{fig_Mse2bitP0}  presents the simulation results for verifying the 2-bit estimator $\hat{\Lambda}_{0+}$ and its bias-corrected version $\hat{\Lambda}_{0+,c}$. For simplicity, we choose $\eta_3 \in \{0.05,\ 0.1,\ 0.25,\ 0.75,\ 1.5,\ 2\}$ and we fix $\eta_2 = 3\eta_3$, $\eta_1  = 3\eta_2$. Although these choices are not optimal, we can see from Figure~\ref{fig_Mse2bitP0} that the estimators  still perform  well for such a wide range of $\eta_3$ values. Compared to 1-bit estimators, the 2-bit estimators are noticeably more accurate and  less sensitive to  parameters. Again, the bias-correction step is useful when the sample size $n$ is not large. \\

\begin{figure}[h!]
\begin{center}
\mbox{
\includegraphics[width=2.2in]{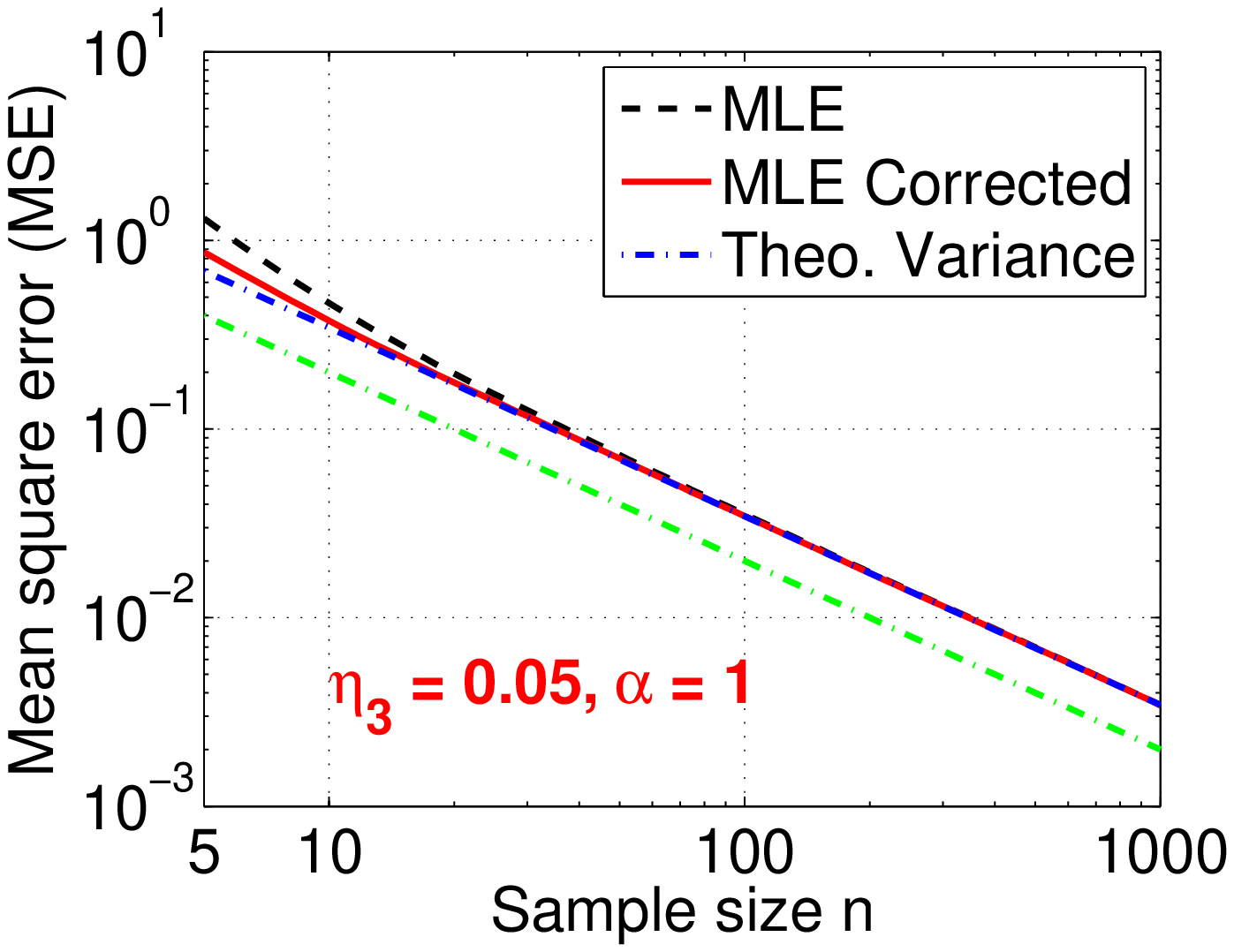}\hspace{0in}
\includegraphics[width=2.2in]{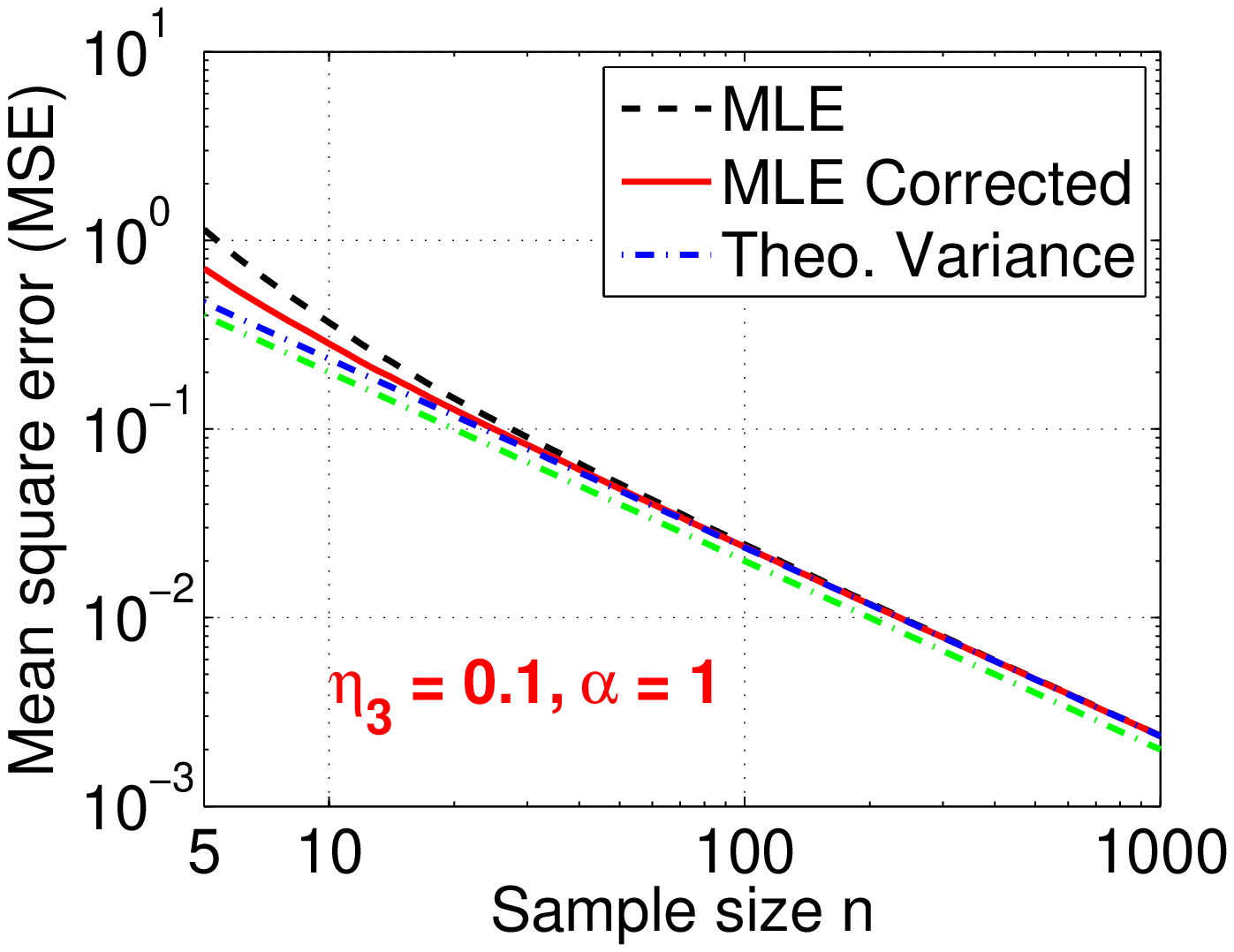}\hspace{0in}
\includegraphics[width=2.2in]{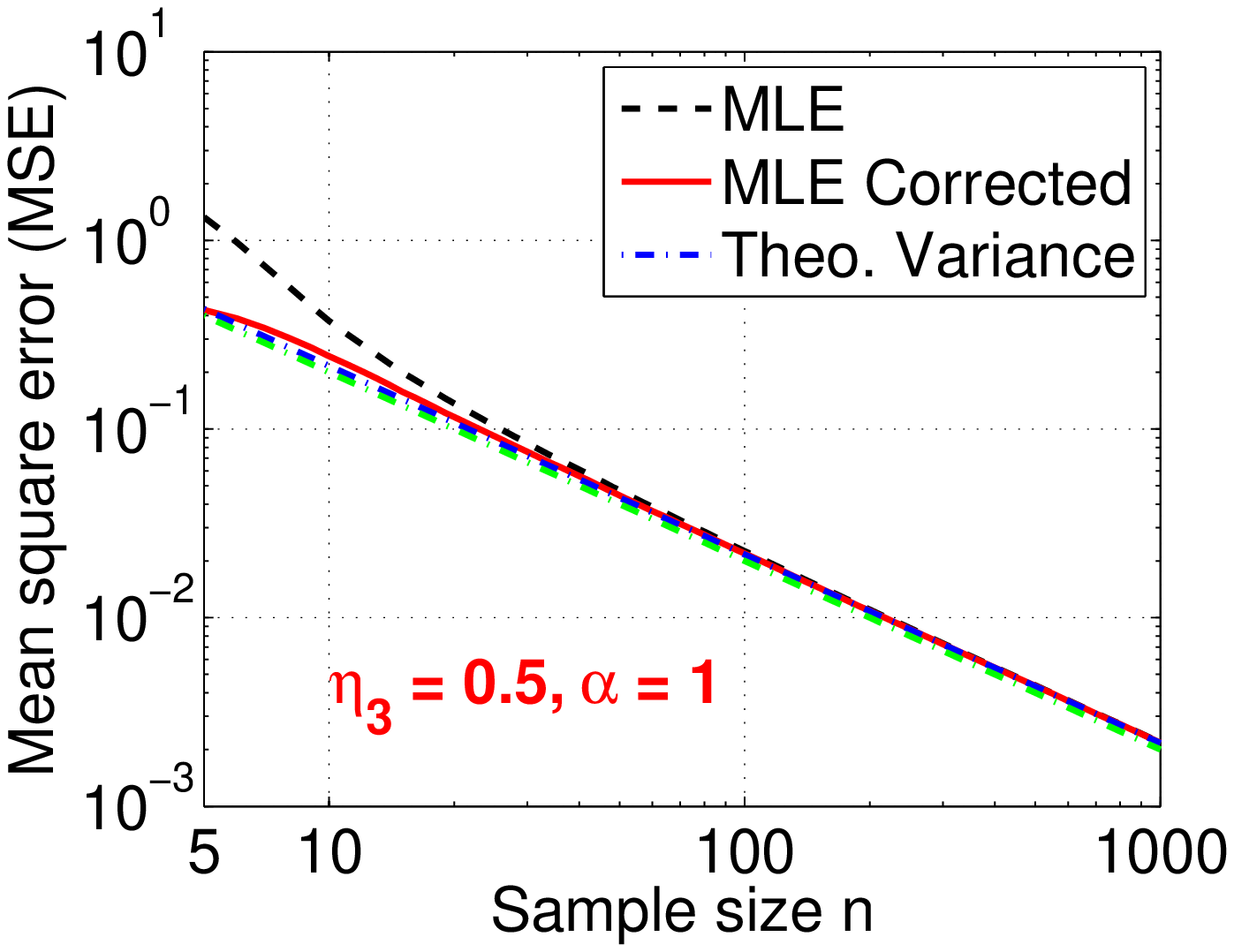}
}

\mbox{
\includegraphics[width=2.2in]{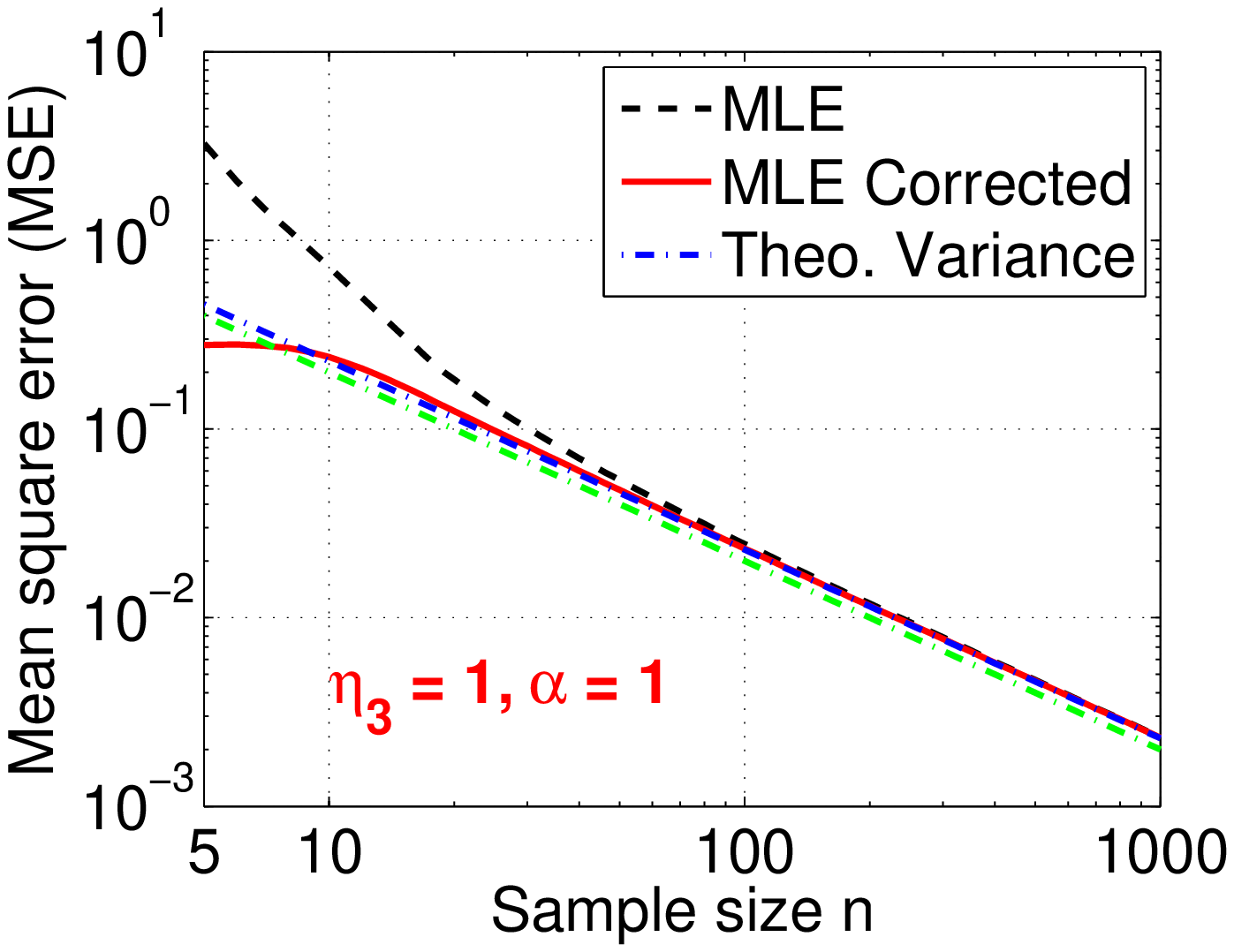}\hspace{0in}
\includegraphics[width=2.2in]{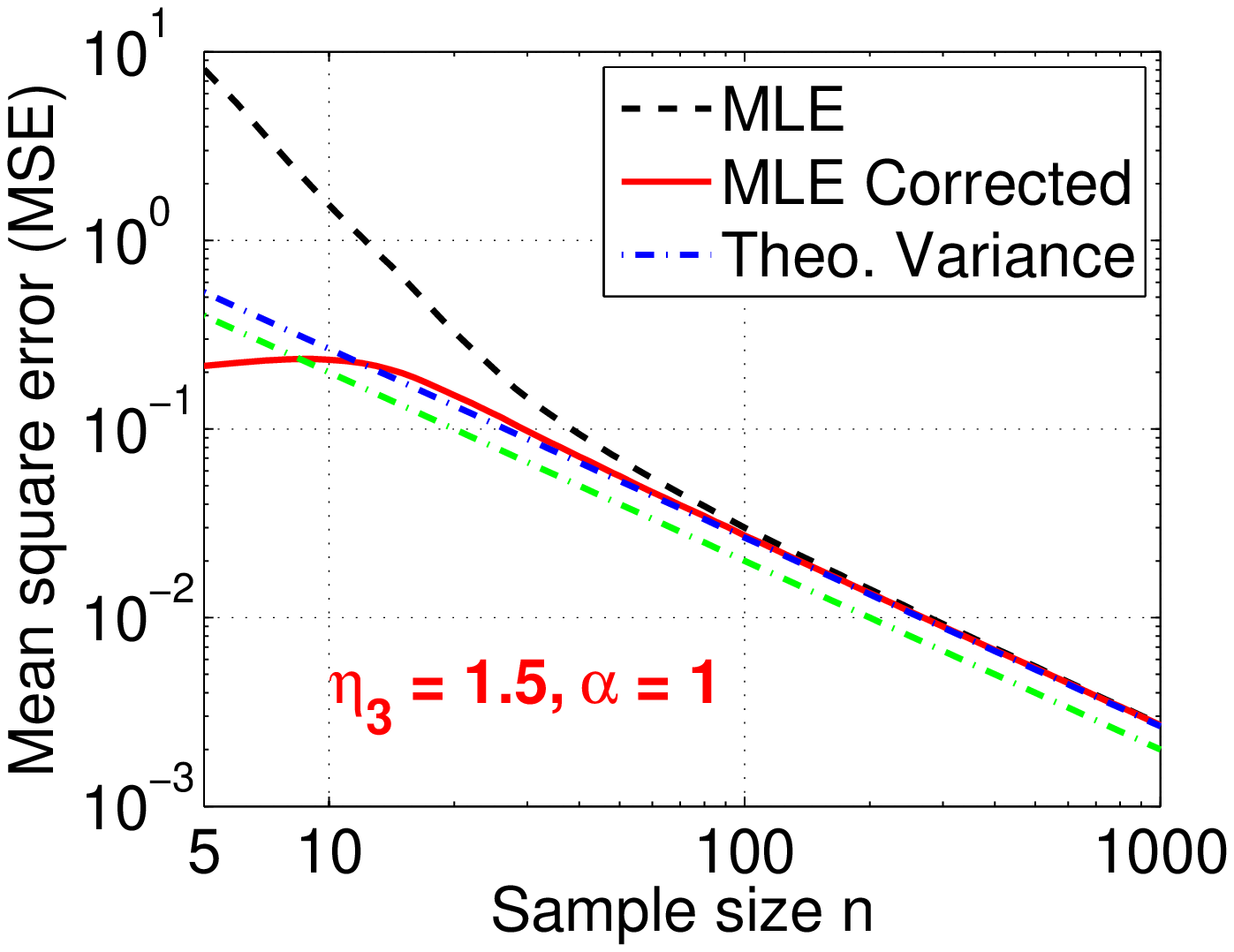}\hspace{0in}
\includegraphics[width=2.2in]{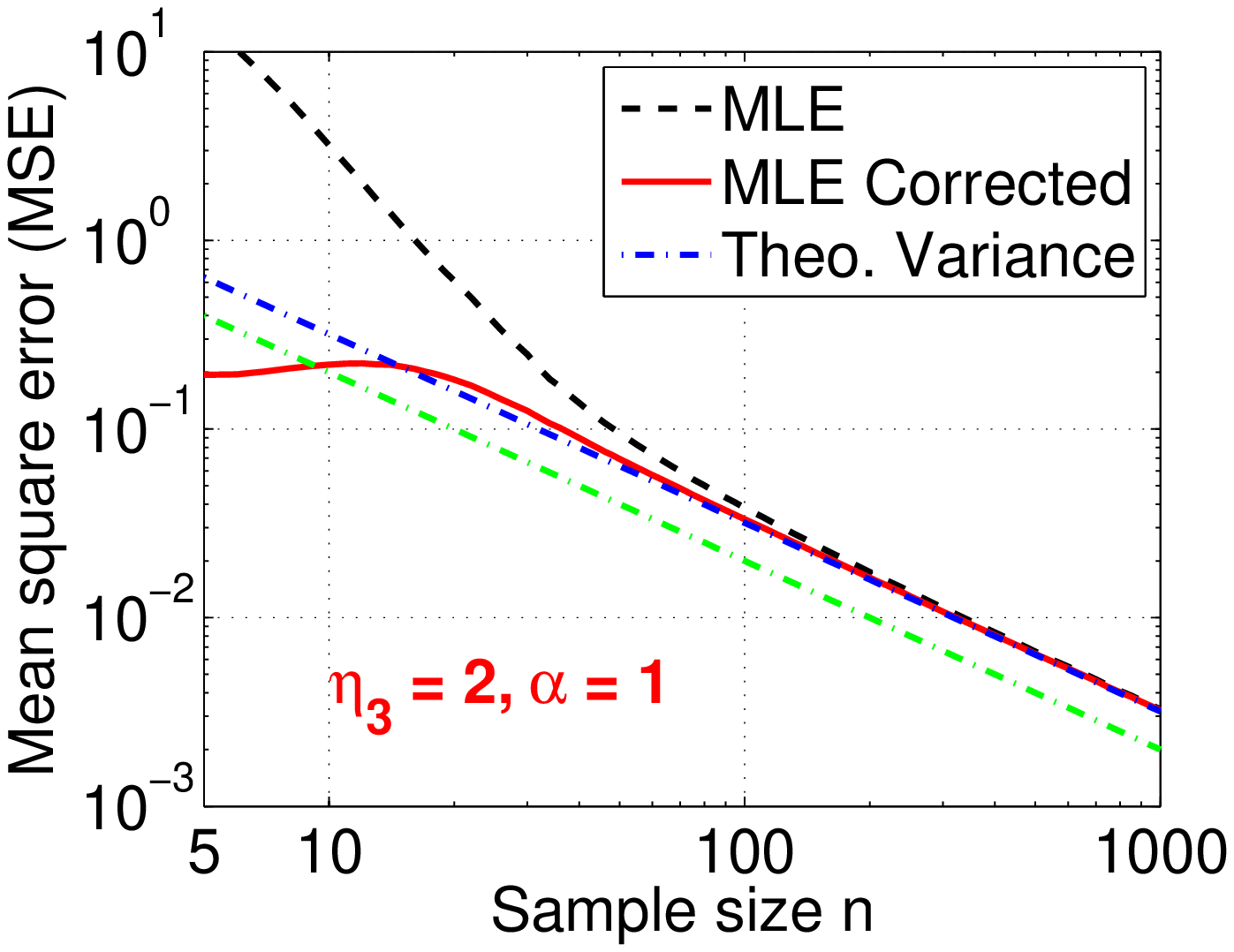}
}

\end{center}
\vspace{-0.3in}
\caption{Mean square errors of the 2-bit estimator $\hat{\Lambda}_{1}$ (dashed curves) and its bias-corrected version $\hat{\Lambda}_{1,c}$ (solid curves), for $\alpha=1$, by using 6 different $\eta_3$ values (one for each panel) and fixing $\eta_2=3\eta_3$, $\eta_1=3\eta_2$. The lowest curve (dashed dot and green if color is available) in each panel represents the optimal variance using full  information, which is $2/n$ for $\alpha=1$.}\label{fig_Mse2bitP1}
\end{figure}

Similar to Figure~\ref{fig_Mse1bitP0}, we can  observe some discrepancies at  large $n$ (as magnified by the log-scale of the y-axis). Again, this is because we simulate the data using $\alpha=0.05$ and we use estimators based on $\alpha=0+$. To remove this effect, we also provide simulations  for $\alpha=1$ in Figure~\ref{fig_Mse2bitP1}.

\clearpage

\subsection{Efficient Computational Procedure for the MLE Solutions}

With the 1-bit scheme, the  cost for computing the MLE is negligible because of the closed-form solution.  With the 2-bit scheme, however, the computational cost might be a concern if we try to find the MLE solution numerically every time (at run time).  A computationally efficient solution is  to tabulate the results.  To see this, we can re-write the log-likelihood function
\begin{align}\notag
l =& \frac{n_1}{n}\log F_\alpha\left(1/\eta_1\right) + \frac{n_2}{n}\log \left[F_\alpha\left(1/\eta_2\right)-F_\alpha\left(1/\eta_1\right) \right]\\\notag
&+ \frac{n_3}{n}\log \left[F_\alpha\left(1/\eta_3\right)-F_\alpha\left(\eta_2\right) \right]
+ \frac{n-(n_1+n_2+n_3)}{n}\log \left[1-F_\alpha\left(\eta_3\right) \right]
\end{align}
This means, we only need to tabulate the results for the combination of $n_1/n, n_2/n, n_3/n$ (which all vary between 0 and 1). Suppose we tabulate $T$ values for each $n_i/n$ (i.e., at an accuracy of $1/T$), then the table size is only $T^3$, which is  merely $10^6$ if we let $T=100$.  \\

\begin{figure}[h!]
\begin{center}
\mbox{
\includegraphics[width=2.2in]{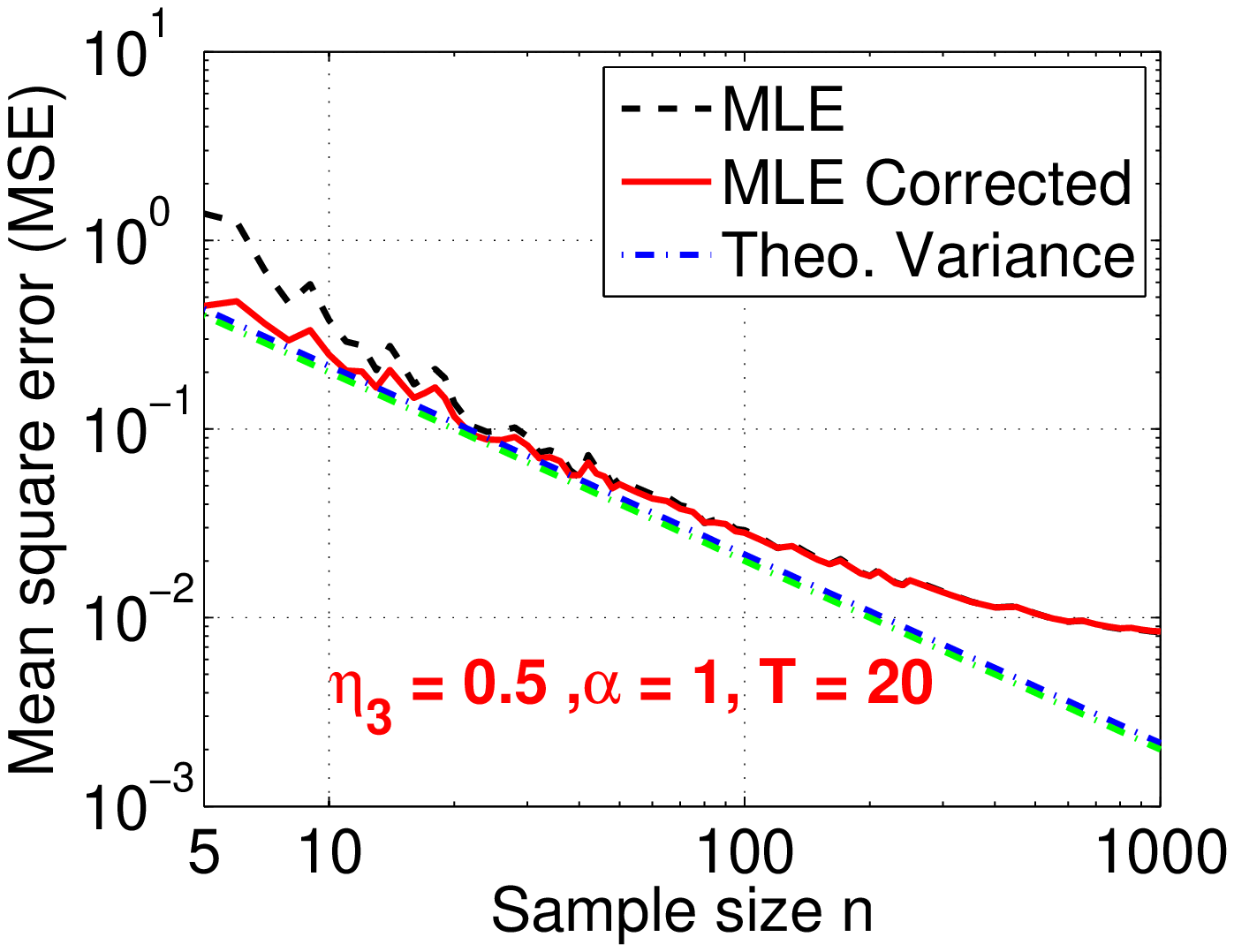}\hspace{0.2in}
\includegraphics[width=2.2in]{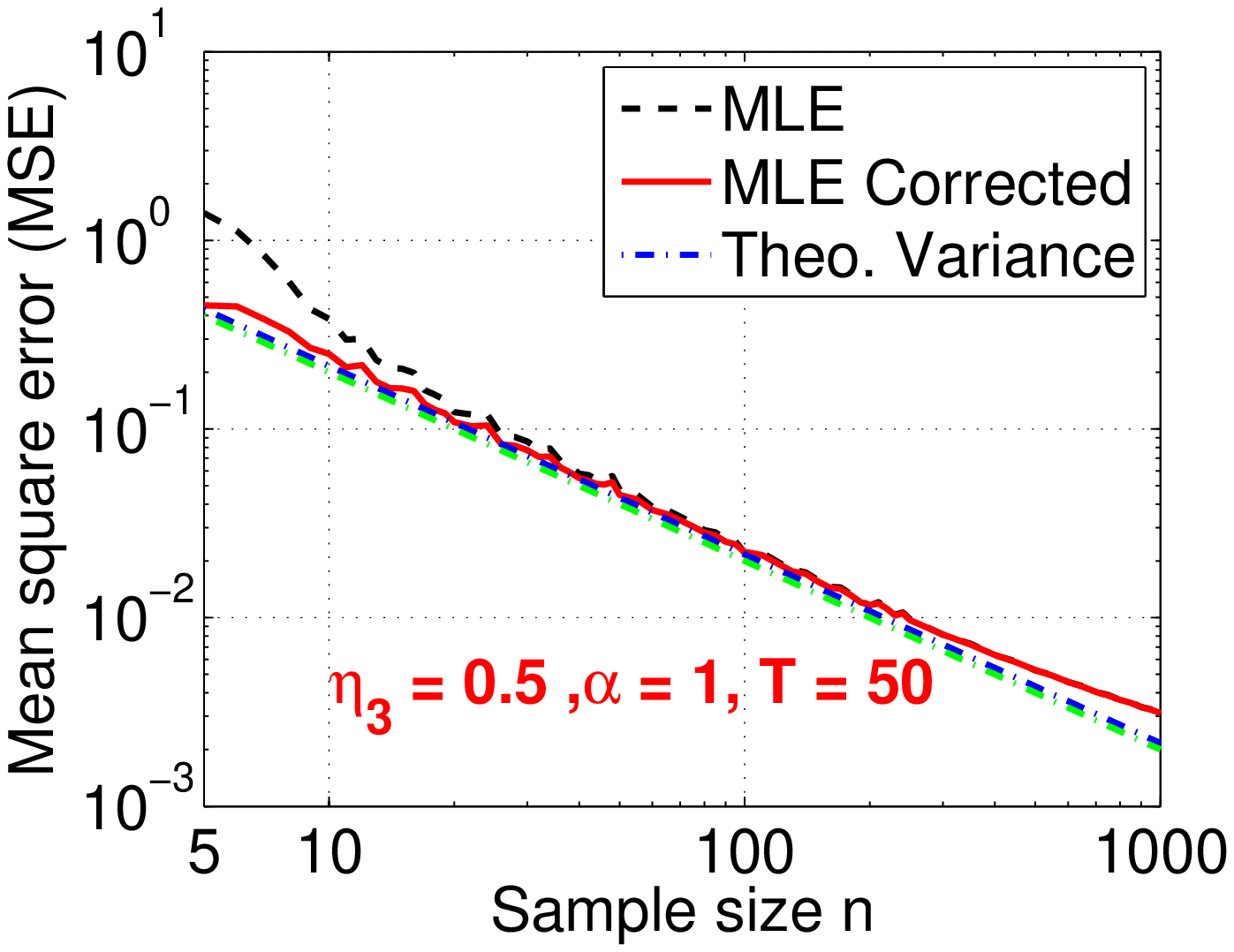}

}

\mbox{
\includegraphics[width=2.2in]{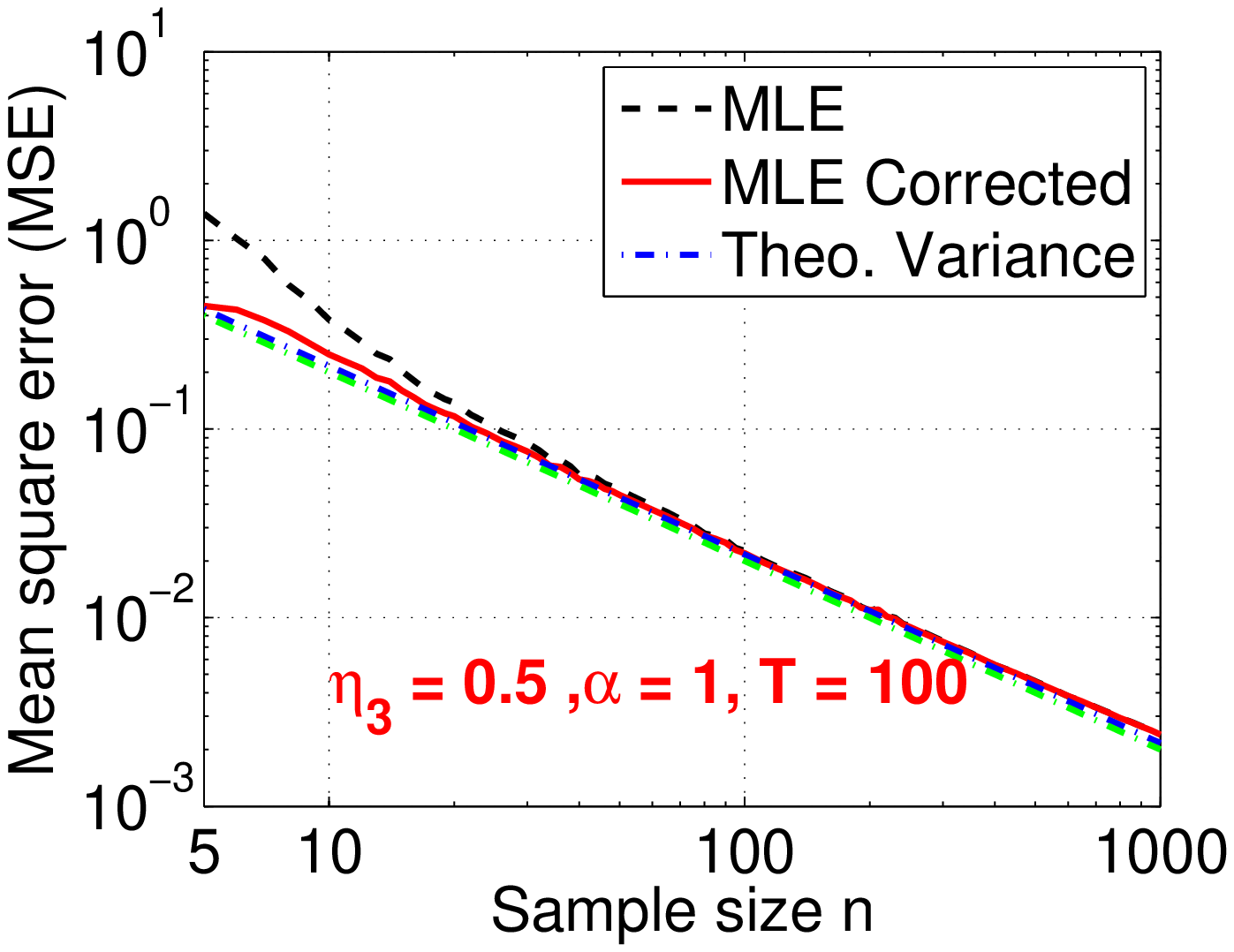}\hspace{0.2in}
\includegraphics[width=2.2in]{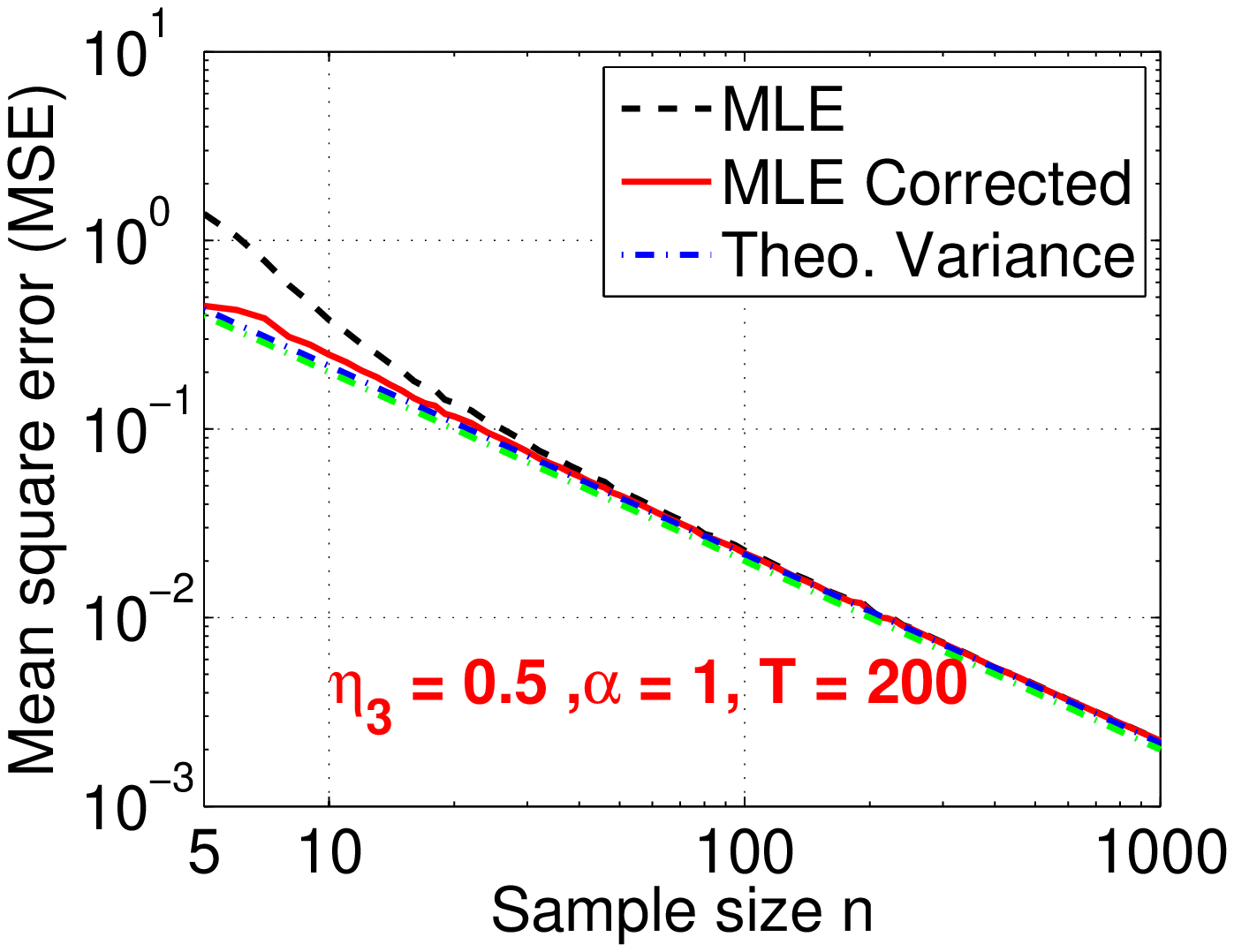}\hspace{0in}
}

\end{center}
\vspace{-0.3in}
\caption{Mean square errors of the 2-bit tabulation-based estimator $\hat{\Lambda}_{1}$ (dashed curves) and its bias-corrected version $\hat{\Lambda}_{1,c}$ (solid curves), for $\alpha=1$ and $T\in\{20,\ 50,\ 100,\ 200\}$ tabulation levels.  by fixing $\eta_3=0.5$, $\eta_2=3\eta_3$, $\eta_1=3\eta_2$. The lowest curve (dashed dot and green if color is available) in each panel represents the optimal variance using full  information, which is $2/n$ for $\alpha=1$.}\label{fig_Mse2bitP1T}
\end{figure}

Here we conduct a simulation study for $\alpha =1$ and $T \in \{20,\ 50,\ 100,\ 200\}$, as presented in Figure~\ref{fig_Mse2bitP1T}.  We let $\eta_3=0.5$, $\eta_2 = 3\eta_3$, $\eta_1 = 3\eta_2$. We can see that the results are already  good when $T=100$ (or even just $T=50$). This confirms the effectiveness of the tabulation scheme. \\

Therefore, tabulation provides an efficient solution to the computational problem for finding the MLE. Here, we have presented only a simple tabulation scheme based on uniform grids. It is possible to improve the scheme by using, for example, adaptive grids.

\newpage

\section{Multi-Bit (Multi-Partition) Coding and Estimation}

Clearly, we can extend this methodology to more than 2 bits.  With more bits, it is more flexible to consider schemes based on $(m+1)$ partitions. For example $m=1$ for the 1-bit scheme, $m=3$ for the 2-bit scheme, and $m=7$ for the 3-bit scheme. We feel $m\leq5$ is practical. The asymptotic variance of the MLE $\hat{\Lambda}_\alpha$ can be expressed as
\begin{align}\notag
&Var\left(\hat{\Lambda}_\alpha\right) = \frac{\Lambda^2_\alpha}{n}V_\alpha(\eta_1, ..., \eta_{m}) + O\left(\frac{1}{n^2}\right),\hspace{0.3in}\text{where}\\\notag
&\frac{1}{V_\alpha(\eta_1, ...., \eta_{m})} =
\frac{1}{\eta_1^2}\frac{f^2_\alpha\left(1/\eta_1\right)}{F_\alpha\left(1/\eta_1\right)}+\frac{1}{\eta_m^2}\frac{f^2_\alpha\left(1/\eta_m\right)}{1-F_\alpha\left(1/\eta_m\right)}
+\sum_{s=1}^{m-1}\frac{\left[f_\alpha\left(1/\eta_{s+1}\right)/\eta_{s+1}-f_\alpha\left(1/\eta_{s}\right)/\eta_s\right]^2}{F_\alpha\left(1/\eta_{s+1}\right)-F_\alpha\left(1/\eta_{s}\right)}
\end{align}

Here, we provide some numerical results for $m=5$, to demonstrate that using more partitions does further reduce the estimation variances and further stabilize the estimates in that the estimation accuracy is not as sensitive to parameters.

\subsection{$\alpha=0+$ and $m=5$}

\begin{figure}[h!]
\begin{center}
\mbox{
\includegraphics[width=2.2in]{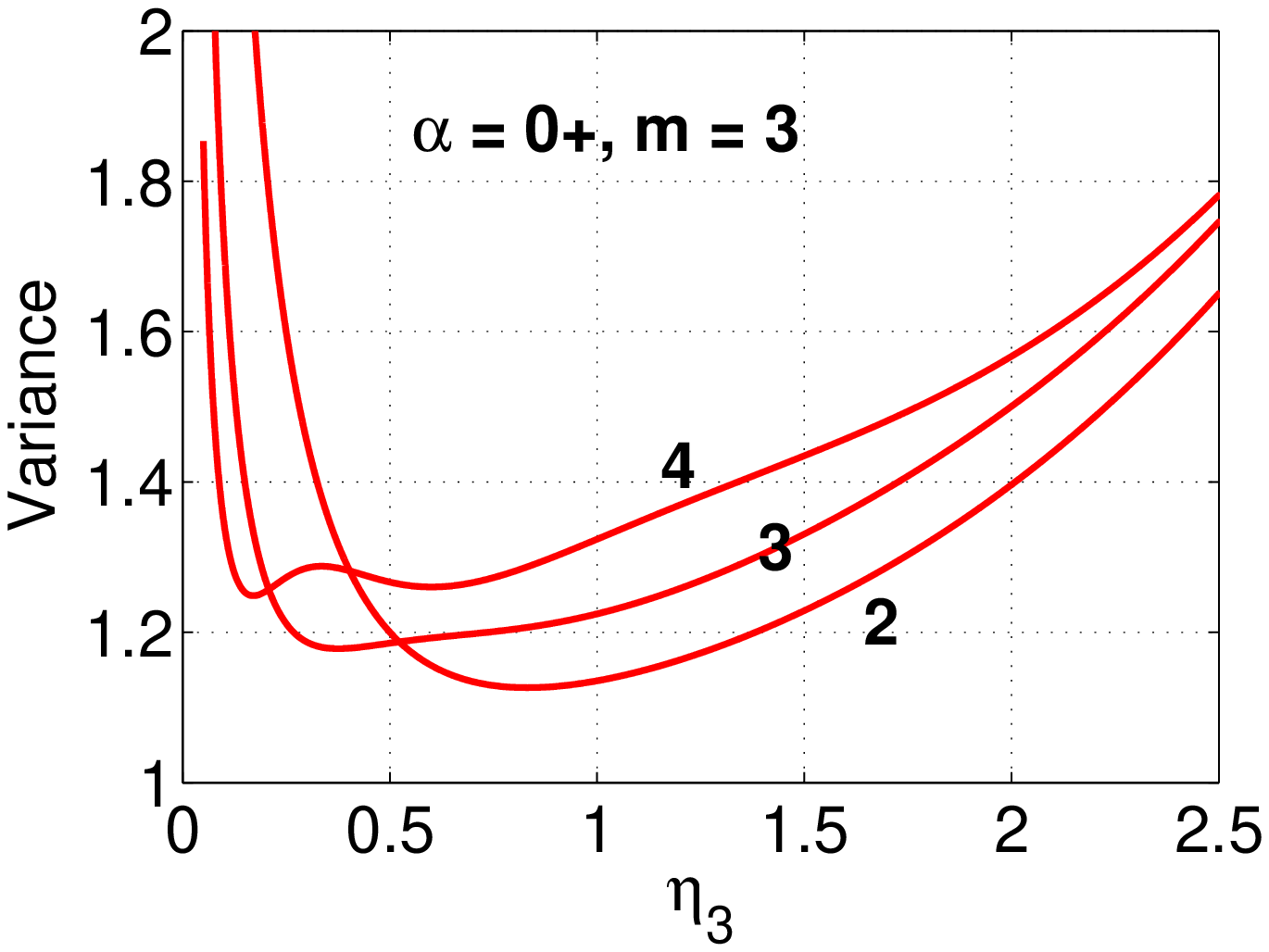}\hspace{0.2in}
\includegraphics[width=2.2in]{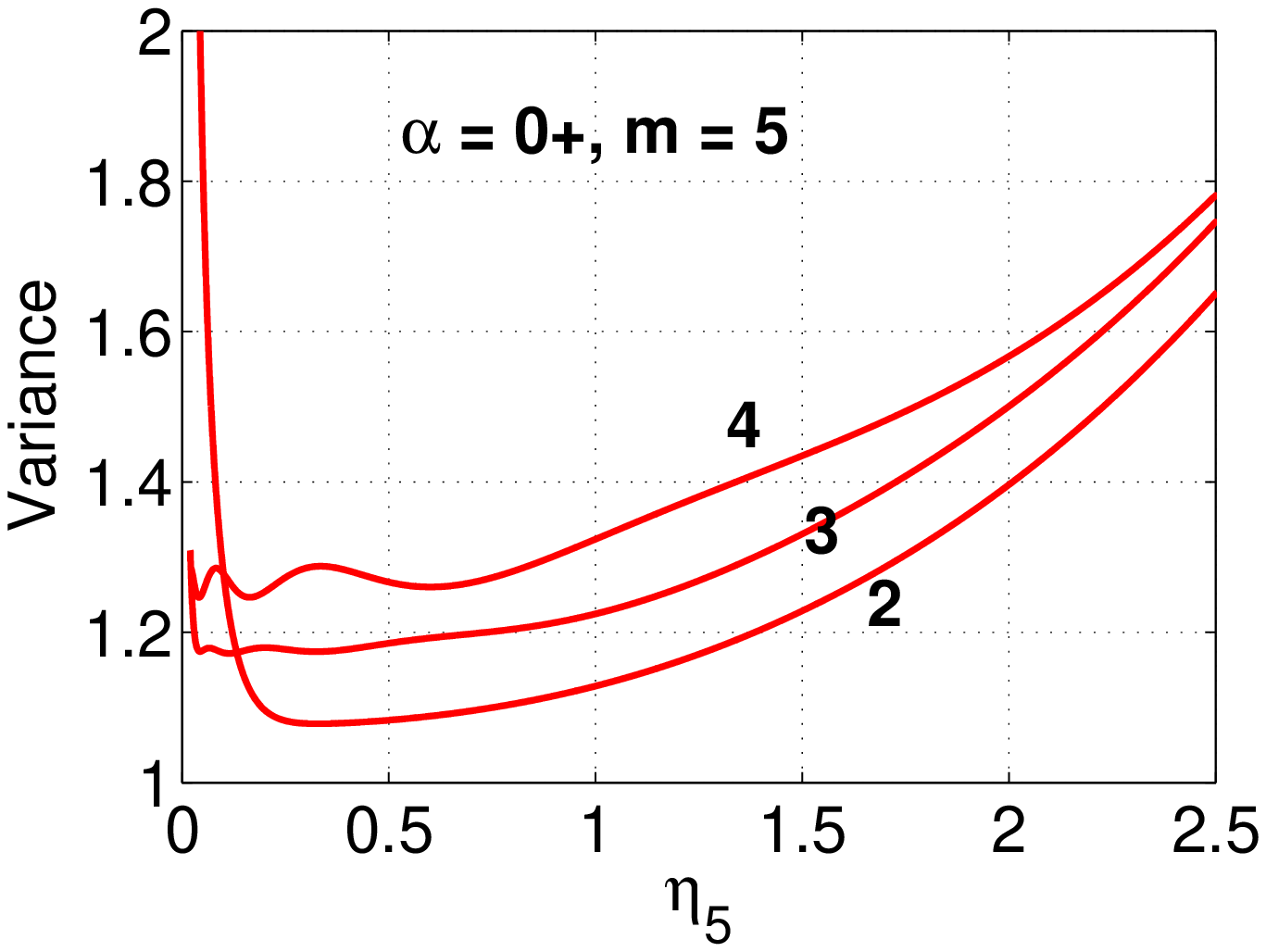}
}
\end{center}
\vspace{-0.3in}
\caption{\textbf{Left} (4-partition): \ $V_{0+}\left(\eta_1,\eta_2,\eta_3\right)$ for varying $\eta_3$ and  $\eta_2 = t \eta_3$, $\eta_1 = t\eta_2$, at $t=2, 3, 4$.
\textbf{Right} (6-partition):\ $V_{0+}\left(\eta_1,\eta_2,\eta_3,\eta_4,\eta_5\right)$ for varying $\eta_5$ and  $\eta_i = t \eta_{i+1}$, at $t=2, 3, 4$. }\label{fig_V5Al0}
\vspace{-0.1in}
\end{figure}

Numerically, the minimum of $V_{0+}\left(\eta_1,\eta_2,\eta_3,\eta_4,\eta_5\right)$ is $1.055$, attained at $\eta_1 = 4.464,\hspace{0.05in}    \eta_2 = 2.871,\hspace{0.05in}    \eta_3 = 1.853,\hspace{0.05in}    \eta_4 = 1.099,\hspace{0.05in}    \eta_5 = 0.499$.
Figure~\ref{fig_V5Al0} (right panel) plots $V_{0+}\left(\eta_1,\eta_2,\eta_3,\eta_4,\eta_5\right)$ for varying $\eta_5$ and $\eta_{i} = t\eta_{i+1}$, $i=4, 3, 2, 1$. For comparison, we also plot (in the left panel) $V_{0+}\left(\eta_1,\eta_2,\eta_3\right)$ for varying $\eta_3$, and $\eta_2=t\eta_3$, $\eta_1=t\eta_2$.  We can see that with more partitions, the performance becomes significantly more robust.

\subsection{$\alpha=1$ and $m=5$}

\begin{figure}[h!]
\begin{center}
\mbox{
\includegraphics[width=2.2in]{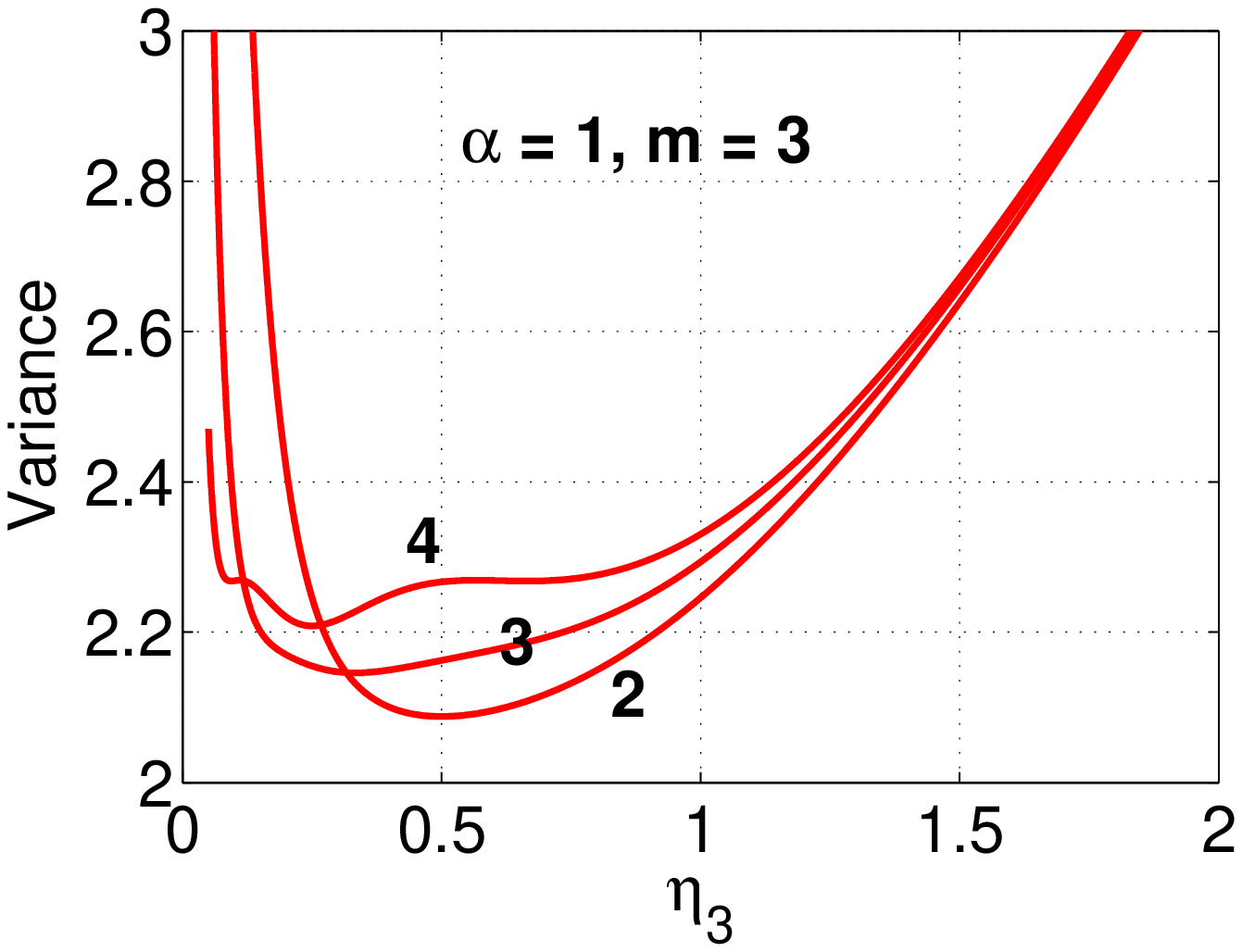}\hspace{0.2in}
\includegraphics[width=2.2in]{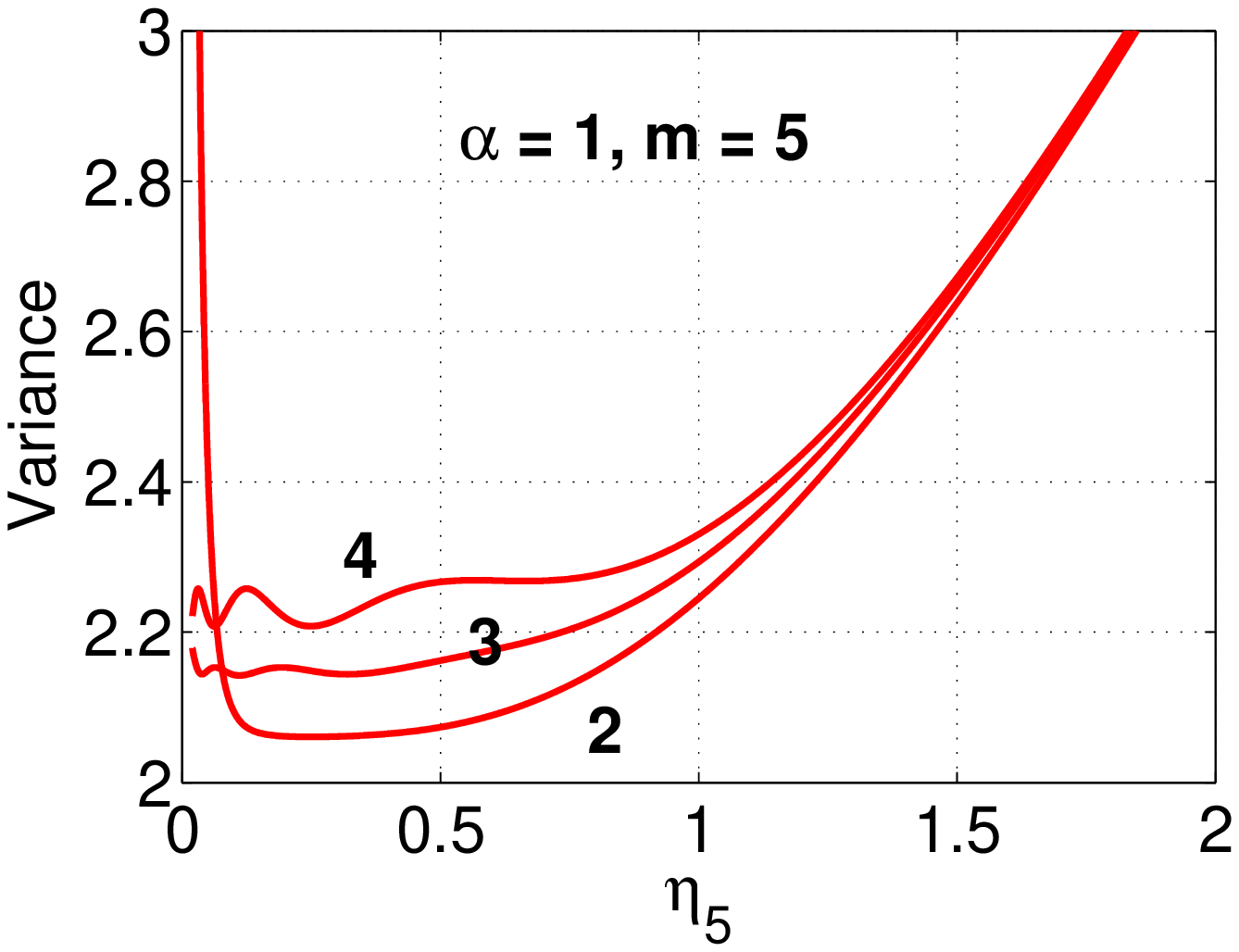}
}

\end{center}
\vspace{-0.3in}
\caption{\textbf{Left} (4-partition): \ $V_{1}\left(\eta_1,\eta_2,\eta_3\right)$ for varying $\eta_3$ and  $\eta_2 = t \eta_3$, $\eta_1 = t\eta_2$, at $t=2, 3, 4$.
\textbf{Right} (6-partition):\ $V_{1}\left(\eta_1,\eta_2,\eta_3,\eta_4,\eta_5\right)$ for varying $\eta_5$ and  $\eta_i = t \eta_{i+1}$, at $t=2, 3, 4$.}\label{fig_V5Al1}\vspace{-0.1in}
\end{figure}

Numerically, the minimum of $V_{1}\left(\eta_1,\eta_2,\eta_3,\eta_4,\eta_5\right)$ is $2.036$, attained at  $\eta_1 = 2.602,\hspace{0.1in}    \eta_2 = 1.498,\hspace{0.1in}    \eta_3 = 1.001,\hspace{0.1in}    \eta_4 = 0.668,\hspace{0.1in}    \eta_5 = 0.385$.
Figure~\ref{fig_V5Al1} (right panel) plots $V_{1}\left(\eta_1,\eta_2,\eta_3,\eta_4,\eta_5\right)$ for varying $\eta_5$ and $\eta_{i} = t\eta_{i+1}$, $i=4, 3, 2, 1$. Again, for comparison, we also plot (in the left panel) $V_{1}\left(\eta_1,\eta_2,\eta_3\right)$ for varying $\eta_3$, and $\eta_2=t\eta_3$, $\eta_1=t\eta_2$. Clearly, using more partitions stabilizes the variances even when the parameters are chosen less optimally.

\vspace{-0.1in}
\subsection{$\alpha=2$ and $m=5$}

Numerically, the minimum of $V_2\left(\eta_1,\eta_2,\eta_3,\eta_4,\eta_5\right)$ is $2.106$, attained at  $\eta_1 = 0.893,\   \eta_2 = 0.339,\    \eta_3 = 0.184,\    \eta_4 = 0.111,\    \eta_5 = 0.068$.
Figure~\ref{fig_V5Al1} (right panel) plots $V_{2}\left(\eta_1,\eta_2,\eta_3,\eta_4,\eta_5\right)$ for varying $\eta_5$ and $\eta_{i} = t\eta_{i+1}$, $i=4, 3, 2, 1$, as well as (left panel) $V_{2}\left(\eta_1,\eta_2,\eta_3\right)$ for varying $\eta_3$, and $\eta_2=t\eta_3$, $\eta_1=t\eta_2$.

\begin{figure}[h!]
\begin{center}
\mbox{
\includegraphics[width=2.2in]{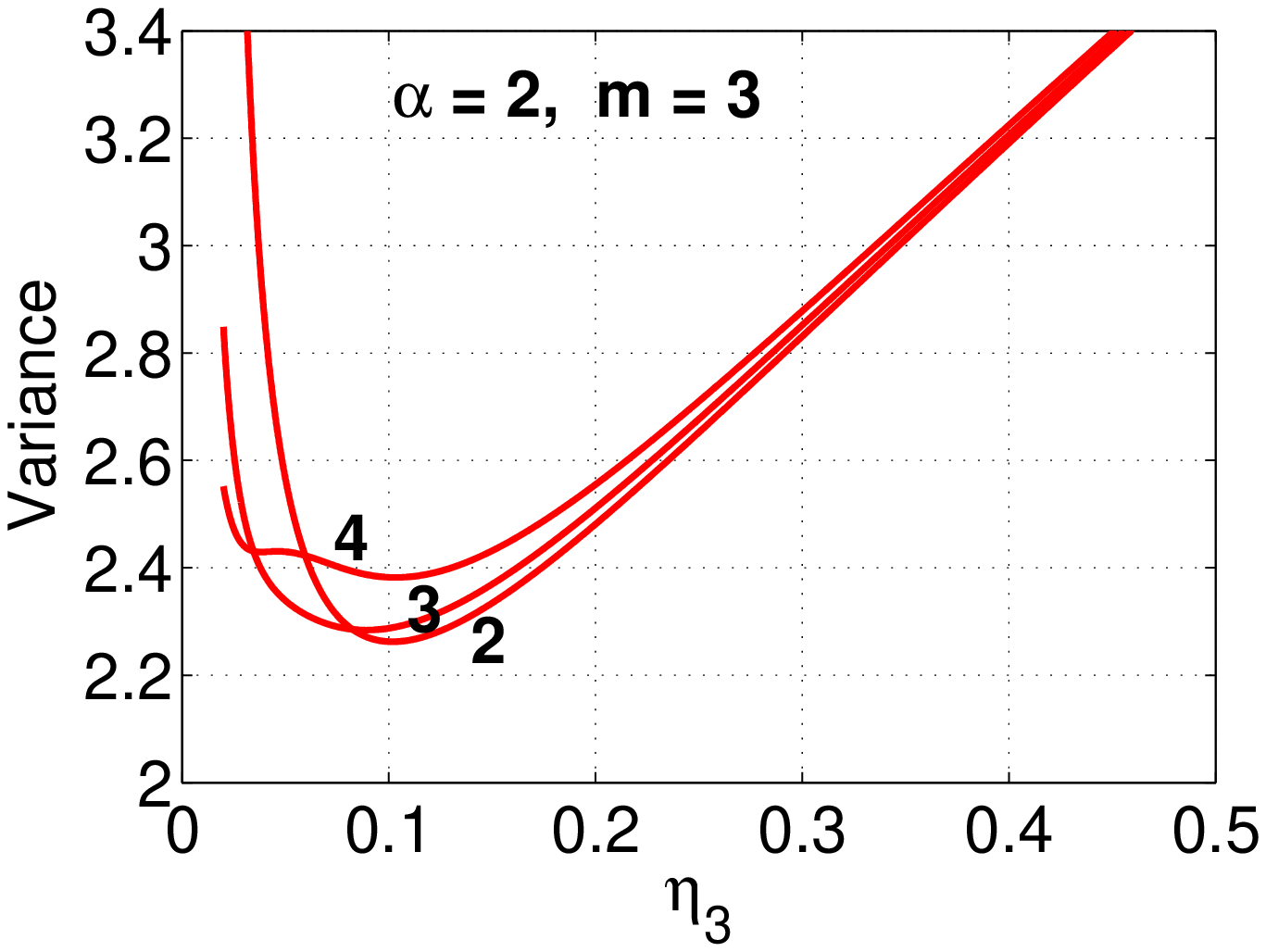}\hspace{0.2in}
\includegraphics[width=2.2in]{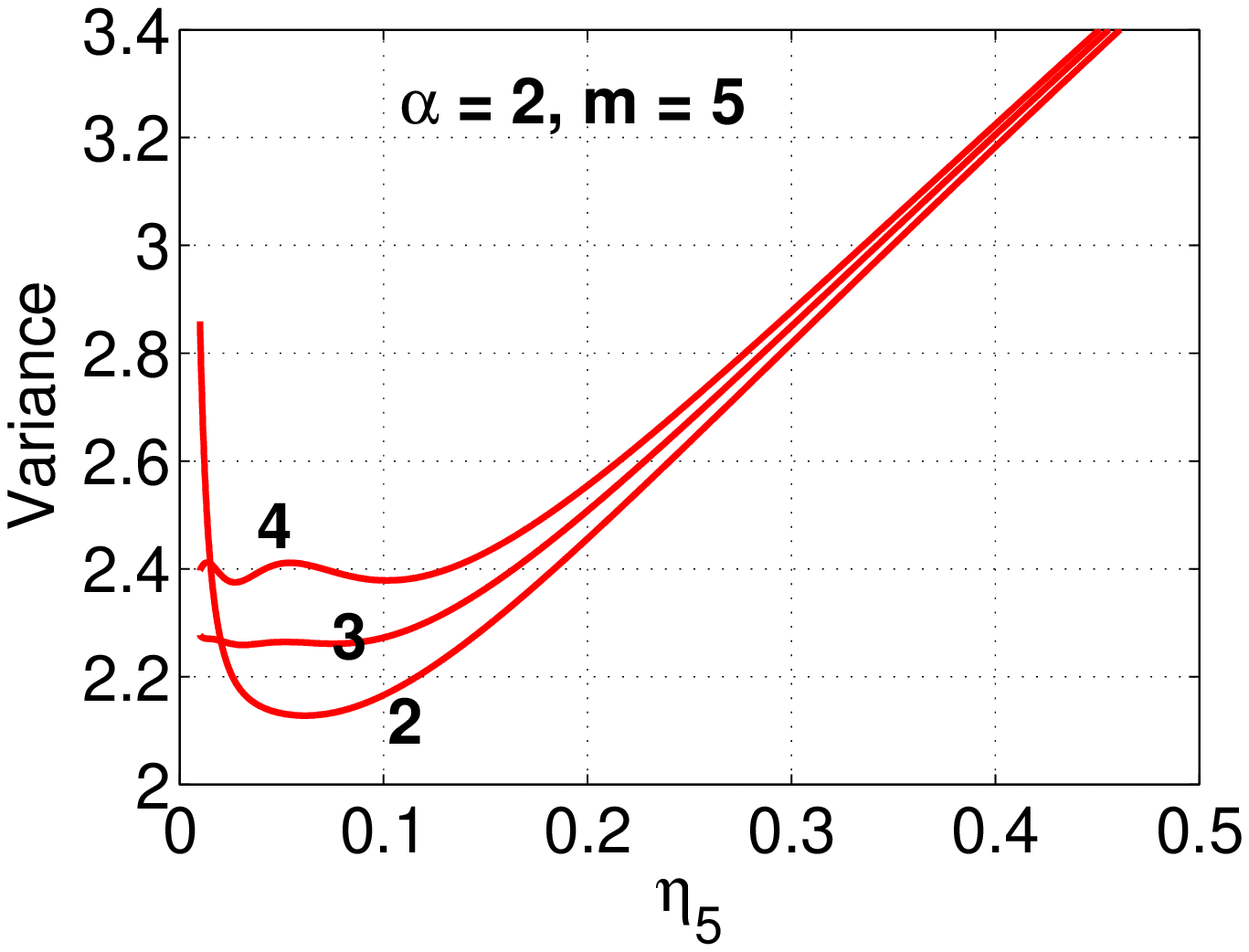}
}
\end{center}
\vspace{-0.3in}
\caption{\textbf{Left} (4-partition): \ $V_{2}\left(\eta_1,\eta_2,\eta_3\right)$ for varying $\eta_3$ and  $\eta_2 = t \eta_3$, $\eta_1 = t\eta_2$, at $t=2, 3, 4$.
\textbf{Right} (6-partition):\ $V_{2}\left(\eta_1,\eta_2,\eta_3,\eta_4,\eta_5\right)$ for varying $\eta_5$ and  $\eta_i = t \eta_{i+1}$, at $t=2, 3, 4$.}\label{fig_V5Al2}
\end{figure}

\section{Extension and Future Work}

Previously, \cite{Report:Li_Konig_AccurateHashing}  used counts and MLE, for improving classical minwise hashing and $b$-bit minwise hashing.  In this paper, we  focus on coding schemes for $\alpha$-stable random projections on individual data vectors. We feel an important line of future work would be the study of coding schemes for analyzing the relation of two or multiple data vectors, which will be  useful, for example, in the context of large-scale machine learning and efficient search/retrieval in massive data.

For example, \cite{Proc:Li_NIPS13} considered  nonnegative  data vectors under the sum-to-one constraint  (i.e., the $l_1$ norm = 1). After applying Cauchy stable random projections  separately on two data vectors, the collision probability of the two  signs of the projected data is essentially monotonic in the $\chi^2$ similarity (which is popular in computer vision). Now the open question is that, suppose we do not know the $l_1$ norms, how we should design coding schemes so that we can still evaluate the $\chi^2$ similarity (or other similarities) using Cauchy random projections.

Another recent paper~\cite{Proc:Li_ICML14} re-visited classical Gaussian random projections (i.e., $\alpha=2$). By assuming unit $l_2$ norms for the data vectors, \cite{Proc:Li_ICML14} developed multi-bit coding schemes and estimators for the correlation between vectors. Can we, using just a few bits,  still estimate the correlation if at the same time we must also estimate the $l_2$ norms?

\section{Conclusion}

Motivated by the recent  work on  ``one scan 1-bit compressed sensing'', we have developed 1-bit and multi-bit coding schemes for estimating the scale parameter of  $\alpha$-stable distributions. These simple coding schemes (even with just 1-bit) perform well in that, if the parameters are chosen appropriately, their variances are actually not  much larger than the variances using full (i.e., infinite-bit) information. In general, using more bits increases the computational cost  or storage cost (e.g., the cost of tabulations), with the benefits of stabilizing the performance so that the estimation variances do not increase much even when the parameters are  far from optimal. In practice, we expect the $(m+1)$-partition scheme,  combined with  tabulation, for $m=3$, 4, or 5, should be overall preferable. Here $m=3$ corresponds to the 2-bit scheme, $m=1$ to the 1-bit scheme.

\newpage

\appendix
\section{Proof of Theorem~\ref{thm_1bit} and Bias Corrections}\label{proof_thm_1bit}

The log-likelihood of the $n = n_1+n_2$ observations is
\begin{align}\notag
l  =& n_1\log F_\alpha\left(C/\Lambda_\alpha\right) + n_2\log \left[1-F_\alpha\left(C/\Lambda_\alpha\right) \right]
\end{align}
and its first derivative is
\begin{align}\notag
l^\prime = \frac{\partial l}{\partial \Lambda_\alpha}
 =& \left(-\frac{C}{\Lambda_\alpha^2}\right)\left( n_1\frac{f_\alpha\left(C/\Lambda_\alpha\right)}{F_\alpha\left(C/\Lambda_\alpha\right)} +
n_2\frac{-f_\alpha\left(C/\Lambda_\alpha\right)}{1-F_\alpha\left(C/\Lambda_\alpha\right)}\right)\\\notag
 =& \left(-\frac{C}{\Lambda_\alpha^2}\right)\left(n_1\frac{f}{F}-n_2\frac{f}{1-F}\right)\\\notag
=&\left(-\frac{C}{\Lambda_\alpha^2}\right)\left(\frac{n_1-nF}{F-F^2}\right)f
\end{align}
For simplicity, we use $F, f, f^\prime, f^{\prime\prime}$ for $F_\alpha\left(C/\Lambda_\alpha\right)$, $f_\alpha\left(C/\Lambda_\alpha\right)$, $f^\prime_\alpha\left(C/\Lambda_\alpha\right)$ and $f^{\prime\prime}_\alpha\left(C/\Lambda_\alpha\right)$, respectively.  Setting $l^\prime=0$ leads to the MLE solution: $\frac{n_1}{n} = F_\alpha\left(C/\Lambda_\alpha\right)$, ie., $\hat{\Lambda}_\alpha = \frac{C}{F_\alpha^{-1}\left(n_1/n\right)}$.

According to classical statistical results~\cite{Article:Bartlett_53,Article:Shenton_63},
\begin{align}\notag
&E\left(\hat{\Lambda}_\alpha \right) = \Lambda_\alpha - \frac{E(l^\prime)^3+E(l^\prime l^{\prime\prime})}{2I^2} + O\left(\frac{1}{n^2}\right),\\\notag
&Var\left(\hat{\Lambda}_\alpha \right) = \frac{1}{I}+O\left(\frac{1}{n^2}\right)\\\notag
\end{align}
where the Fisher Information $I = E(l^\prime)^2 = -E(l^{\prime\prime})$. Here the derivatives $l^\prime$, $l^{\prime\prime}$,  $l^{\prime\prime\prime}$  are with respect to $\Lambda_\alpha$. Thus, we need to computer the derivatives of $l$ and evaluate their expectations.

By property of binomial distribution, we have $E(n_1) = nF_1$ and
\begin{align}\notag
&E(n_1-E(n_1)) = 0\\\notag
&E(n_1-E(n_1))^2 = nF(1-F)\\\notag
&E(n_1-E(n_1))^3 = nF(1-F)(1-2F)
\end{align}
Obviously
\begin{align}\notag
E(l^\prime) =\left(-\frac{C}{\Lambda_\alpha^2}\right)\left(\frac{nF-nF}{F-F^2}\right) f= 0
\end{align}
Furthermore,
\begin{align}\notag
&E(l^\prime)^2 =\left(-\frac{C}{\Lambda_\alpha^2}\right)^2f^2\left(\frac{nF(1-F)}{(F-F^2)^2}\right)= n\frac{C^2}{\Lambda_\alpha^4}\frac{f^2}{F(1-F)} = I\\\notag
&E(l^\prime)^3 =\left(-\frac{C}{\Lambda_\alpha^2}\right)^3f^3\left(\frac{nF(1-F)(1-2F)}{(F-F^2)^3}\right)= -n\frac{C^3}{\Lambda_\alpha^6}\frac{f^3(1-2F)}{F^2(1-F)^2}
\end{align}

Next we work on the second derivative
\begin{align}\notag
l^{\prime\prime} =&-\frac{2}{\Lambda_\alpha}l^\prime + \left(-\frac{C}{\Lambda_\alpha^2}\right)^2\left(\frac{n_1-nF}{F-F^2}\right)f^\prime+\left(-\frac{C}{\Lambda_\alpha^2}\right)^2f\left(\frac{-n_1f+2n_1fF-nfF^2}{(F-F^2)^2}\right)\\\notag
=&-\frac{2}{\Lambda_\alpha}l^\prime + \frac{f^\prime}{f}\left(-\frac{C}{\Lambda_\alpha^2}\right) l^\prime  +
\left(-\frac{C}{\Lambda_\alpha^2}\right)\frac{f}{1-F}l^\prime  - \left(-\frac{C}{\Lambda_\alpha^2}\right)^2n_1\left(\frac{f^2}{F^2(1-F)}\right)
\end{align}
\begin{align}\notag
E(l^{\prime\prime}) =\left(-\frac{C}{\Lambda_\alpha^2}\right)^2f\left(\frac{-nfF+2nfF^2-nfF^2}{(F-F^2)^2}\right) = -n\frac{C^2}{\Lambda_\alpha^4}\frac{f^2}{F(1-F)} = -I
\end{align}
and higher-order derivatives
\begin{align}\notag
l^{\prime\prime\prime} =&\left[-\frac{2}{\Lambda_\alpha}l^\prime +\left(-\frac{C}{\Lambda_\alpha^2}\right) l^\prime\left(\frac{f^\prime}{f}  +
\frac{f}{1-F}\right)  - \left(-\frac{C}{\Lambda_\alpha^2}\right)^2\left(\frac{n_1f^2}{F^2(1-F)}\right)\right]'\\\notag
=&\frac{2}{\Lambda_\alpha^2}l^\prime-\frac{2}{\Lambda_\alpha}l^{\prime\prime}+\left(\frac{2C}{\Lambda_\alpha^3}\right) l^\prime\left(\frac{f^\prime}{f}  +
\frac{f}{1-F}\right)+ \left(-\frac{C}{\Lambda_\alpha^2}\right) l^{\prime\prime}\left(\frac{f^\prime}{f}  +\frac{f}{1-F}\right)\\\notag
&+\left(-\frac{C}{\Lambda_\alpha^2}\right)^2 l^\prime\left(\frac{f^{\prime\prime}f-(f^\prime)^2}{f^2}  +\frac{f^\prime(1-F)+f^2}{(1-F)^2}\right)\\\notag
&+\left(\frac{4C^2}{\Lambda_\alpha^5}\right)\left(\frac{n_1f^2}{F^2(1-F)}\right) - \left(-\frac{C}{\Lambda_\alpha^2}\right)^3\left(\frac{2n_1ff^\prime(F^2-F^3)-n_1f^3(2F-3F^2)}{F^4(1-F)^2}\right)
\end{align}
\begin{align}\notag
E(l^\prime l^{\prime\prime}) = E\left[
-\frac{2}{\Lambda_\alpha}(l^\prime)^2 +\left(-\frac{C}{\Lambda_\alpha^2}\right) (l^\prime)^2\left(\frac{f^\prime}{f}  +
\frac{f}{1-F}\right)  - \left(-\frac{C}{\Lambda_\alpha^2}\right)^2\left(\frac{f^2}{F^2(1-F)}\right)n_1l^\prime
\right]
\end{align}
\begin{align}\notag
E(n_1l^\prime) =  \left(-\frac{C}{\Lambda_\alpha^2}\right)\left(\frac{E\left(n_1(n_1-nF)\right)}{F-F^2}\right)f = \left(-\frac{C}{\Lambda_\alpha^2}\right)\left(\frac{nF(1-F)}{F-F^2}\right)f = \left(-\frac{C}{\Lambda_\alpha^2}\right)nf
\end{align}
\begin{align}\notag
E(l^\prime l^{\prime\prime}) =& \left(
-\frac{2}{\Lambda_\alpha} +\left(-\frac{C}{\Lambda_\alpha^2}\right)\left(\frac{f^\prime}{f}  +
\frac{f}{1-F}\right)\right)n\frac{C^2}{\Lambda_\alpha^4}\frac{f^2}{F(1-F)}  - \left(-\frac{C}{\Lambda_\alpha^2}\right)^3\left(\frac{nf^3}{F^2(1-F)}\right)\\\notag
=&\left(
-\frac{2}{\Lambda_\alpha} +\left(-\frac{C}{\Lambda_\alpha^2}\right)\left(\frac{f^\prime}{f} \right)\right)n\frac{C^2}{\Lambda_\alpha^4}\frac{f^2}{F(1-F)}  - \left(-\frac{C}{\Lambda_\alpha^2}\right)^3\left(\frac{nf^3(1-2F)}{F^2(1-F)^2}\right)
\end{align}
\begin{align}\notag
E(l^\prime)^3+E(l^\prime l^{\prime\prime}) = \left(
-\frac{2}{\Lambda_\alpha} +\left(-\frac{C}{\Lambda_\alpha^2}\right)\left(\frac{f^\prime}{f} \right)\right)n\frac{C^2}{\Lambda_\alpha^4}\frac{f^2}{F(1-F)} = \left(
-\frac{2}{\Lambda_\alpha} +\left(-\frac{C}{\Lambda_\alpha^2}\right)\left(\frac{f^\prime}{f} \right)\right)I
\end{align}
Therefor, the bias-correction term is
\begin{align}\notag
\frac{E(l^\prime)^3+E(l^\prime l^{\prime\prime}) }{I^2} =& \frac{\left(-\frac{2}{\Lambda_\alpha} +\left(-\frac{C}{\Lambda_\alpha^2}\right)\left(\frac{f^\prime}{f} \right)\right)}{n\frac{C^2}{\Lambda_\alpha^4}\frac{f^2}{F(1-F)}}
=-\frac{\Lambda_\alpha}{n}\frac{\left(2 +\frac{C}{\Lambda_\alpha}\frac{f^\prime}{f}\right)}{\frac{C^2}{\Lambda_\alpha^2}\frac{f^2}{F(1-F)}}
=-\frac{\Lambda_\alpha}{n}\frac{\left(2 +z\frac{f^\prime}{f}\right)}{z^2\frac{f^2}{F(1-F)}}\\\notag
=& -\frac{\Lambda_\alpha}{n}\frac{n_1}{n}\left(1-\frac{n_1}{n}\right)\frac{2 +z\frac{f^\prime}{f}}{z^2f^2}
\end{align}
where we denote $z =\frac{1}{\eta}= \frac{C}{\Lambda_\alpha}$. Note that since $\l^\prime=0$, we have $z = \frac{C}{\Lambda_\alpha}= F_{\alpha}^{-1}(n_1/n)$, $F_\alpha(z) = \frac{n_1}{n}$. \\

Next, we derive more explicit expressions for $\alpha=0+$, $\alpha=1$, and $\alpha=2$.\\

When $\alpha\rightarrow0+$,
\begin{align}\notag
F_{0+}(z) = e^{-1/z},\hspace{0.05in}  f_{0+}(z) = \frac{1}{z^2}e^{-1/z},\hspace{0.05in}  f^\prime_{0+}(z) = \frac{-2}{z^3}e^{-1/z} + \frac{1}{z^4}e^{-1/z},\hspace{0.05in} F^{-1}_{0+}(z) = \frac{1}{\log 1/z}
\end{align}

\begin{align}\notag
&2+z\frac{f^\prime(z)}{f(z)} =\frac{1}{z}  = \frac{1}{F_{\alpha}^{-1}(n_1/n)} = \log n/n_1\\\notag
&z^2f^2(z) = \frac{1}{z^2}e^{-2/z} = \frac{1}{z^2} e^{-2/F^{-1}_\alpha(n_1/n)} = \frac{\log^2(n/n_1)}{ (n/n_1)^2}
\end{align}

\begin{align}\notag
\frac{E(l^\prime)^3+E(l^\prime l^{\prime\prime}) }{I^2}  = -\frac{\Lambda_\alpha}{n}\frac{n_1}{n}\left(1-\frac{n_1}{n}\right)\frac{2 +z\frac{f^\prime}{f}}{z^2f^2}  = -\frac{\Lambda_\alpha}{n}\frac{n_1}{n}\left(1-\frac{n_1}{n}\right) \frac{(n/n_1)^2}{\log n/n_1} = -\frac{\Lambda_\alpha}{n}\frac{\left(n/n_1-1\right)}{\log n/n_1}
\end{align}
Recall that $l^\prime = 0\Rightarrow \frac{n_1}{n} = F_\alpha\left(C/\Lambda_\alpha\right)\Rightarrow\frac{C}{\Lambda_\alpha} = F_\alpha^{-1}(n_1/n)=\frac{1}{\log n/n_1}\Rightarrow e^{\Lambda_\alpha/C} = n/n_1 $.

Therefore, the bias-corrected MLE for $\alpha\rightarrow0+$ is
\begin{align}\notag
&\hat{\Lambda}_{0+,c} =\frac{C\log n/n_1}{1+ \frac{1/n_1-1/n}{2\log n/n_1}  }
\end{align}

When $\alpha=1$,  by properties of Cauchy distribution, we know
\begin{align}\notag
F_1(z) = \frac{2}{\pi}\tan^{-1}z,\hspace{0.1in} f_1(z) = \frac{2}{\pi}\frac{1}{1+z^2},\hspace{0.1in} f_1^\prime(z) = \frac{2}{\pi}\frac{-2z}{(1+z^2)^2},\hspace{0.1in} F_1^{-1}(z) = \tan\frac{\pi}{2}z
\end{align}
\begin{align}\notag
2+z\frac{f^\prime(z)}{f(z)} = 2+\frac{-2z^2}{1+z^2} = \frac{2}{1+z^2},\hspace{0.3in}
z^2 f^2(z)= \frac{4}{\pi^2}\frac{z^2}{(1+z^2)^2}
\end{align}
Note that $z = \frac{C}{\Lambda_\alpha} = F_\alpha^{-1}(n_1/n) = \tan \frac{\pi}{2}\frac{n_1}{n}$, $\tan^{-1}z = \frac{\pi}{2}\frac{n_1}{n}$. We have
\begin{align}\notag
\frac{2 +z\frac{f^\prime}{f}}{z^2f^2}  = \frac{\pi^2}{2}\frac{1+z^2}{z^2} = \frac{\pi^2}{2}\left(1+\frac{1}{z^2}\right) = \frac{\pi^2}{2}\left(1+\frac{1}{\tan^2\frac{\pi}{2}\frac{n_1}{n}}\right)
\end{align}
\begin{align}\notag
\frac{E(l^\prime)^3+E(l^\prime l^{\prime\prime}) }{I^2}  = -\frac{\Lambda_\alpha}{n}\frac{n_1}{n}\left(1-\frac{n_1}{n}\right)\frac{2 +z\frac{f^\prime}{f}}{z^2f^2}
= -\frac{\Lambda_\alpha}{n}\frac{n_1}{n}\left(1-\frac{n_1}{n}\right)\frac{\pi^2}{2}\left(1+\frac{1}{\tan^2\frac{\pi}{2}\frac{n_1}{n}}\right)
\end{align}

Therefore, the bias-corrected MLE for $\alpha=1$ is
\begin{align}\notag
\hat{\Lambda}_{1,c} = \frac{\frac{C}{\tan\frac{\pi}{2}\frac{n_1}{n}}}{1+\frac{1}{n}\frac{\pi^2}{4}\frac{n_1}{n}\left(1- \frac{n_1}{n}\right)\left(1+\frac{1}{\tan^2 \frac{\pi}{2}\frac{n_1}{n}}\right)
}
\end{align}

\vspace{0.1in}

When $\alpha=2$, since $S(2,1)\sim \sqrt{2}\times N(0,1)$, i.e.,  $|s_\alpha|^2\sim 2 \chi^2_1$, we have
\begin{align}\notag
&F_2(z) = F_{\chi^2_1}(z/2) = 2\Phi(\sqrt{z/2})-1,\\\notag
&F_2^{-1}(t) = 2F_{\chi^2_1}^{-1}(t) = 2\left[\Phi^{-1}\left(\frac{t+1}{2}\right)\right]^2\\\notag
&z = F_2^{-1}(n_1/n) = 2\left[\Phi^{-1}\left(\frac{n_1/n+1}{2}\right)\right]^2\\\notag
&f_2(z) = f_{\chi^2_1}(z/2)/2 = \frac{1}{2\sqrt{2\pi}}\frac{1}{\sqrt{z/2}}e^{-z/4}= \frac{1}{2\sqrt{\pi z}}e^{-z/4}\\\notag
&f_2^\prime(z) = -\frac{1}{4\sqrt{\pi}z^{3/2}}e^{-z/4} - \frac{1}{8\sqrt{\pi}z^{1/2}}e^{-z/4}\\\notag
&z\frac{f_2^\prime(z)}{f_2(z)} = -\frac{1}{2}-\frac{z}{4},\hspace{0.2in}z^2 f^2(z) = \frac{z}{4 \pi}e^{-z/2}
\end{align}

\begin{align}\notag
\frac{E(l^\prime)^3+E(l^\prime l^{\prime\prime}) }{I^2}  = -\frac{\Lambda_\alpha}{n}\frac{n_1}{n}\left(1-\frac{n_1}{n}\right)\frac{2 +z\frac{f^\prime}{f}}{z^2f^2}
=-\frac{\Lambda_\alpha}{n}\frac{n_1}{n}\left(1-\frac{n_1}{n}\right)\pi e^{z/2} \left(\frac{6}{z}-1\right)
\end{align}

Therefore, the bias-corrected MLE for $\alpha=2$ is
\begin{align}\notag
\hat{\Lambda}_{2,c} = \frac{\frac{C}{2F^{-1}_{\chi^2_1}(n_1/n)}}{1 +  \frac{\pi}{n}\frac{n_1}{n}\left(1-\frac{n_1}{n}\right)  \left(\frac{3}{F^{-1}_{\chi^2_1}(n_1/n) }-1\right) e^{F^{-1}_{\chi^2_1}(n_1/n)}}
\end{align}

\section{Proof of Theorem~\ref{thm_bounds}}~\label{proof_thm_bounds}

The task is to prove the  following two  bounds:
\begin{align}\notag
&\mathbf{Pr}\left(\hat{\Lambda}_\alpha\geq(1+\epsilon)\Lambda_\alpha\right)\leq \exp\left(-n\frac{\epsilon^2}{G_{R,\alpha,C,\epsilon}}\right),\hspace{0.02in} \epsilon\geq 0\\\notag
&\mathbf{Pr}\left(\hat{\Lambda}_\alpha\leq (1-\epsilon)\Lambda_\alpha\right)\leq \exp\left(-n\frac{\epsilon^2}{G_{L,\alpha,C,\epsilon}}\right),\hspace{0.02in} 0\leq\epsilon\leq1
\end{align}

The proof is based on the expression of the MLE estimator $\hat{\Lambda}_\alpha = C/F^{-1}_\alpha(n_1/n)$, the fact that $n_1\sim Binomial(n,F_\alpha(1/\eta))$, and Chernoff's original tail bounds~\cite{Article:Chernoff_52} for the binomial distribution.

For the right tail bound, we have
\begin{align}\notag
&\mathbf{Pr}\left(\hat{\Lambda}_\alpha\geq(1+\epsilon)\Lambda_\alpha\right)\\\notag
=&\mathbf{Pr}\left(\frac{C}{F_\alpha^{-1}(n_1/n)} \geq (1+\epsilon)\Lambda_\alpha\right)\\\notag
=&\mathbf{Pr}\left(\frac{n_1}{n}\leq F_\alpha\left(\frac{C}{(1+\epsilon)\Lambda_\alpha}\right)\right)\\\notag
=&\mathbf{Pr}\left(\frac{n_1}{n}\leq F_\alpha\left(\frac{1}{(1+\epsilon)\eta}\right)\right)\\\notag
\leq&\left[\frac{F_\alpha(1/\eta)}{F_\alpha(1/(1+\epsilon)\eta)}\right]^{nF_\alpha(1/(1+\epsilon)\eta)}
\times\left[\frac{1-F_\alpha(1/\eta)}{1-F_\alpha(1/(1+\epsilon)\eta)}\right]^{n-nF_\alpha(1/(1+\epsilon)\eta)}\\\notag
=&\exp\left(-n\frac{\epsilon^2}{G_{R,\alpha,C,\epsilon}}\right)
\end{align}
where
\begin{align}\notag
&\frac{\epsilon^2}{G_{R,\alpha,C,\epsilon}} = -F_\alpha(1/(1+\epsilon)\eta)\log\left[\frac{F_\alpha(1/\eta)}{F_\alpha(1/(1+\epsilon)\eta)}\right]%\\\notag
-(1-F_\alpha(1/(1+\epsilon)\eta))\log\left[\frac{1-F_\alpha(1/\eta)}{1-F_\alpha(1/(1+\epsilon)\eta)}\right]\\\notag
\end{align}

Next, for the left tail bound, we have
\begin{align}\notag
&\mathbf{Pr}\left(\hat{\Lambda}_\alpha\leq(1-\epsilon)\Lambda_\alpha\right)\\\notag
=&\mathbf{Pr}\left(\frac{C}{F_\alpha^{-1}(n_1/n)} \leq(1-\epsilon)\Lambda_\alpha\right)\\\notag
=&\mathbf{Pr}\left(\frac{n_1}{n}\geq F_\alpha\left(\frac{C}{(1-\epsilon)\Lambda_\alpha}\right)\right)\\\notag
=&\mathbf{Pr}\left(\frac{n_1}{n}\geq F_\alpha\left(\frac{1}{(1-\epsilon)\eta}\right)\right)\\\notag
\leq&\left[\frac{F_\alpha(1/\eta)}{F_\alpha(1/(1-\epsilon)\eta)}\right]^{nF_\alpha(1/(1-\epsilon)\eta)}%\\\notag
\times\left[\frac{1-F_\alpha(1/\eta)}{1-F_\alpha(1/(1-\epsilon)\eta)}\right]^{n-nF_\alpha(1/(1-\epsilon)\eta)}\\\notag
=&\exp\left(-n\frac{\epsilon^2}{G_{L,\alpha,C,\epsilon}}\right)
\end{align}
where
\begin{align}\notag
&\frac{\epsilon^2}{G_{L,\alpha,C,\epsilon}} = -F_\alpha(1/(1-\epsilon)\eta)\log\left[\frac{F_\alpha(1/\eta)}{F_\alpha(1/(1-\epsilon)\eta)}\right]%\\\notag
-(1-F_\alpha(1/(1-\epsilon)\eta))\log\left[\frac{1-F_\alpha(1/\eta)}{1-F_\alpha(1/(1-\epsilon)\eta)}\right]\\\notag
\end{align}

%\newpage

\section{Proof of Theorem~\ref{thm_2bit}}\label{proof_thm_2bit}

With the  2-bit scheme, we need to  introduce 3 threshold values: $C_1\leq C_2\leq C_3$, and   define
\begin{align}\notag
&p_1 = \mathbf{Pr}\left(z_j \leq C_1\right) = F_\alpha\left(C_1/\Lambda_\alpha\right)\\\notag
&p_2 = \mathbf{Pr}\left(C_1<z_j \leq C_2\right) = F_\alpha\left(C_2/\Lambda_\alpha\right) - F_\alpha\left(C_1/\Lambda_\alpha\right)\\\notag
&p_3 = \mathbf{Pr}\left(C_2<z_j \leq C_3\right) = F_\alpha\left(C_3/\Lambda_\alpha\right) - F_\alpha\left(C_2/\Lambda_\alpha\right)\\\notag
&p_4 = \mathbf{Pr}\left(z_j > C_3\right) = 1 - F_\alpha\left(C_3/\Lambda_\alpha\right)
\end{align}
and
\begin{align}\notag
&n_1 =  \sum_{j=1}^n 1\{z_j\leq C_1\},\hspace{0.5in}
n_2 =  \sum_{j=1}^n 1\{C_1<z_j\leq C_2\}\\\notag
&n_3 =  \sum_{j=1}^n 1\{C_2<z_j\leq C_3\},\hspace{0.5in}
n_4 =  \sum_{j=1}^n 1\{z_j> C_3\}
\end{align}

The log-likelihood of these $n = n_1+n_2+n_3+n_4$ observations can be expressed as
\begin{align}\notag
l =& n_1\log p_1 + n_2 \log p_2 +  n_3\log p_3 + n_4 \log p_4\\\notag
=& n_1\log F_\alpha\left(C_1/\Lambda_\alpha\right) + n_2\log \left[F_\alpha\left(C_2/\Lambda_\alpha\right)-F_\alpha\left(C_1/\Lambda_\alpha\right) \right] \\\notag
+& n_3\log \left[F_\alpha\left(C_3/\Lambda_\alpha\right)-F_\alpha\left(C_2/\Lambda_\alpha\right) \right]+ n_4\log \left[1-F_\alpha\left(C_3/\Lambda_\alpha\right) \right]\\\notag
=& n_1 \log F_1 +n_2\log (F_2 - F_1) + n_3\log(F_3 - F_2) + n_4 \log (1-F_3)
\end{align}
To seek the MLE  of $\Lambda_\alpha$, we need to compute the first derivative:
\begin{align}\notag
l^\prime = \frac{\partial l}{\partial \Lambda_\alpha} =&  n_1\frac{f_\alpha\left(C_1/\Lambda_\alpha\right)}{F_\alpha\left(C_1/\Lambda_\alpha\right)}\left(-\frac{C_1}{\Lambda_\alpha^2}\right) +
n_2\frac{f_\alpha\left(C_2/\Lambda_\alpha\right)\left(-\frac{C_2}{\Lambda_\alpha^2}\right)-f_\alpha\left(C_1/\Lambda_\alpha\right)\left(-\frac{C_1}{\Lambda_\alpha^2}\right)}{F_\alpha\left(C_2/\Lambda_\alpha\right)-F_\alpha\left(C_1/\Lambda_\alpha\right)}\\\notag
+&
n_3\frac{f_\alpha\left(C_3/\Lambda_\alpha\right)\left(-\frac{C_3}{\Lambda_\alpha^2}\right)-f_\alpha\left(C_2/\Lambda_\alpha\right)\left(-\frac{C_2}{\Lambda_\alpha^2}\right)}{F_\alpha\left(C_3/\Lambda_\alpha\right)-F_\alpha\left(C_2/\Lambda_\alpha\right)}
+n_4\frac{-f_\alpha\left(C_3/\Lambda_\alpha\right)}{1-F_\alpha\left(C_3/\Lambda_\alpha\right)}\left(-\frac{C_3}{\Lambda_\alpha^2}\right)\\\notag
=&n_1\frac{\left(-\frac{C_1}{\Lambda_\alpha^2}\right)f_1}{F_1}+n_2\frac{\left(-\frac{C_2}{\Lambda_\alpha^2}\right)f_2 -\left(-\frac{C_1}{\Lambda_\alpha^2}\right)f_1 }{F_2-F_1}
+n_3\frac{\left(-\frac{C_3}{\Lambda_\alpha^2}\right)f_3 -\left(-\frac{C_2}{\Lambda_\alpha^2}\right)f_2 }{F_3-F_2}
+n_4\frac{-\left(-\frac{C_3}{\Lambda_\alpha^2}\right)f_3 }{1-F_3}
\end{align}
Since $E(n_1) = nF_1$, $E(n_2) = n(F_2-F_1)$, $E(n_3) = n(F_3-F_2)$, $E(n_4) = n(1-F_3)$, we have
\begin{align}\notag
E(l^\prime) =&\left(-\frac{C_1}{\Lambda_\alpha^2}\right)f_1+\left(-\frac{C_2}{\Lambda_\alpha^2}\right)f_2 -\left(-\frac{C_1}{\Lambda_\alpha^2}\right)f_1
+\left(-\frac{C_3}{\Lambda_\alpha^2}\right)f_3 -\left(-\frac{C_2}{\Lambda_\alpha^2}\right)f_2
-\left(-\frac{C_3}{\Lambda_\alpha^2}\right)f_3  = 0
\end{align}

Next, we compute the Fisher Information,
\begin{align}\notag
l^{\prime\prime}=&n_1\frac{\left(-\frac{C_1}{\Lambda_\alpha^2}\right)^2f_1^\prime F_1 - \left(-\frac{C_1}{\Lambda_\alpha^2}\right)^2(f_1)^2}{F_1^2}\\\notag
+&n_2\frac{\left[\left(-\frac{C_2}{\Lambda_\alpha^2}\right)^2f_2^\prime -\left(-\frac{C_1}{\Lambda_\alpha^2}\right)^2f_1^\prime \right](F_2-F_1) - \left[\left(-\frac{C_2}{\Lambda_\alpha^2}\right)f_2 -\left(-\frac{C_1}{\Lambda_\alpha^2}\right)f_1 \right]^2}{(F_2-F_1)^2}\\\notag
+&n_3\frac{\left[\left(-\frac{C_3}{\Lambda_\alpha^2}\right)^2f_3^\prime -\left(-\frac{C_2}{\Lambda_\alpha^2}\right)^2f_2^\prime \right](F_3-F_2) - \left[\left(-\frac{C_3}{\Lambda_\alpha^2}\right)f_3 -\left(-\frac{C_2}{\Lambda_\alpha^2}\right)f_2 \right]^2}{(F_3-F_2)^2}\\\notag
+&n_4\frac{\left[ -\left(-\frac{C_3}{\Lambda_\alpha^2}\right)^2f_3^\prime \right](1-F_3) - \left[-\left(-\frac{C_3}{\Lambda_\alpha^2}\right)f_3 \right]^2}{(1-F_3)^2}
\end{align}

\begin{align}\notag
&-I = E(l^{\prime\prime})=\frac{\left(-\frac{C_1}{\Lambda_\alpha^2}\right)^2f_1^\prime F_1 - \left(-\frac{C_1}{\Lambda_\alpha^2}\right)^2(f_1)^2}{F_1}\\\notag
+&\frac{\left[\left(-\frac{C_2}{\Lambda_\alpha^2}\right)^2f_2^\prime -\left(-\frac{C_1}{\Lambda_\alpha^2}\right)^2f_1^\prime \right](F_2-F_1) - \left[\left(-\frac{C_2}{\Lambda_\alpha^2}\right)f_2 -\left(-\frac{C_1}{\Lambda_\alpha^2}\right)f_1 \right]^2}{(F_2-F_1)}\\\notag
+&\frac{\left[\left(-\frac{C_3}{\Lambda_\alpha^2}\right)^2f_3^\prime -\left(-\frac{C_2}{\Lambda_\alpha^2}\right)^2f_2^\prime \right](F_3-F_2) - \left[\left(-\frac{C_3}{\Lambda_\alpha^2}\right)f_3 -\left(-\frac{C_2}{\Lambda_\alpha^2}\right)f_2 \right]^2}{(F_3-F_2)}\\\notag
+&\frac{\left[ -\left(-\frac{C_3}{\Lambda_\alpha^2}\right)^2f_3^\prime \right](1-F_3) - \left[-\left(-\frac{C_3}{\Lambda_\alpha^2}\right)f_3 \right]^2}{(1-F_3)}\\\notag
=&-\frac{\left(-\frac{C_1}{\Lambda_\alpha^2}\right)^2(f_1)^2}{F_1}
-\frac{\left[\left(-\frac{C_2}{\Lambda_\alpha^2}\right)f_2 -\left(-\frac{C_1}{\Lambda_\alpha^2}\right)f_1 \right]^2}{(F_2-F_1)}
-\frac{\left[\left(-\frac{C_3}{\Lambda_\alpha^2}\right)f_3 -\left(-\frac{C_2}{\Lambda_\alpha^2}\right)f_2 \right]^2}{(F_3-F_2)}
-\frac{\left[-\left(-\frac{C_3}{\Lambda_\alpha^2}\right)f_3 \right]^2}{(1-F_3)}\\\notag
\end{align}

The asymptotic bias is
\begin{align}\notag
&E\left(\hat{\Lambda}_\alpha \right) = \Lambda_\alpha - \frac{E(l^\prime)^3+E(l^\prime l^{\prime\prime})}{2I^2} + O\left(\frac{1}{n^2}\right)
\end{align}

For convenience, we re-write $l^\prime$ and $l^{\prime\prime}$ as follows.
\begin{align}\notag
l^\prime=&\left[n_1-nF_1\right]\frac{\left(-\frac{C_1}{\Lambda_\alpha^2}\right)f_1}{F_1}+\left[n_2-n(F_2-F_1)\right]\frac{\left(-\frac{C_2}{\Lambda_\alpha^2}\right)f_2 -\left(-\frac{C_1}{\Lambda_\alpha^2}\right)f_1 }{F_2-F_1}\\\notag
+&\left[n_3-n(F_3-F_2)\right]\frac{\left(-\frac{C_3}{\Lambda_\alpha^2}\right)f_3 -\left(-\frac{C_2}{\Lambda_\alpha^2}\right)f_2 }{F_3-F_2}
+\left[n_4-n(1-F_3)\right]\frac{-\left(-\frac{C_3}{\Lambda_\alpha^2}\right)f_3 }{1-F_3}\\\notag
=&\sum_{i=1}^4 z_i p_i^\prime/p_i, \hspace{0.3in} \text{where } \ z_i = n_i - np_i, \hspace{0.1in} p_i^\prime = \frac{\partial p_i}{\partial \Lambda_\alpha}
\end{align}
\begin{align}\notag
l^{\prime\prime}=&\left[n_1-nF_1\right]\frac{\left(-\frac{C_1}{\Lambda_\alpha^2}\right)^2f_1^\prime F_1 - \left(-\frac{C_1}{\Lambda_\alpha^2}\right)^2(f_1)^2}{F_1^2}\\\notag
+&\left[n_2-n(F_2-F_1)\right]\frac{\left[\left(-\frac{C_2}{\Lambda_\alpha^2}\right)^2f_2^\prime -\left(-\frac{C_1}{\Lambda_\alpha^2}\right)^2f_1^\prime \right](F_2-F_1) - \left[\left(-\frac{C_2}{\Lambda_\alpha^2}\right)f_2 -\left(-\frac{C_1}{\Lambda_\alpha^2}\right)f_1 \right]^2}{(F_2-F_1)^2}\\\notag
+&\left[n_3-n(F_3-F_2)\right]\frac{\left[\left(-\frac{C_3}{\Lambda_\alpha^2}\right)^2f_3^\prime -\left(-\frac{C_2}{\Lambda_\alpha^2}\right)^2f_2^\prime \right](F_3-F_2) - \left[\left(-\frac{C_3}{\Lambda_\alpha^2}\right)f_3 -\left(-\frac{C_2}{\Lambda_\alpha^2}\right)f_2 \right]^2}{(F_3-F_2)^2}\\\notag
+&\left[n_4-n(1-F_3)\right]\frac{\left[ -\left(-\frac{C_3}{\Lambda_\alpha^2}\right)^2f_3^\prime \right](1-F_3) - \left[-\left(-\frac{C_3}{\Lambda_\alpha^2}\right)f_3 \right]^2}{(1-F_3)^2}\\\notag
-&n\frac{\left(-\frac{C_1}{\Lambda_\alpha^2}\right)^2(f_1)^2}{F_1}
-n\frac{ \left[\left(-\frac{C_2}{\Lambda_\alpha^2}\right)f_2 -\left(-\frac{C_1}{\Lambda_\alpha^2}\right)f_1 \right]^2}{(F_2-F_1)}
-n\frac{ \left[\left(-\frac{C_3}{\Lambda_\alpha^2}\right)f_3 -\left(-\frac{C_2}{\Lambda_\alpha^2}\right)f_2 \right]^2}{(F_3-F_2)}
-n\frac{\left[-\left(-\frac{C_3}{\Lambda_\alpha^2}\right)f_3 \right]^2}{(1-F_3)}\\\notag
=& \sum_{i=1}^4 z_i \frac{p_i^{\prime\prime}p_i - (p_i^\prime)^2}{p_i^2} - I ,\hspace{0.2in} \text{where } \ p_i^{\prime\prime} =\frac{\partial^2 p_i}{\partial \Lambda_\alpha^2}\end{align}
We will take advantage of the central comments of multinomial:
\begin{align}\notag
&E( (n_i - np_i)^2) = np_i(1-p_i)\\\notag
&E( (n_i - np_i)(n_j-np_j)) = -np_ip_j \hspace{0.2in} (i\neq j)\\\notag
&E( (n_i - np_i)^3) = np_i(1-p_i)(1-2p_i)\\\notag
&E( (n_i - np_i)^2(n_j-np_j)) = -np_ip_j(1-2p_i) \hspace{0.2in} (i\neq j)\\\notag
&E( (n_i - np_i)(n_j-np_j)(n_k-np_k)) = 2np_ip_jp_k \hspace{0.2in} (i\neq j\neq k)\\\notag
\end{align}
and the following expansion,
\begin{align}\notag
(a+b+c+d)^2 =& a^3+b^3+c^3+d^3+3a^2b+3a^2c+3a^2d+3ab^2+3b^2c+3b^2d+3ac^2+3bc^2\\\notag
&+3c^2d+3ad^2+3bd^2+3cd^2+6abc+6abd+6acd+6bcd
\end{align}

We are now ready to compute $E(l^\prime)^3$. Because
\begin{align}\notag
l^\prime=\sum_{i=1}^4 z_i p_i^\prime/p_i, \hspace{0.3in} \text{where } \ z_i = n_i - np_i, \hspace{0.1in} p_i^\prime = \frac{\partial p_i}{\partial \Lambda_\alpha}
\end{align}
we need to compute
\begin{align}\notag
&E\left( \left(z_1\frac{p_1^\prime}{p_1}\right)^2\left( z_2\frac{p_2^\prime}{p_2} +z_3\frac{p_3^\prime}{p_3} + z_4\frac{p_4^\prime}{p_4} \right) \right)\\\notag
=&-np_1(1-2p_1)\frac{(p_1^\prime)^2}{p_1^2}\left(p_2\frac{p_2^\prime}{p_2} + p_3\frac{p_3^\prime}{p_3}+ p_4\frac{p_4^\prime}{p_4}\right)\\\notag
=&-n(1-2p_1)\frac{(p_1^\prime)^2}{p_1}\left(p_2^\prime + p_3^\prime+ p_4^\prime\right)\\\notag
=& n(1-2p_1)\frac{(p_1^\prime)^3}{p_1}
\end{align}
and
\begin{align}\notag
&E\left( z_1\frac{p_1^\prime}{p_1}  z_2\frac{p_2^\prime}{p_2} z_3\frac{p_3^\prime}{p_3} \right)
=2np_1^\prime p_2^\prime p_3^\prime
\end{align}
Thus
\begin{align}\notag
E(l^\prime)^3 =& \sum_{i=1}^4 n (1-p_i)(1-2p_i)\frac{(p_i^\prime)^3}{p_i^2}+ 3 \sum_{i=1}^4n(1-2p_i)\frac{(p_i^\prime)^3}{p_i}\\\notag
+&12n p_1^\prime p_2^\prime p_3^\prime + 12n p_1^\prime p_2^\prime p_4^\prime + 12n p_2^\prime p_3^\prime p_4^\prime + 12n p_1^\prime p_3^\prime p_4^\prime \\\notag
=&n\sum_{i=1}^4\frac{(p_i^\prime)^3}{p_i^2}-4n\sum_{i=1}^4 (p_i^\prime)^3
+12n p_1^\prime p_2^\prime p_3^\prime + 12n p_1^\prime p_2^\prime p_4^\prime + 12n p_2^\prime p_3^\prime p_4^\prime + 12n p_1^\prime p_3^\prime p_4^\prime \\\notag
\end{align}
Next, we compute $E\left(l^\prime l^{\prime\prime}\right)$.
\begin{align}\notag
E\left(l^\prime l^{\prime\prime}\right) =& E\left[\left(\sum_{i=1}^4 z_i \frac{p_i^\prime}{p_i}\right) \left(\sum_{i=1}^4 z_i \frac{p_i^{\prime\prime}p_i - (p_i^\prime)^2}{p_i^2}\right)\right]\\\notag
=&E\left[\sum_{i=1}^4 z_i^2 \frac{p_i^{\prime\prime}p_i^\prime p_i - (p_i^\prime)^3}{p_i^3} + \sum_{i\neq j} z_iz_j\frac{p_i^\prime}{p_i}\frac{p_j^{\prime\prime}p_j - (p_j^\prime)^2}{p_j^2} \right]\\\notag
=&n\sum_{i=1}^4 (1-p_i)\frac{p_i^{\prime\prime}p_i^\prime p_i - (p_i^\prime)^3}{p_i^2} -n\sum_{i\neq j} p_i^\prime\frac{p_j^{\prime\prime}p_j - (p_j^\prime)^2}{p_j}\\\notag
=&n\sum_{i=1}^4 \frac{p_i^{\prime\prime}p_i^\prime p_i - (p_i^\prime)^3}{p_i^2}-n\sum_{i=1}^4 p_i^{\prime\prime}p_i^\prime + n\sum_{i=1}^4 \frac{(p_i^\prime)^3}{p_i} -n\sum_{i\neq j} p_i^\prime p_j^{\prime\prime} +n\sum_{i\neq j}^4 \frac{p_i^\prime(p_j^\prime)^2}{p_j}\\\notag
=&n\sum_{i=1}^4 \frac{p_i^{\prime\prime}p_i^\prime}{p_i} - n\sum_{i=1}^4\frac{(p_i^\prime)^3}{p_i^2}+ n\left(\sum_{i=1}^4 p_i^\prime\right)\left(\sum_{i=1}^4\frac{(p_i^\prime)^2}{p_i}\right)-n\left(\sum_{i=1}^4p_i^\prime\right)\left(\sum_{i=1}^4 p_i^{\prime\prime}\right)\\\notag
=&n\sum_{i=1}^4 \frac{p_i^{\prime\prime}p_i^\prime}{p_i} - n\sum_{i=1}^4\frac{(p_i^\prime)^3}{p_i^2}
\end{align}
\begin{align}\notag
E\left(l^\prime l^{\prime\prime}\right) + E\left(l^\prime \right)^3 =& n\sum_{i=1}^4 \frac{p_i^{\prime\prime}p_i^\prime}{p_i} -4n\sum_{i=1}^4 (p_i^\prime)^3
+12n p_1^\prime p_2^\prime p_3^\prime + 12n p_1^\prime p_2^\prime p_4^\prime + 12n p_2^\prime p_3^\prime p_4^\prime + 12n p_1^\prime p_3^\prime p_4^\prime \\\notag
=&n\sum_{i=1}^4 \frac{p_i^{\prime\prime}p_i^\prime}{p_i}
\end{align}
To see this, we can use the fact that $1 = \sum_{i=1}^4 p_i$, $0  = \sum_{i=1}^4 p_i^\prime $, and
\begin{align}\notag
0 = \left(\sum_{i=1}^4 p_i^\prime\right)^3 =& \sum_{i=1}^4 (p_i^\prime)^3 + 3\sum_{i=1}^4 (p_i^\prime)^2(-p_i^\prime) + 6p_1^\prime p_2^\prime p_3^\prime +6 p_1^\prime p_2^\prime p_4^\prime + 6p_2^\prime p_3^\prime p_4^\prime + 6 p_1^\prime p_3^\prime p_4^\prime\\\notag
=&-2\sum_{i=1}^4 (p_i^\prime)^3 + 6p_1^\prime p_2^\prime p_3^\prime +6 p_1^\prime p_2^\prime p_4^\prime + 6p_2^\prime p_3^\prime p_4^\prime + 6 p_1^\prime p_3^\prime p_4^\prime
\end{align}
Therefore,
\begin{align}\notag
&E\left(l^\prime l^{\prime\prime}\right) + E\left(l^\prime \right)^3 =n\sum_{i=1}^4 \frac{p_i^{\prime\prime}p_i^\prime}{p_i} \\\notag
=&n \frac{\left(-\frac{C_1}{\Lambda_\alpha^2}\right)f_1 \left[2\frac{C_1}{\Lambda_\alpha^3}f_1 + \left(-\frac{C_1}{\Lambda_\alpha^2}\right)^2f_1^\prime\right]}{F_1}\\\notag
+&n \frac{\left[\left(-\frac{C_2}{\Lambda_\alpha^2}\right)f_2 -\left(-\frac{C_1}{\Lambda_\alpha^2}\right)f_1 \right] \left[2\frac{C_2}{\Lambda_\alpha^3}f_2 + \left(-\frac{C_2}{\Lambda_\alpha^2}\right)^2f_2^\prime- 2\frac{C_1}{\Lambda_\alpha^3}f_1 - \left(-\frac{C_1}{\Lambda_\alpha^2}\right)^2f_1^\prime\right]}{F_2-F_1}\\\notag
+&n \frac{\left[\left(-\frac{C_3}{\Lambda_\alpha^2}\right)f_3 -\left(-\frac{C_2}{\Lambda_\alpha^2}\right)f_2 \right] \left[2\frac{C_3}{\Lambda_\alpha^3}f_3 + \left(-\frac{C_3}{\Lambda_\alpha^2}\right)^2f_3^\prime- 2\frac{C_2}{\Lambda_\alpha^3}f_2 - \left(-\frac{C_2}{\Lambda_\alpha^2}\right)^2f_2^\prime\right]}{F_3-F_2}\\\notag
+&n \frac{\left(-\frac{C_3}{\Lambda_\alpha^2}\right)f_3  \left[ 2\frac{C_3}{\Lambda_\alpha^3}f_3 + \left(-\frac{C_3}{\Lambda_\alpha^2}\right)^2f_3^\prime\right]}{1-F_3}
\end{align}
Because
\begin{align}\notag
I = &n\frac{\left(-\frac{C_1}{\Lambda_\alpha^2}\right)^2(f_1)^2}{F_1}
+n\frac{ \left[\left(-\frac{C_2}{\Lambda_\alpha^2}\right)f_2 -\left(-\frac{C_1}{\Lambda_\alpha^2}\right)f_1 \right]^2}{(F_2-F_1)}
+n\frac{ \left[\left(-\frac{C_3}{\Lambda_\alpha^2}\right)f_3 -\left(-\frac{C_2}{\Lambda_\alpha^2}\right)f_2 \right]^2}{(F_3-F_2)}
+n\frac{\left[-\left(-\frac{C_3}{\Lambda_\alpha^2}\right)f_3 \right]^2}{(1-F_3)}
\end{align}
and
\begin{align}\notag
\frac{E\left(l^\prime l^{\prime\prime}\right) + E\left(l^\prime \right)^3}{2I^2} = \frac{\Lambda_\alpha}{n}\left(-\frac{1}{B}+\frac{D}{2B^2}\right)
\end{align}
we have
\begin{align}\notag
E\left(\hat{\Lambda}_\alpha\right)  = \Lambda_\alpha - \frac{\Lambda_\alpha}{n}\left(-\frac{1}{B}+\frac{D}{2B^2}\right) + O\left(\frac{1}{n^2}\right)
=\Lambda_\alpha\left(1+\frac{1}{nB}-\frac{D}{2nB^2}\right)+ O\left(\frac{1}{n^2}\right)
\end{align}
which leads to a bias-corrected estimator
\begin{align}\notag
\hat{\Lambda}_{\alpha,c}=\frac{\hat{\Lambda}_\alpha}{1+\frac{1}{nB}-\frac{D}{2nB^2}}
\end{align}
where
\begin{align}\notag
B = &\frac{\left(-\frac{C_1}{\Lambda_\alpha}\right)^2f_1^2}{F_1}
+\frac{ \left[\left(-\frac{C_2}{\Lambda_\alpha}\right)f_2 -\left(-\frac{C_1}{\Lambda_\alpha}\right)f_1 \right]^2}{F_2-F_1}
+\frac{ \left[\left(-\frac{C_3}{\Lambda_\alpha}\right)f_3 -\left(-\frac{C_2}{\Lambda_\alpha}\right)f_2 \right]^2}{F_3-F_2}
+\frac{\left(-\frac{C_3}{\Lambda_\alpha}\right)^2f_3^2}{1-F_3}
\end{align}
and
\begin{align}\notag
D=& \frac{\left(-\frac{C_1}{\Lambda_\alpha}\right)^3f_1f_1^\prime}{F_1}
+ \frac{\left[\left(-\frac{C_2}{\Lambda_\alpha}\right)f_2 -\left(-\frac{C_1}{\Lambda_\alpha}\right)f_1 \right]
\left[ \left(-\frac{C_2}{\Lambda_\alpha}\right)^2f_2^\prime - \left(-\frac{C_1}{\Lambda_\alpha}\right)^2f_1^\prime\right]}{F_2-F_1}\\\notag
+& \frac{\left[\left(-\frac{C_3}{\Lambda_\alpha}\right)f_3 -\left(-\frac{C_2}{\Lambda_\alpha}\right)f_2 \right]
\left[\left(-\frac{C_3}{\Lambda_\alpha}\right)^2f_3^\prime - \left(-\frac{C_2}{\Lambda_\alpha}\right)^2f_2^\prime\right]}{F_3-F_2}
+\frac{\left(-\frac{C_3}{\Lambda_\alpha}\right)^3f_3  f_3^\prime}{1-F_3}
\end{align}

{
\bibliographystyle{abbrv}
\bibliography{../bib/mybibfile}

}

\end{document}